\documentclass[%
nofootinbib,
 amsmath,amssymb,
 aps,
 jcp,
]{revtex4-2}

\usepackage{graphicx}
\usepackage{dcolumn}

\usepackage{bm}
\usepackage[english]{babel}

\usepackage[utf8]{inputenc}
\usepackage[T1]{fontenc}
\usepackage{mathtools}
\usepackage{placeins}

\usepackage{lipsum}

\usepackage{letltxmacro}

\LetLtxMacro{\ORIGselectlanguage}{\selectlanguage}
\makeatletter
\DeclareRobustCommand{\selectlanguage}[1]{%
  \@ifundefined{alias@\string#1}
    {\ORIGselectlanguage{#1}}
    {\begingroup\edef\x{\endgroup
       \noexpand\ORIGselectlanguage{\@nameuse{alias@#1}}}\x}%
}
\newcommand{\definelanguagealias}[2]{%
  \@namedef{alias@#1}{#2}%
}
\makeatother

\definelanguagealias{en}{english}

\usepackage{physics}
\usepackage{subcaption}

\newcommand{\rref}[1]{Eq.\ (\ref{#1})}
\newcommand{\rrefsa}[1]{Eqs.\ (\ref{#1})}
\newcommand{\rrefsb}[1]{(\ref{#1})}
\newcommand{\bd}[1]{\textbf{#1}} 
\newcommand{\un}[1]{\underline{#1}} 
\newcommand{\lrr}[1]{\left({#1}\right)} 
\newcommand{\lrs}[1]{\left[{#1}\right]} 
\newcommand{\bA}{{\bf A}}
\newcommand{\ubA}{\underline{{\bf A}}}
\newcommand{\ubJ}{\underline{\textbf{J}}}
\newcommand{\bB}{{\bf B}}
\newcommand{\ubB}{\underline{{\bf B}}}
\newcommand{\bM}{{\bf M}}

\newcommand{\bu}{{\bf u}}
\newcommand{\bv}{{\bf v}}
\newcommand{\bx}{{\bf x}}
\newcommand{\br}{{\bf r}}

\newcommand{\bs}{{\bf s}}
\newcommand{\ba}{{\bf a}}

\newcommand{\be}{{\bf e}}
\newcommand{\bI}{{\bf I}}
\newcommand{\bJ}{{\bf J}}
\newcommand{\Jij}{{\bf J}_{ij}}
\newcommand{\Jpa}{\textbf{J}_{pa}}

\newcommand{\uJij}{{\underline{\bf J}}_{ij}}
\newcommand{\uJpp}{{\underline{\bf J}}_{pp}}

\newcommand{\bL}{{\bf L}}

\newcommand{\adj}{\mbox{adj}}
\newcommand{\ur}[1]{\mathrm{#1}}
\newcommand{\Rintegral}{\int_{-\infty}^{+\infty} \ur{d}\textbf{r} \,} 
\newcommand{\nr}{\addtocounter{equation}{1}\tag{\theequation}} 
\newcommand{\erfi}{\ur{erfi}} 
\usepackage{bm}
\usepackage{mathtools}
\usepackage{physics}
\newcommand*\diff{\mathop{}\!\mathrm{d}}

\newcommand{\mx}[1]{\mathbf{#1}}
\newcommand{\phikl[1]}{\matrixel{\phi_k}{#1}{\phi_l}}

\usepackage[unicode]{hyperref}
\hypersetup{
   unicode=true,          
   plainpages=false,
   colorlinks=true,       
   linkcolor=black,          
   citecolor=black,        
}

\begin{abstract}
 \noindent
Methods for calculating lower bounds to the exact energy using the variance of the upper bound energy are discussed and explored. All the matrix elements of the Hamiltonian squared are collected and considered, and those for which no known solutions could be found in the literature are derived for an explicitly correlated Gaussian (ECG) basis set. Analytical Solutions are determined for two-electron, mono-nuclear systems, in addition to a one-dimensional integral expression which has use in polyatomic calculations. The newly derived integral expressions have been implemented in the integral library of the QUANTEN computer program.

\end{abstract}

\begin{document}

\title{Integrals for lower bounds to the exact energy} 
\author{Robbie T. Ireland} \affiliation{Institute of Chemistry, ELTE, Eötvös Loránd University, 
Pázmány Péter sétány 1/A, Budapest, H-1117, Hungary}
\affiliation{School of Chemistry, University of Glasgow,
University Avenue, G12 8QQ, Glasgow, United Kingdom}

\maketitle

\section{Introduction}  \label{sec:intro}
\noindent
Since the successful application of Schrödinger equation to the hydrogen atom in the mid 1920's, few-body systems have been of great interest to physicists and chemists. The ability to describe few-particle systems such as small atoms and molecules however becomes problematic due to a lack of analytical solutions. As such, quantum chemistry requires the use of approximations in order to simplify quantum problems, allowing for approximate solutions to be obtained for the physical properties of the systems in question. 

In atomic and molecular physics, one property of great interest is the energy of different eigenstates. The energy of a particular state can be calculated by approximating the wave function through linear and non-linear parameterisation. This approximate wave function is called the trial wave function for the system, and it can be shown that the energy expectation value associated with this trial wave function serves as an upper bound to the exact energy\cite{Ritz1909, MacDonalnd1933}.  Calculating `tight' upper bounds to the exact energy is well known and occurs frequently in modern quantum chemistry. However, the upper-bound value itself does not give information on how close it is to the exact energy.  If a lower bound could also be calculated then this, along with the upper-bound value, would give an interval within which the exact energy can be found. Of course, the lower-bound value is only meaningful if the lower bound is of the same quality as the upper bound. 

As is true with upper bounds, improving the basis set with respect to the variational principle improves the lower-bound value, though the convergence of lower bounds is slower. Nevertheless, convergence of both upper and lower bounds leads to ever-tighter intervals, within which the exact energy lies. Two such methods for calculating lower bounds are the Weinstein criterion \cite{Weinstein1934} and Temple's lower bound \cite{Temple2007}, presented by D. Weinstein and G. Temple respectively. Weinstein's bound tends to give lower bounds that are of too low a quality compared to the upper bounds, whereas Temple's bound is regarded to give better quality lower bounds \cite{suzuki_stochastic_1998, ZsuzsannaToth2018}.  Both the Weinstein criterion and Temple's bound make use of the variance of the upper bound energy, ${\sigma_{E}}^2$, which requires the matrix elements of the Hamiltonian operator squared, $\hat{H}^2$. 

The trial wave function can be written as a linear combination of basis functions in the many-particle state. The choice of basis function can impact the ease of computation and the accuracy of the results, where the basis functions in this report are taken to be explicitly correlated Gaussian functions, ECGs. For a single ECG basis function, the correlation between all pairs of particles is considered which at times can return more accurate results compared to the widely used Slater determinant basis.  

Many of the analytical expressions of the matrix elements of $\hat{H^2}$ are already available \cite{Stanke2016, Zaklama2020, Ferenc2020}, but some contain integrals that require further consideration; the evaluation of these integrals has been the main focus of this research. Whilst the primary focus of these integrals is to calculate lower bounds to the exact energy, the integrals also have relevance in other applications, such as in relativistic calculations. As such, the derived expressions have been implemented in the QUANTEN computer program \cite{MolecularQuantumDynamicsResearchGroup} .

This report will first give an overview of the relevant literature, including an account of upper bounds in quantum mechanics (Sec. \ref{subsec:upperbounds}) and the  variational method (Sec. \ref{sec:svm}). The use of upper bounds in calculating lower bounds is then discussed (Sec. \ref{sec:LowerBound}) and a description of ECG basis sets, the basis functions of this report, is then provided (Sec. \ref{sec:ECG}). Sections \ref{sec:notations} and \ref{sec:comexp} provide a list of notations and commonly used expressions used in the derivations. Section \ref{Sec:IntHamaSq} provides a description of both the electronic and molecular Hamiltonians, as well as their constituent operators. The necessary integrals are calculated in Sec. \ref{ElecHamInt}, which are then written to describe the $n=2$-electron case in Sec. \ref{sec:neqtwo}. The integrals for which no solutions could be found in literature are then solved analytically in Sec. \ref{sec:MatrixEl}. An outlook for the use of these integrals in obtaining numerical results for lower bounds, and their other uses in quantum chemistry, is then provided in Sec. \ref{sec:conc}. 

\section{Theoretical Background}

\subsection{Introduction to Upper and Lower Bounds} \label{subsec:Motivations}
\noindent
Consider a set $P$ (this could be, for example, the set of real numbers, the set of integers etc...); we define some subset of $P$ to be the subset $S$ and let $x$ be the elements of $S$. Then, we define any \textbf{upper bound} of $S$ to be an element $a \in P$ such that 
\begin{align*}
    a \ge x \ ,
\end{align*}
for all $x$ $\in$ $S$. Note that there can be many different values of $a$, all of which are contained in the set $P$. For example, if we have the set of numbers $ S = \{2,3,4\} $ which is a subset of the integers, $P=\mathbb{I}$, then the integers 5, 6, and 7 could all be considered as upper bounds to this set. The supremum of the subset $S$, $\ur{sup}(S)$, is the lowest upper bound that is greater than or equal to all of the values of the subset $S$; if $a$ describes all the possible upper bounds to $S$, then the supremum of $S$ is the value $y$ such that
\begin{align*}
    \ur{sup}(S) = y \le a \hspace{0.5cm} \ur{for} \hspace{0.5cm}  y \in P \ . \nr
\end{align*}
For example, if we consider the same subset of the integers $\{2,3,4\}$, then the supremum would be 4, as this is the least element of the integers that is no less than any of the elements of the subset.

We can also define any \textbf{lower bound} to $S$ to be an element $b \in P$ such that 
\begin{align*}
    b \le x \ , 
\end{align*}
for all $x$ $\in$ $S$. Again, there can be many different values for $b$, all of which belong to the set $P$. If $b$ describes all the possible lower bounds to $S$, the infimum of the subset $S$, $\ur{inf}(S)$ is then defined to be the value $z$ such that
\begin{align*}
    \ur{inf}(S) = z \ge b \hspace{0.5cm} \ur{for} \hspace{0.5cm} z \in P \ , \nr
\end{align*}
which explicitly states that the infimum is the greatest lower bound. 

As shall be discussed in the following sections, we can apply upper and lower bounds to characterise the numerical results obtained through approximating the exact energy.
\subsection{Upper Bounds In Quantum Mechanics} \label{subsec:upperbounds} 
\noindent
Quantum-mechanical systems can be well approximated by a trial wave function $\ket{\Psi_{\ur{Trial}}}$ which is an approximation to the exact wave function for the system. The energy expectation value corresponding to our trial function is the functional $E[\Psi_{\ur{Trial}}]$, defined by the Rayleigh quotient\cite{Atkins2018,Horn1985,Bransden2000}, 
\begin{align}
    E[\Psi_{\ur{Trial}}] = \frac{\matrixel{\Psi_{\ur{Trial}}}{\hat{H}}{\Psi_{\ur{Trial}}}}{\braket{\Psi_{\ur{Trial}}}{\Psi_{\ur{Trial}}}} \ . \label{eq:expectation}
\end{align}
From now on, we will write $E[\Psi_{\ur{Trial}}]$ as $E$ for clarity. Let $\ket{\Psi_{\ur{Trial}}}$ be the trial wave function for some system with Hamiltonian $\hat{H}$. We can then expand the trial wave function in terms of the exact eigenfunctions of the Hamiltonian,
\begin{align}
    \ket{\Psi_{\ur{Trial}}} = \sum_{k = 1} c_k \ket{\Psi_k} \nr \ , \label{eq:TrailExact}
\end{align}
where $\ket{\Psi_k}$ are the exact eigenfunctions with coefficients $c_k \in \mathbb{C}$, and $k$ is some positive integer. The corresponding exact energy eigenvalues of the system are then denoted $E_k$, where the lowest eigenvalue corresponds to the ground-state energy, $E_1$. Then, by writing the Rayleigh quotient, \rref{eq:expectation}, in terms of the linear combination in \rref{eq:TrailExact},
\begin{align*}
    E = \frac{ \matrixel{\sum_{k = 1} c_k \Psi_k }{\hat{H}}{{\sum_{l = 1} c_l} \Psi_l} }{\braket{\sum_{k = 1} c_k \Psi_k}{\sum_{l = 1} c_l \Psi_l}} 
    = \frac{ \sum_{k, \, l = 1} c_k^* c_l \matrixel{\Psi_k }{\hat{H}}{\Psi_l }}{\sum_{k, \, l = 1} c_k^* c_l \braket{\Psi_k}{\Psi_l}} \ , \nr
\end{align*}
noting that $\matrixel{\Psi_k}{\hat{H}}{\Psi_l} = E_l \braket{\Psi_k}{\Psi_l}$, where $l$ is also a positive integer. The basis functions are exact eigenfunctions of the Hamiltonian operator, which is a Hermetian operator. Hence, the basis functions are orthogonal to each other meaning $\braket{\Psi_k}{\Psi_l} = \delta_{kl} $ for properly-normalised functions, where $\delta_{kl}$ is the Kronecker delta. As such, the only non-zero term in the summation is the $k = l$ term. Then,
\begin{align*}
    E = \frac{\sum_{k = 1} |c_k|^2 E_{k}}{\sum_{k = 1} |c_k|^2} \ . \nr
\end{align*}
As the ground state energy is the lowest eigenvalue, the exact eigenvalue $E_k$ is such that $E_k \ge E_1$ for $k \ge 1$. Hence, it follows that
\begin{align*}
    E = \frac{\sum_{k = 1} |c_k|^2 E_{k}}{\sum_{k = 1} |c_k|^2} \ge \frac{\sum_{k = 1} |c_k|^2}{\sum_{k = 1} |c_k|^2}E_1 = E_1 \ , \nr
\end{align*}
which leads to
\begin{align*}
    E \ge E_1 \ . \nr \label{eq:ritztheorem}
\end{align*}
Equation (\ref{eq:ritztheorem}) tells us that the expectation value obtained from $\ket{\Psi_{\ur{Trial}}}$ is an upper bound to the exact ground-state energy. 
\subsection{The Variational Method} \label{sec:svm}
\noindent
The `tightness' of the upper bound (how close the upper bound is to the exact energy) is determined by the linear and non-linear parameters of the basis functions and is improved through the variational method. This involves creating a trial wave function of $N_{\ur{b}}$ basis functions, which we will take to be an approximation to one of the eigenfunctions of the Hamiltonian.  This basis set spans the subspace $\mathcal{V}_{N_{b}}$, a subspace of Hilbert space, and the energy expectation value is given by the Rayleigh quotient, \rref{eq:expectation}. By varying the linear and non-linear parameters, we minimise the Rayleigh quotient and in doing so find the parameters that return lowest value of $E$. 

The generalised Ritz theorem states that the expectation value of $\hat{H}$ corresponding to $\ket{\Psi_{\ur{Trial}}}$ is stationary in the neighbourhood of its eigenvalues \cite{suzuki_stochastic_1998, Mayer2003}. More explicitly, for any infinitesimal change in the trial wave function, $\ket{\delta\Psi_\ur{{Trial}}}$, the associated change in the energy expectation value is zero, $\delta E = 0$. So, from \rref{eq:expectation},
\begin{align*}
    \delta E &= \delta\lrs{\frac{\matrixel{\Psi_{\ur{Trial}}}{\hat{H}}{\Psi_{\ur{Trial}}}}{\braket{\Psi_{\ur{Trial}}}{\Psi_{\ur{Trial}}}}} = 0  \\
    &\implies \frac{\matrixel{\delta \Psi_{\ur{Trial}}}{\hat{H}}{\Psi_{\ur{Trial}}}\braket{\Psi_{\ur{Trial}}}{\Psi_{\ur{Trial}}} - \braket{\delta \Psi_{\ur{Trial}}}{\Psi_{\ur{Trial}}}\matrixel{\Psi_{\ur{Trial}}}{\hat{H}}{\Psi_{\ur{Trial}}}}{\braket{\Psi_{\ur{Trial}}}{\Psi_{\ur{Trial}}}^2} + c.c = 0 \ . \nr   \label{eq:eigenproof}
\end{align*}
The term $c.c$ represents the complex conjugate of the fraction above. In order to make both the real and imaginary parts equal to zero in \rref{eq:eigenproof}, both terms on the left hand side are made equal to zero independently. Hence, recognising the Rayleigh quotient in \rref{eq:eigenproof} and assuming that the trial wave function is normalised to one, we obtain 
\begin{align*}
    &\frac{\matrixel{\delta \Psi_{\ur{Trial}}}{\hat{H}}{\Psi_{\ur{Trial}}} - E\braket{\delta\Psi_{\ur{Trial}}}{\Psi_{\ur{Trial}}}}{\braket{\Psi_{\ur{Trial}}}} = 0 \\
    \implies &\matrixel{\delta \Psi_{\ur{Trial}}}{\hat{H} - E}{\Psi_{\ur{Trial}}} = 0 \ . \nr \label{eq:eigenproof1}
\end{align*} 

\subsubsection{Linear Variational Method}
\noindent
We now write the trial wavefunction $\ket{\Psi_{\ur{Trial}}}$ in terms of the linear combination 
\begin{align*}
    \ket{\Psi_{\ur{Trial}}} = \sum_{k = 1}^{N_{\ur{b}}} c_k \ket{\Phi_k} \nr \ , \label{eq:trial}
\end{align*}
for basis functions $\ket{\Phi_k}$, of which there are a total of $N_b$, and linear parameters (coefficients) $c_k \in \mathbb{C}$. We note that these basis functions are not necessarily orthogonal, though they can be made orthogonal to each other if necessary by employing an appropriate orthogonalisation procedure, such as the Gram-Schmidt orthogonalisation procedure \cite{Schmidt1907, James1992}. Inserting \rref{eq:trial} into \rref{eq:eigenproof1} then gives
\begin{align*}
    &\matrixel{\sum_{k = 1}^{N_{\ur{b}}} \delta c_k \Phi_k}{\hat{H} - E}{\sum_{l = 1}^{N_{\ur{b}}} c_l \Phi_l} 
    = \sum_{k=1}^{N_{b}} \sum_{l=1}^{N_{b}} \delta c_{k}^* \lrs{\matrixel{\Phi_k}{\hat{H}}{\Phi_l} - E \braket{\Phi_k}{\Phi_l}} c_l = 0 \ . \nr
\end{align*}
for positive integers $k$ and $l$. We then define the matrix elements
\begin{align*}
    H_{kl} &= \matrixel{\Phi_k}{\hat{H}}{\Phi_l} \, , \\
    S_{kl} &= \braket{\Phi_k}{\Phi_l} \ ,
\end{align*}
where  we recall that the basis functions are not necessarily orthogonal.  Using these matrix elements, 
\begin{align*}
    \sum_{k=1}^{N_{b}}\sum_{l=1}^{N_{b}} \delta c_{k}^* \lrs{ H_{kl} - E{S_{kl}} } c_{l}= 0 \ . \nr
\end{align*}
For any infinitesimal change in $c_k$, it must follow that
\begin{align*}
    \sum_{k=1}^{N_{b}}\sum_{l=1}^{N_{b}}\lrs{ H_{kl} - E{S_{kl}} } c_{l}= 0 \ . \nr
\end{align*}
By introducing matrix notation we arrive at the following matrix eigenvalue problem, also known as the Schr\"odinger equation for an overlapping basis,
\begin{align}
    \bd{H}\bd{c} = E\bd{S}\bd{c} \ , \label{eq:MolecSchroEq}
\end{align}
where $\bd{H}$ is the $N_{\ur{b}} \times N_{\ur{b}}$ Hamiltonian matrix, $\bd{S}$ is the $N_{\ur{b}} \times N_{\ur{b}}$ overlap matrix and $\bd{c}$ is an $N_{\ur{b}}$-dimensional vector with entries that are the coefficients of the linear combination in \rref{eq:trial}.

The parameters of the basis set can be optimised and used to  build the Hamilton matrix. We then diagonalise the Hamiltonian matrix which provides the eigenvalues and the coefficients, which are contained within the coefficient vector. These are then used to optimise the parameters again, and the process is repeated until the value of $E$ obtained does not improve upon further optimisation of the parameters, at which point the value $E$ has converged to what is taken to be the upper bound to the exact energy. 

There are various methods by which the parameters can be optimised. These include direct optimisation of the non-linear parameters, which involves calculating the gradient of the Rayleigh quotient, or stochastic optimisation of the non-linear parameters \cite{Mitroy2013}. 

\subsection{Lower Bounds in Quantum Mechanics} \label{sec:LowerBound}
\begin{figure}[ht]
    \centering
    \includegraphics[scale=0.7]{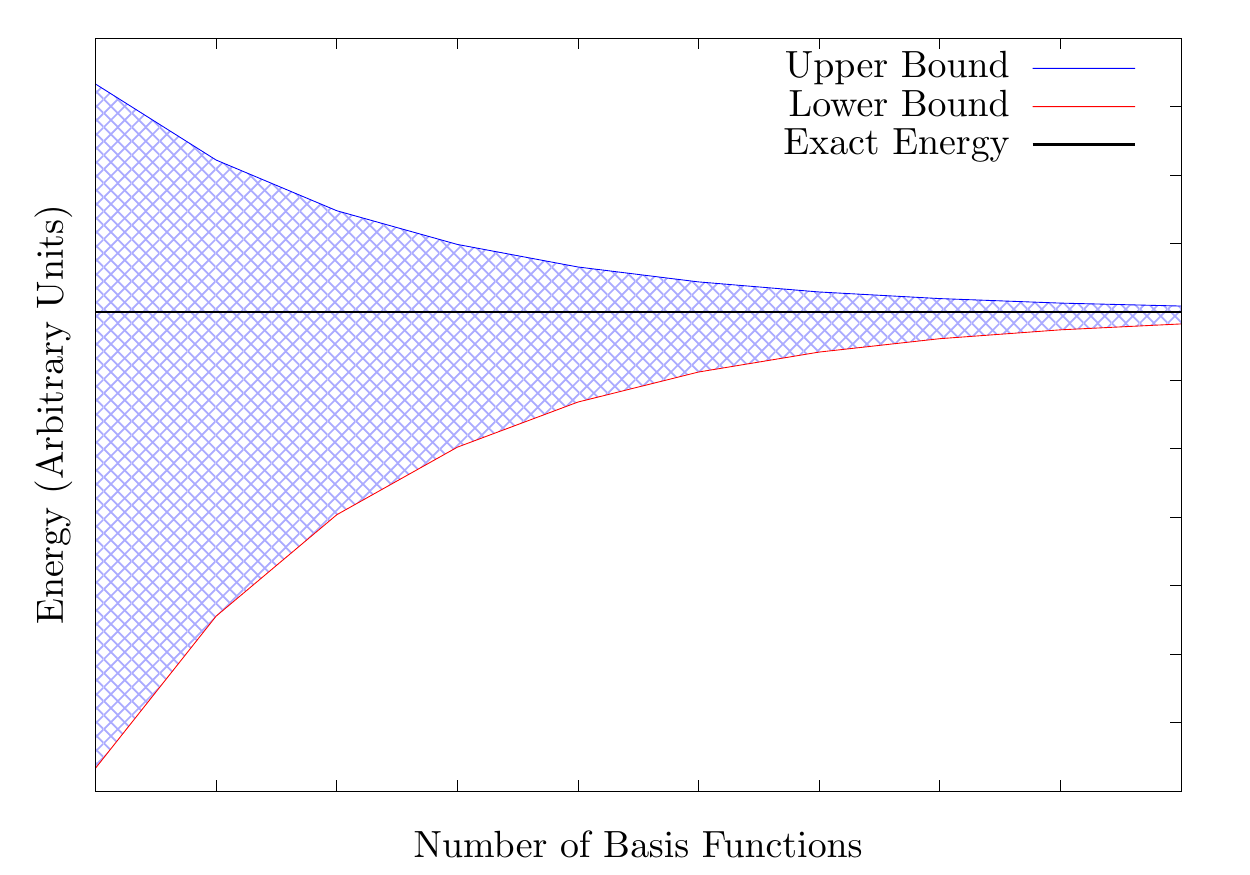}
    \caption{An Illustrative figure for the motivation behind the present work. The thin red line shows the concept of a lower bound to the exact energy, and the shaded area shows the energy range within which the exact energy resides.
    }    
    \label{fig:UpLowFig}
\end{figure}

\noindent
One of the main motivations behind quantum mechanics in chemistry is to understand the stability of the atom, and it has been shown that lower bounds to the exact energy are a requirement for this stability \cite{Lieb2005}.  We are interested in these lower bounds, but from a different perspective. We have discussed a method for calculating and improving upper bounds to the exact energy, where the iterative procedure results in the upper bound converging to a particular value. Upper bounds to the exact energy $E_{n}$ of a given state are readily computed using the variational method, and being able to quote these with lower-bound values would then give an interval within which the exact energy would lie in, as can be seen in figure \ref{fig:UpLowFig}. 

There are already fundamental theoretical results for calculating lower bounds of matrix eigenvalues. However, whilst upper bounds are readily calculated, good lower bounds to the exact energy are not as routinely computed \cite{Pollak2020}. The methods that this report will focus on make use of the variance of the energy expectation value, ${\sigma_E}^2$, which can be given as
\begin{align} 
    {\sigma_E}^2 = \frac{\matrixel{\Psi_{\ur{Trial}}}{(\hat{H}-E)^2}{\Psi_{\ur{Trial}}}}{\braket{\Psi_{\ur{Trial}}}{\Psi_{\ur{Trial}}}} = \frac{\matrixel{\Psi_{\ur{Trial}}}{\hat{H}^2}{\Psi_{\ur{Trial}}}}{\braket{\Psi_{\ur{Trial}}}{\Psi_{\ur{Trial}}}} - E^2 \label{eq:var} \ ,
\end{align}
where the upper bound to the exact energy, $E$, is defined as in \rref{eq:expectation} \cite{suzuki_stochastic_1998}. \\

\subsubsection{The Weinstein Criterion}
\noindent
One method for determining lower bounds to the exact energy is by using the Weinstein criterion \cite{Weinstein1934}. This is derived by using \rref{eq:var} and then writing the trial function in terms of the exact eigenfunctions of the system in question, \rref{eq:TrailExact}. The exact eigenfunctions form a complete orthonormal set, meaning that $\sum_{k, \, l = 1}\braket{\Psi_k}{\Psi_l} = \delta_{kl}$.  As such, we can write the variance of the upper bound energy ${\sigma_E}^2$, \rref{eq:var}, as 
\begin{align*}
    &{\sigma_E}^2 = \sum_{k, \, l = 1} c_k^* c_l \matrixel{\Psi_k}{(\hat{H}-E)^2}{\Psi_l} \\
    \implies &{\sigma_E}^2 = \sum_{k, \, l = 1} c_k^* c_l \matrixel{\Psi_k}{(\hat{H}^2 - 2E\hat{H} + E^2)}{\Psi_l} \\
    \implies &{\sigma_E}^2 = \sum_{k, \, l = 1} c_k^* c_l \lrs{{\matrixel{\Psi_k}{\hat{H}^2}{\Psi_l} - 2E\matrixel{\Psi_k}{\hat{H}}{\Psi_l} + E^2 \braket{\Psi_k}{\Psi_l}}} \ . \nr
\end{align*}
Using the fact that $\hat{H}\ket{\Psi_l} = E_l\ket{\Psi_l}$, where $E_l$ is the $l^{\ur{th}}$ exact eigenvalue, and the fact that the trial wave function is properly normalised,
\begin{align*}
    {\sigma_E}^2 = \sum_{k = 1} |c_k|^2  \lrs{E_k^2 - 2E E_k + E^2} = \sum_{k = 1} |c_k|^2 \lrs{(E_k - E)^2} \ . \nr \label{eq:WeinDer1}
\end{align*}
We now define $E_{n}$ to be the exact eigenvalue to which the upper bound $E$ is closest, meaning
\begin{align*}
    |E_n - E| \le |E_k - E| \nr \label{eq:WeinCon} \ ,
\end{align*}
where $k$ is an arbitrary positive integer. As such
\begin{align*}
    \sum_{k = 1} |c_k|^2(E_{n} - E)^2 \le \sum_{k = 1} |c_k|^2(E_k - E)^2 = {\sigma_E}^2 \ , \nr
\end{align*}
meaning, for $\sum_{k = 1} |c_k|^2 = 1$,
\begin{align*}
    \sigma_E \ge \pm(E_n - E) \ . \nr \label{eq:WeinDer2}
\end{align*}
Rearranging \rref{eq:WeinDer2} for the exact eigenvalue $E_n$ gives
\begin{align*}
    E - \sigma_E \le E_{n} \le E + \sigma_E \ . \nr \label{eq:WeinDer3}
\end{align*}
This leads to the Weinstein criterion \cite{suzuki_stochastic_1998,Weinstein1934, Goedecker1991}, which states that within the interval $[E - \sigma_E, E+\sigma_E]$ there is (as a minimum) one exact eigenvalue. In the case of a system with a ground-state energy $E_1$, the lower bound to this eigenvalue according to the Weinstein criterion would be 
\begin{align}
    E-\sigma_E \le E_1 \ \label{eq:WeinLB}.
\end{align}
so long as \rref{eq:WeinCon} is true.
\subsubsection{Temple's Bound}
\noindent
Another lower bound is Temple's bound\cite{Temple2007}, which also makes use of \rref{eq:var}; the original result can be expressed as
\begin{align}
    \matrixel{\Psi_{\ur{Trial}}}{\hat{H}}{\Psi_{\ur{Trial}}} - \frac{\matrixel{\Psi_{\ur{Trial}}}{\hat{H}^2}{\Psi_{\ur{Trial}}} -\matrixel{\Psi_{\ur{Trial}}}{\hat{H}}{\Psi_{\ur{Trial}}}^2}{\beta - \matrixel{\Psi_{\ur{Trial}}}{\hat{H}}{\Psi_{\ur{Trial}}}} \le E_1  \ ,  \label{HarrellLB}
\end{align}
where $\braket{\Psi_{\ur{Trial}}}{\Psi_{\ur{Trial}}}=1$ and $\beta$ is a scalar defined as
\begin{align*}
    \beta = \inf\lrs{\ur{sp}(\hat{H}) \, \backslash \, \{E_1\}} \, . \nr \label{eq:BetaDef}
\end{align*}
This can be interpreted explicitly as follows; the Hamiltonian $\hat{H}$ has a spectrum of energy eigenvalues $E_1$, $E_2$, ... which we denote as $\ur{sp}(\hat{H})$. Then, $\beta$ is the infimum of this set of eigenvalues (see Sec. \ref{subsec:Motivations}) excluding the ground-state energy, $E_1$. We can see that the numerator of \rref{HarrellLB} defines the variance of the energy expectation value, giving 
\begin{align}
    E - \frac{{\sigma_E}^2}{\beta - E} \le E_1 \, .  \label{TempLB}
\end{align}
This defines the lower bound to the ground-state energy eigenvalue provided that
\begin{align}
    E < \beta \le E_2. \label{betaLB}
\end{align}
We can prove \rref{TempLB} \cite{suzuki_stochastic_1998} provided that \rref{betaLB} is true by considering 
\begin{align*}
    E_1 - E + \frac{{\sigma_E}^2}{\beta - E} &= \frac{1}{\beta - E}\lrr{E_1 (\beta - E) - E (\beta - E) + {\sigma_E}^2} \\
    &= \frac{1}{\beta - E}\lrr{E_1\beta - E_1E - E\beta + E^2 + \frac{\matrixel{\Psi_{\ur{Trial}}}{\hat{H}^2}{\Psi_{\ur{Trial}}}}{\braket{\Psi_{\ur{Trial}}}{\Psi_{\ur{Trial}}}} -E^2} \ , \nr \label{eq:TempDer0}
\end{align*} 
where we have made use of \rref{eq:var} in the second line. We  can then simplify this expression by adding ${E_1}^2$ inside the round brackets of \rref{eq:TempDer0} and then subtracting it, giving 
\begin{align*}
    E_1 - E + \frac{{\sigma_E}^2}{\beta - E} &= \frac{1}{\beta - E}\lrr{E_1\beta - E_1E - E\beta + {E_1}^2 + \frac{\matrixel{\Psi_{\ur{Trial}}}{\hat{H}^2}{\Psi_{\ur{Trial}}}}{\braket{\Psi_{\ur{Trial}}}{\Psi_{\ur{Trial}}}} -{E_1}^2} \\
    &= \frac{1}{\beta - E}\lrr{(E_1 + \beta)(E_1 - E) + \frac{\matrixel{\Psi_{\ur{Trial}}}{\hat{H}^2}{\Psi_{\ur{Trial}}}}{\braket{\Psi_{\ur{Trial}}}{\Psi_{\ur{Trial}}}} -{E_1}^2} \\
    &= \frac{1}{\beta - E}\lrr{-(E_1 + \beta)(E - E_1) + \frac{\matrixel{\Psi_{\ur{Trial}}}{\hat{H}^2}{\Psi_{\ur{Trial}}}}{\braket{\Psi_{\ur{Trial}}}{\Psi_{\ur{Trial}}}} - {E_1}^2} \ . \nr \label{eq:TempDer1}
\end{align*}
By expanding the trial wave function in terms of the exact eigenfunctions $\ket{\Psi_k}$ using \rref{eq:TrailExact}, we can write
\begin{align*}
    \frac{\matrixel{\Psi_{\ur{Trial}}}{\hat{H}^2}{\Psi_{\ur{Trial}}}}{\braket{\Psi_{\ur{Trial}}}{\Psi_{\ur{Trial}}}} -{E_1}^2 &= \frac{\sum_{k, \, l = 1}c_k^* c_l \matrixel{\Psi_k}{\hat{ H}^2}{\Psi_l}}{\sum_{k, \, l = 1}c_k^* c_l\braket{\Psi_k}{\Psi_l}} - {E_1}^2 \\
    &= \frac{\sum_{k, \, l = 1}c_k^* c_l \matrixel{\Psi_k}{\hat{ H}E_l}{\Psi_l}}{\sum_{k, \, l = 1}c_k^* c_l\braket{\Psi_k}{\Psi_l}} - {E_1}^2 \\
    &=\frac{\sum_{k=1} |c_k|^2 {E_k}^2}{\sum_{k = 1}|c_k|^2} - {E_1}^2 \\
    &= \frac{\sum_{k = 2} |c_k|^2 ({E_k}^2 - {E_1}^2)}{\sum_{k = 1} |c_k|^2} \ . \nr \label{eq:TempDer2}
\end{align*}
Similarly, we can write
\begin{align*}
    &E - E_1 = \frac{\matrixel{\Psi_{\ur{Trial}}}{\hat{H}}{\Psi_{\ur{Trial}}}}{\braket{\Psi_{\ur{Trial}}}{\Psi_{\ur{Trial}}}} -{E_1}  =\frac{\sum_{k, \, l = 1 } c_k^* c_l \matrixel{\Psi_k}{\hat{ H}}{\Psi_l}}{\sum_{k, \, l = 1} c_k^* c_l\braket{\Psi_k}{\Psi_l}} - {E_1} \\
    &\implies E - E_1 = \frac{\sum_{k = 2} |c_k|^2(E_k - E_1)}{\sum_{k = 1} |c_k|^2} \ . \nr \label{eq:TempDer3}
\end{align*}
Substituting \rref{eq:TempDer2} and \rref{eq:TempDer3} into \rref{eq:TempDer1}, we have
\begin{align*}
    E_1 - E + \frac{{\sigma_E}^2}{\beta - E} &= \frac{1}{\beta - E} \lrs{\frac{\sum_{k=2} |c_k|^2\lrr{-(E_1 + \beta)(E_k - E_1) + ({E_k}^2 - {E_1}^2)}}{\sum_{k = 1} |c_k|^2}} \\
    & = \frac{1}{\beta - E} \lrs{\frac{\sum_{k=2} |c_k|^2 \lrr{ -E_k\beta  + E_1\beta - E_1E_k + {E_1}^2 - {E_1}^2 + {E_k}^2 } }{\sum_{k = 1} |c_k|^2}}  \\
    &= \frac{1}{\beta - E} \lrs{\frac{\sum_{k=2} |c_k|^2 (E_k - \beta) (E_k - E_1) }{\sum_{k = 1} |c_k|^2}} \ . \nr \label{eq:TempDer4}
\end{align*}
We know that $E_1$ is the lowest possible eigenvalue, and so $E_k - E_1 \ge 0 $. We then require that $E_k - \beta \ge 0$ (which is true for $\beta$ defined by \rref{betaLB} where $k \ge 2$) and that $\beta - E > 0$. Then,  provided that $E < \beta \le E_2 $, the right-hand side is non-negative and so
\begin{align*}
    E_1 - E + \frac{{\sigma_E}^2}{\beta - E} \ge 0 \implies E_1 \ge E - \frac{{\sigma_E}^2}{\beta - E} \ , \nr
\end{align*}
as required. 

\subsubsection{Further Notes on Lower Bounds} \label{sec:FurtherNotes}
\noindent
Like with upper bounds, increasing the size of the basis set results in the lower bounds of Weinstein and Temple converging to the exact energy, which can be shown by considering the variance of the upper bound energy, \rref{eq:var}. As the number of basis functions increases, $\ket{\Psi_{\ur{Trial}}}$  approaches the exact normalised wave function of the system being approximated, which results in the variance tending to zero. For zero variance in Eqs. (\ref{eq:WeinLB}) and (\ref{TempLB}), and using the fact the the upper bound energy $E$ approaches the exact energy, we see that both Weinstein's and Temple's lower bounds also converge to the exact energy. This convergence is slower than that of upper bounds due to the variance depending on the matrix elements of the Hamiltonian squared; upper bounds require the matrix elements of $\hat{H}$, the errors in which can be either positive or negative and so can cancel, leading to faster convergence. The errors in the matrix elements of $\hat{H}^2$, however, are positive and therefore compound, resulting in a slower convergence \cite{Pollak2019,Pollak2020,Nakashima2008}. 

A. F. Stevenson generalised the results of Weinstein and Temple with the expression 
\begin{align}
    E_S = \alpha - \sqrt{(\alpha - E)^2 + {\sigma_E}^2} < E_1 \hspace{0.5cm} \ur{if} \hspace{0.5cm} E_1 < \alpha < \frac{1}{2}(E_1 + E_2)  \ , \label{eq:StevGen}
\end{align}
where $E_S$ is the generalisation of the lower bound and $E_2$ is the exact energy of the first excited state, which is generally not known \cite{Stevenson1938, Stevenson1938b}. Weinstein's lower bound is derived for $\alpha = E$, and Temple's for $\alpha = \frac{1}{2}(E_S + E_2)$. Many other generalisations and variations of Temple's bound have been made, including methods involving the minimisation of the variance ${\sigma_E}^2$  by H. Kleindienst and W. Altmann \cite{Kleindienst1976} and by combining Temple's bound with the the inner projection method \cite{Bazley, Hill1979}. More recently, E. Pollak has made modifications to Temple's bound which have been applied to various model systems, in particular the quartic oscillator \cite{Pollak2019a,Pollak2019,Pollak2020,Nakashima2008}.

Lehmann bounds provide a set of bounds which can be thought of to `bracket' exact eigenvalues with upper and lower bounds, first proposed by N. J. Lehmann in 1949 \cite{Lehmann1949, Lehmann1950, Ovtchinnikov2011}. L\"owdin's bracketing function, proposed in 1965 by P. O. L\"owdin \cite{Per-OlovLowdin1965}, also acts in a similar bracketing fashion, which both the Weinstein Criterion and Temple's lower bound can be related to. The bracketing function is the function $E_L = f(E)$  such that the values $E$ and $E_L$ will bracket at least one exact energy eigenvalue of the Hamiltonian for a particular system. Hence, if $E$ is taken to be the upper bound, which is equal to $\matrixel{\Psi_{\ur{Trial}}}{\hat{H}}{\Psi_{\ur{Trial}}}$ for $\braket{\Psi_{\ur{Trial}}}{\Psi_{\ur{Trial}}} = 1$, then $E_L$ serves as the lower bound according to L\"owdin. The full form of the bracketing function is
\begin{align}
    E_L = f(E) = \bra{\Psi_{\ur{Trial}}}\hat{H}+\hat{H}\hat{T}\hat{H}\ket{\Psi_{\ur{Trial}}} \ ,  \label{eq:Lowdins}
\end{align}
where $\hat{H}$ is the Hamiltonian for the system. The operator $\hat{T}$ is known as the reduced resolvent which can be defined symbolically as \footnote{The full form of the reduced resolvent is $\hat{T} = \hat{P}\frac{1}{\alpha \hat{O} + \hat{P}(E - \hat{H})\hat{P}}\hat{P}$, where $\alpha$ is a non-zero scalar.}, 
\begin{align}
    \hat{T} = \frac{\hat{P}}{E - \hat{H}} \ ,  \label{eq:ReducedResolvent}
\end{align}
where  $\hat{P}$ is the projection operator defined as $\hat{P} = \hat{I} - \hat{O}$ with identity operator $\hat{I}$. The operator $\hat{O}$ is the associated projection operator, $\hat{O} = \ket{\Psi_{\ur{Trial}}}\bra{\Psi_{\ur{Trial}}}$. The difficulty in using the function $f(E)$ arises from the denominator of \rref{eq:ReducedResolvent} \cite{ZsuzsannaToth2018, Szabados2014}. 

L\"owdin's bracketing function is mostly used within the context of perturbation theory, where the Hamiltonian of a system $\hat{H}$ can be written in terms of an unperturbed Hamiltonian $\hat{H}_0$ and some perturbation $\hat{V}$ such that $\hat{H} = \hat{H}_0 + \hat{V} $. In the instances where the basis set is very large, such as when using Slater determinants, perturbation theory can be used to estimate the eigenvalues and so upper and lower bounds can provide useful information \cite{Per-OlovLowdin1965, Lowdin1962}. 
\subsection{Explicitly Correlated Gaussian Functions \label{sec:ECG}}
\noindent
There are many choices of basis sets that can be used in computational chemistry which can be applied to different computational methods within a given set of approximations \cite{Atkins2018,Atkins2013}. One such basis set is the Hartree product which involves writing the wave function for a multi-particle system as a product of wave functions for each particle, 
\begin{align}
    \Psi(q_1, \, q_2, \, ... \, q_{N_{\ur{p}}}) = \psi_{\alpha}(q_1)\psi_{\beta}(q_2)...\psi_{\nu}(q_{N_{\ur{p}}}) \ , \label{eq:orbapp}
\end{align}
for $N_{\ur{p}}$ particles, where the  variable $q_i$ associated with the $i^{\ur{th}}$ particle  is itself a combination of the spatial coordinates $\br_i$ and discrete spin coordinates. The subscripts $\alpha, \beta, ..., \nu$ represent the set of quantum numbers of the wave function $\psi$ with coordinates $q_i$. However, for systems of identical particles, the (anti-)symmetric behaviour of (fermions) bosons under exchange needs to be taken into account; this is done by considering the symmetrisation operator $\hat{\mathcal{S}}$ when the particles are bosons,
\begin{align}
    \hat{\mathcal{S}} = \frac{1}{\sqrt{N_{\ur{p}}!}}\sum_{p=1}^{N_{\ur{p}}!}\hat{P} \ , \label{eq:symop}
\end{align}
and the antisymmetrisation operator $\hat{\mathcal{A}}$ when the particles are fermions,
\begin{align}
    \hat{\mathcal{A}} = \frac{1}{\sqrt{N_{\ur{p}}!}}\sum_{p=1}^{N_{\ur{p}}!}\epsilon_p \hat{P} \ . \label{eq:antsymop}
\end{align}
The summations of Eqs. (\ref{eq:symop}) and (\ref{eq:antsymop}) are over the total number of permutations of identical particles; for $N_{\ur{p}}$ identical particles, there are $N_{\ur{p}}!$ permutations. The operator $\hat{P}$ represents the permutation operator, and $p$ represents the $p^{\ur{th}}$ permutation, for $p = 1, \, 2, \, ...,\, N_{\ur{p}}!$ \cite{Matyus2019, Matyus2012}. The $\epsilon_p$ term is the parity of the permutation, which is equal to $-1$ for odd permutations and $+1$ for even permutations. The totally anti-symmetric wave function $\ket{\Psi_{\ur{A}}}$ in terms of the single particle wave functions of \rref{eq:orbapp} can then be written in terms of the \textit{Slater determinant}, 
\begin{align*} 
    \ket{\Psi_{\ur{A}}} =  \frac{1}{\sqrt{N_{\ur{p}}!}}
    \begin{vmatrix*} 
     \psi_{\alpha}(q_1) & \psi_{\beta}(q_1) & \cdots & \psi_{\nu}(q_1) \\
     \psi_{\alpha}(q_2) & \psi_{\beta}(q_2) & \cdots & \psi_{\nu}(q_2) \\
     \vdots & \vdots & \ddots & \vdots \\
     \psi_{\alpha}(q_{N_{\ur{p}}}) & \psi_{\beta}(q_{N_{\ur{p}}}) & \cdots & \psi_{\nu}(q_{N_{\ur{p}}}) 
    \end{vmatrix*} \ .  \nr \label{eq:slatdet}
\end{align*}
This wave function is anti-symmetric, as swapping the coordinates of two particles $q_i$ is equivalent to interchanging the two corresponding rows, which changes the sign of the determinant \cite{Hartree1927,Slater1930a,Slater1930,Bransden2000}. Slater determinants are often used in standard quantum chemistry methodologies such as the Hartree-Fock method, Configurational Interaction (CI) and Multi Configurational Self-Consistent Field (MCSCF).

The present work considers few-particle systems with the aim of achieving sub-1 ppb convergence of the electronic or molecular energy and is in the context of (ultra-)high resolution spectroscopy. Using Slater determinant bases, it is possible to describe even large systems with $\ur{m}$E$_{\ur{h}}$ accuracy, but our interest is in few-particle systems with the aim of achieving sub-$\ur{n}$E$_{\ur{h}}$ convergence of the electronic or molecular energy for small systems of the lightest elements. Within this domain of molecular physics, the sub-$\ur{n}$E$_\text{h}$ regime can be reached with explicitly correlated Gaussian functions \cite{Mitroy2013, J.Rychlewski2003, Bubin2013}.  Such basis sets have functions containing $r_{ij} = |\br_i - \br_j|$ terms, where $r_{ij}$ is the distance between the $i^{\ur{th}}$ and $j^{\ur{th}}$ particles. 

\begin{figure}[t]
\subcaptionbox{Uncorrelated Gaussian}{\includegraphics[width=0.50\textwidth]{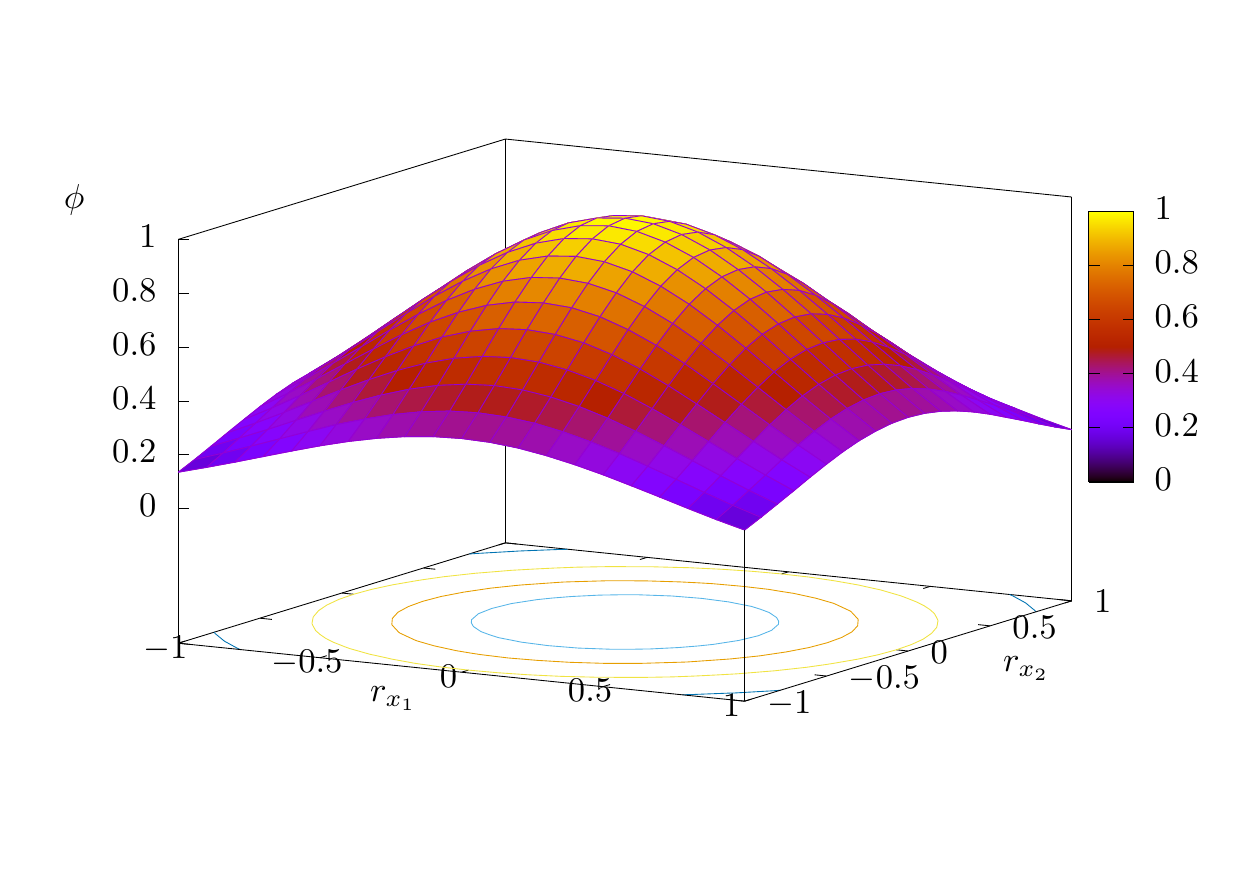}}%
\hfill 
\subcaptionbox{Correlated Gaussian}{\includegraphics[width=0.50\textwidth]{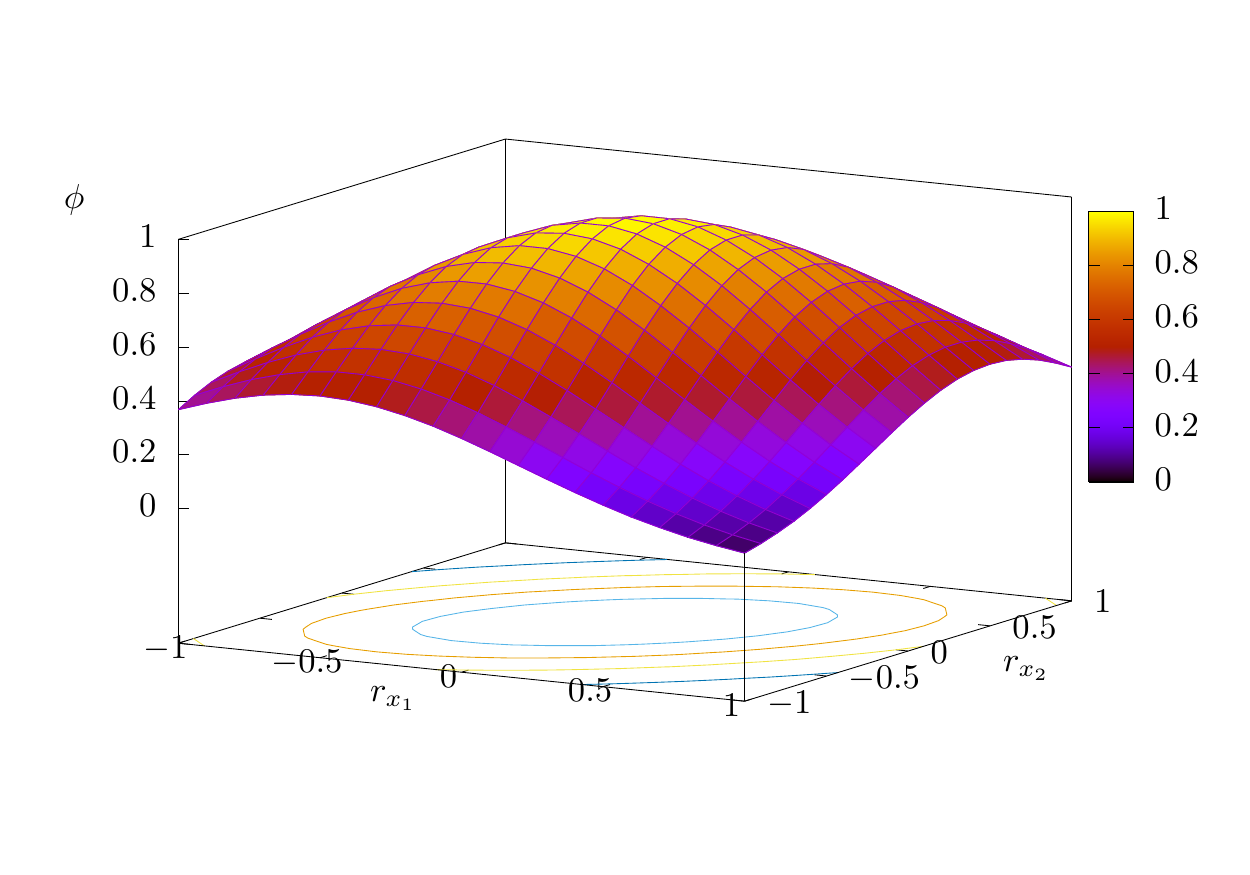}}%
\caption{\bd{(a)} A one-dimensional uncorrelated Gaussian function for a two-particle system. \textbf{(b)} A one-dimensional correlated Gaussian function for a two-particle system.}
\label{fig:un/corgaus}
\end{figure}

The use of correlated functions in few-body problems first appeared in the early years of quantum mechanics, particularly by E. Hylleraas in his studies of the helium atom \cite{Hylleraas1,Hylleraas2,Hylleraas3,Hylleraas4,Hylleraas5}. However, it was in 1960 when S. Boys \cite{Boys1960} and K. Singer \cite{Singer1960} became the first to propose a wave function for $N_{\ur{p}}$ particles that involved an exponential with an argument containing a squared correlation factor (i.e. a Gaussian function). The functions in this type of basis set are therefore called explicitly correlated Gaussian (ECG) functions, which are the basis functions used in this report. The correlation factors take into account the interactions between all of the particles in the system. Compared to the widely-used Slater determinants, which do not consider explicitly the correlation between particles in the basis functions, greater accuracy in the energy can be achieved when using ECGs. This is presented in Figs. \ref{fig:un/corgaus} (a) and (b) , where Fig. \ref{fig:un/corgaus} (a) shows a one-dimensional uncorrelated Gaussian function for two particles, in which it can be seen that ignoring the interactions between particles leads to a spherical symmetry in the plot. However, in order to achieve sub-$\ur{n}$E$_\ur{h}$ accuracy, the explicit correlations between particles must be taken into account. The affect of this in one dimension can be seen in Fig. \ref{fig:un/corgaus} (b), where the spherical symmetry present in Fig. \ref{fig:un/corgaus} (a) becomes distorted. 

ECGs are often used in quantum-mechanical few-body problems not only due to the benefits of considering the correlation between particles, but also due to the relative ease with which the matrix elements can be calculated, the algebraic difficulty of which does not increase for $N_{\ur{p}} \ge 3$. Despite ECG basis sets being one of the easiest basis sets to handle, it has only been since the 1990's that the use of ECG's has become (more) common place \cite{Mitroy2013}. In spite of their advantages, ECG functions do not have the correct behaviour in the $r_{ij} \rightarrow 0$ and $r_{ij} \rightarrow \infty$ limits. In the $r_{ij} \rightarrow 0$ limit, the exact wave function should have a cusp \cite{Kato1957,Mayer2003}. However, the ECG wave functions are smooth at these coalescence points. As $r_{ij} \rightarrow \infty$, the exact wave function should decay exponentially. Whilst ECG wave functions do decay in this limit, the decay is much faster due to the quadratic argument \cite{Mitroy2013, Mayer2003}. 

We will consider basis functions which are an (anti-)symmetrised product of two functions, a spatial function $\phi_k$ and a spin function $\chi_k$, such that
\begin{align}
    \ket{\Phi_k} = \hat{\xi} \ket{\phi_k\chi_k} \ , \label{eq:anstaz} 
\end{align}
where $\hat{\xi}$ is the (anti-)symmetrisation operator. Outwith the Born--Oppenheimer approximation, the spatial functions are eigenfunctions of the square of the total spatial angular momentum operator, $\hat{N}^2$, the operator corresponding to the projection of the total spatial angular momentum onto a particular axis, $\hat{N}_{M}$, and the parity operator, $\mathcal{\hat{P}}$. The spin functions are eigenfunctions of the square of the total spin angular momentum operator, $\hat{S}^2_a$, and the operator representing the projection of the total spin angular momentum onto a particular axis, $\hat{S_{a}}_{M}$, where $a$ represents the type of particle (electron, proton etc...) \cite{Matyus2012}. 

The spatial functions are the functions of primary interest in this report, as it is with these functions that the matrix elements are calculated. This is due to the spin-independence of the Hamiltonian. If we consider our trial wave function $\ket{\Psi_{\ur{Trial}}}$ as the linear combination of basis functions in \rref{eq:anstaz}, then the energy expectation value is
\begin{align*}
    \matrixel{\Psi_{\ur{Trial}}}{\hat{H}}{\Psi_{\ur{Trial}}} = \matrixel{\sum_{k=1}^{N_{b}}c_{k}\Phi_k }{\hat{H}}{\sum_{l=1}^{N_{b}}c_{l}\Phi_l} = \sum_{k=1}^{N_{b}}c_{k}^*\sum_{l=1}^{N_{b}}c_{l}\matrixel{\hat{\xi} \left\{\phi_k\chi_k \right\} }{\hat{H}}{\hat{\xi} \left\{\phi_l\chi_l \right\}} \ , \nr \label{eq:SpinSpaceExp1}
\end{align*}
where we have used \rref{eq:trial} followed by \rref{eq:anstaz}. We can then combine the two $\hat{\xi}$ operators, as $\comm{\hat{\xi}}{\hat{H}} = 0$, and use its quasi-idempotent property, $\hat{\xi}^2 = \sqrt{N_{\ur{p}}!} \, \hat{\xi}$. Then, \rref{eq:SpinSpaceExp1} can be written as
\begin{align*}
    \matrixel{\Psi_{\ur{Trial}}}{\hat{H}}{\Psi_{\ur{Trial}}} &=\sum_{k, \, l=1}^{N_{b}}c_{k}^* c_{l} \sqrt{N_{\ur{p}}!} \matrixel{\phi_k\chi_k}{\hat{H}}{\hat{\xi} \left \{\phi_l\chi_l \right \} } \ .  \nr 
\end{align*}
We can then write the (anti-)symmetrisation operator as
\begin{align*}
    \hat{\xi} = \frac{1}{\sqrt{N_{\ur{p}}!}} \sum_{p = 1}^{N_{\ur{p}}!} \epsilon_p \hat{P} \ ,  \nr
\end{align*}
where $\epsilon_p$ is equal to $-1$ for an odd number of permutations of identical fermions, and $+1$ otherwise. The operator $\hat{P}$ is the permutation operator which operates on both the spatial and spin functions. Using this definition,
\begin{align*}
    \matrixel{\Psi_{\ur{Trial}}}{\hat{H}}{\Psi_{\ur{Trial}}}  &= \sum_{k, \, l=1}^{N_{b}}c_{k}^*c_{l} \sum_{p = 1}^{N_{\ur{p}}!} \epsilon_p \matrixel{\phi_k\chi_k}{\hat{H}}{\lrr{\hat{P}\phi_l}\lrr{\hat{P}\chi_l}} \\
    &= \sum_{k, \, l=1}^{N_{b}} \sum_{p = 1}^{N_{\ur{p}}!} c_{k}^*c_{l} \epsilon_p \matrixel{\chi_k}{\hat{P}}{\chi_l} \matrixel{\phi_k}{\hat{H}\hat{P}}{\phi_l} \ , \nr \label{eq:SpinSpaceExp4}
\end{align*}
where in the final line we have used the fact that the Hamiltonian is spin independent and so does not operate on the spin functions. The spin functions can be written as a linear combination of the uncoupled many-spin spin functions, where the coefficients are the Clebsch--Gordan coefficients \cite{Nakamura_2010}. As such, the spin integrals of \rref{eq:SpinSpaceExp4} can be solved with relative ease compared to the spatial integrals which are substantially more involved, as will be observed in Sec. \ref{sec:MatrixEl}.
For the spatial functions ${\phi_k}$, the functions that will be used are ECGs. Whilst there are many kinds of ECG functions, we will primarily use floating, or shifted, ECGs (FECGs) \cite{Mitroy2013,Stanke2016}, which take the form
\begin{align}
    {\phi_k} = \exp[-(\textbf{r}-\textbf{s}_k)^T\textbf{\underline{A}}_k(\textbf{r}-\textbf{s}_k)] \ , \label{ECGEQN}
\end{align}
where $\textbf{r}$ represents a particular Cartesian position in the configuration space and $\textbf{s}_k$ represents a shift from that position; both vectors are $3N_{\ur{p}}$ -- dimensional, where $N_{\ur{p}}$ is the number of particles in the system. The matrix $\ubA_k$ is a $3N_{\ur{p}}\times3N_{\ur{p}}$ positive-definite, symmetric matrix (${\ubA_k}^T$ = $\ubA_k$) whose entries are adjustable non-linear parameters. We define $\ubA_k$ as $\ubA_k = \bd{A}_k \otimes \bd{I}_3$, where $\bd{I}_3$ is the $3 \times 3$ identity matrix and $\bd{A}_k$ is an $N_{\ur{p}} \times N_{\ur{p}}$ matrix. Finding the Kronecker product of $\bd{A}_k$ with $\bd{I}_3$ ensures that the $x, y, z$ Cartesian components for every particle, which is contained in the $\bd{r}$ vector, have the same coefficients (which occurs when $\ubA_k$ and $\bd{r}$ are multiplied together in the argument of the exponential).
\section{Identifying Necessary Integrals }
\noindent
We begin this chapter by stating the notations and commonly used expressions which are used throughout this work. 
\subsection*{Notations\label{sec:notations}}
\begin{align*} 
    \underline{\bd{A}}_k &= {\bf A}_k \otimes {\bf I}_3 \\
    {\bf A}_{lk} &= {\bf A}_l + {\bf A}_k \\
    {\bf B}_{lk} &= {\bf A}_l  {\bf A}_k \\
    {\bf e}_l &= \underline{{\bf A}}_l {\bf s}_l \\
    {\bf e}_{lk} &= {\bf e}_{l} + {\bf e}_k \\
    \eta_{lk} &= {\bf s}_l^T \underline{{\bf A}}_l {\bf s}_l + {\bf s}_k^T \underline{{\bf A}}_k {\bf s}_k  \\
    \bd{s} &= \ubA_{kl}^{-1}\bd{e}_{lk} \\
    \gamma_{lk} &= {\bf e}_{lk}^T \underline{{\bf A}}_{lk}^{-1} {\bf e}_{lk}- \eta_{lk} \\
    J_{ij,pq} &= \delta_{pq}\left(\delta_{pi}+\delta_{pj}\right)-\left(\delta_{pi}\delta_{qj}+\delta_{qi}\delta_{pj}\right)\left( 1-\delta_{pq}\right)\\
    K_{ij,pq} &= \delta_{pq} \left( \delta_{ip}-\delta_{jp}\right)\\
    \beta_{lkij} &= \be_{lk}^T  \underline{\bA}_{lk}^{-1} \underline{\bJ}_{ij} \underline{\bA}_{lk}^{-1} \be_{lk}  \\
    a_{lkij} &= \Tr \left( \bJ_{ij} \bA_{lk}^{-1}  \right) \\
\end{align*}
\subsection*{Commonly Used Expressions}
\label{sec:comexp}
\begin{align*}
    &|\bA \otimes \bI_n | = | \bA |^n \ \nr \label{eq:directdet}  \\
    &\Tr{ \bA \otimes \bI_n} = n \Tr{\bA} \nr \label{eq:directtr} \\
    &\int_{-\infty}^{\infty} d\mx{r} \, \phi_k \phi_l = \textbf{S}_{kl} = \exp[\gamma_{kl}]\pi^{\frac{3n}{2}}|\bA_{kl}|^{-\frac{3}{2}}  \nr \label{eq:Overlap} \\
    &\phi_k\phi_l = \exp[-\eta_{kl}]\exp[-\textbf{r}^T\textbf{\underline{A}}_{kl}\textbf{r} +2\textbf{e}^T_{kl}\textbf{r}] \nr \label{eq:ProductECG}  \\
    &\nabla = \nabla_1 + \nabla_2 + \nabla_3 + ... + \nabla_{N_{\ur{p}}} = \sum_{i=1}^{N_{\ur{p}}} \nabla_i \nr \label{eq:NablaSum} \\
    &\matrixel{\phi_k}{\nabla^2}{\phi_l} = \braket{\nabla^2\phi_k}{\phi_l} \nr \label{eq:IntPartsResult} \\
    &\nabla\phi_k = -2\ubA_k(\textbf{r} - \textbf{s}_k)\phi_k \nr \label{eq:nabphi} \\ 
    &\nabla^2\phi_k = \nabla^T\nabla\phi_k = 4\lrs{(\bd{r}-\bd{s}_k)^T\ubA_k\ubA_k(\bd{r}-\bd{s}_k)  - 6\Tr{\bd{A}_k}}\phi_k \nr \label{eq:NabSqPhi} \\
\end{align*}
\subsection{The Hamiltonian Squared} \label{Sec:IntHamaSq}
\noindent
We now know that in order to calculate lower bounds we not only require an upper bound to the exact energy, but also the variance of the upper bound. As such, the matrix elements of $\hat{H}^2$ are required (see \rref{eq:var}).  If the Born--Oppenheimer approximation is used, then the electronic and nuclear wave functions are treated separately on account of the mass of the nuclei being far greater than that of the electrons. This report is concerned only with the electronic problem, which is solved in the presence of the fixed nuclei \cite{Born1927}. As such, the Hamiltonian of the system can be referred to as the electronic Hamiltonian, $\hat{H}_{\ur{el}}$, when using the Born--Oppenheimer approximation. If the electrons and the nuclei are handled together as an isolated quantum system, then the Hamiltonian can be referred to as the molecular Hamiltonian, $\hat{H}_{\ur{mol}}$. 

Like with classical mechanics, the Hamiltonian of the system can be written as the sum of the kinetic energy and potential energy operators. For an isolated system, the molecular Hamiltonian $\hat{H}_{\ur{mol}}$ (also known as the Coulomb Hamiltonian) can be written as \cite{Atkins2018}
\begin{align}
    \hat{H}_{\ur{mol}} = \hat{T}_{\ur{e}} + \hat{T}_{\ur{n}} + \hat{V}_{\ur{ee}} + \hat{V}_{\ur{nn}} + \hat{V}_{\ur{ne}} \ . \label{MolecHam}
\end{align}
The term $\hat{T}_{\ur{e}}$ corresponds to the kinetic energy of all the electrons; $\hat{T}_{\ur{n}}$ corresponds to the kinetic energy of the nuclei; $\hat{V}_{\ur{ee}}$ describes the interactions between the electrons; $\hat{V}_{\ur{nn}}$ describes the interactions between the nuclei; $\hat{V}_{\ur{ne}}$ describes the interactions between the electrons and the nuclei. Note that \rref{MolecHam} ignores small additional terms, such as the effects of the spin of electrons and nuclei that enter if relativistic effects are also accounted for. Each term takes the following form in atomic units;
 \\
\begin{align*}
    \hat{T}_{\ur{e}} &= -\frac{1}{2}\sum_{i=1}^n \nabla_i^2 \ , \nr \label{eq:EKin} \\ 
    \hat{T}_{\ur{n}} &= -\frac{1}{2}\sum_{a=1}^N\frac{1}{M_a}\nabla_a^2 \ , \nr \label{eq:NKin} \\ 
    \hat{V}_{\ur{ee}} &= \frac{1}{2}\sum_{i=1}^n\sum_{j \ne i}^n 
    \frac{1}{|\textbf{r}_i-\textbf{r}_j|} = \frac{1}{2}\sum_{i=1}^n\sum_{j \ne i}^n \frac{1}{r_{ij}} \ , \nr \label{eq:EEPot} \\ 
   \hat{V}_{\ur{nn}} &= \frac{1}{2}\sum_{a=1}^N\sum_{b \ne a}^N \frac{Z_aZ_b}{|\textbf{R}_a - \textbf{R}_b|} = \frac{1}{2}\sum_{a=1}^N\sum_{b \ne a}^N \frac{Z_aZ_b}{r_{ab}} \ , \nr \label{eq:NNPot} \\ 
    \hat{V}_{\ur{ne}} &= 
-\sum_{i=1}^n\sum_{a=1}^N\frac{Z_a}{|\textbf{r}_i-\textbf{R}_a|} = -\sum_{i=1}^n\sum_{a=1}^N\frac{Z_a}{r_{ia}} \ . \nr \label{eq:NEPot} \\ \end{align*}
The indices $i$ and $j$ correspond to electrons (of which there are a total of $n$), whilst the indices $a$ and $b$ correspond to nuclei (of which there are a total of $N$). We note that $\nabla = \lrr{\frac{\partial}{\partial x}, \, \frac{\partial}{\partial y}, \, \frac{\partial}{\partial z}}  $ is the nabla operator, $\textbf{r}_i$ is the Cartesian position of the $i^{\ur{th}}$ electron, $Z_a$ is the charge of the $a^{\ur{th}}$ nucleus and $\textbf{R}_a$ is the Cartesian position of the $a^{\ur{th}}$ nucleus with mass $M_a$. We have also made use of $r_{ij} =|\textbf{r}_i - \textbf{r}_j| $, $r_{ab} =|\textbf{R}_a - \textbf{R}_b| $ and $r_{ia} = |\textbf{r}_i - \textbf{R}_a|$. 
Within the Born--Oppenheimer approximation the descriptions for the electronic and the nuclear motion are separated. The electronic Hamiltonian, $\hat{H}_{\ur{el}}$, is then
\begin{align}
    \hat{H}_{\ur{el}} = \hat{T}_{\ur{e}} + \hat{V}_{\ur{ee}} + \hat{V}_{\ur{nn}} + \hat{V}_{\ur{ne}} \ . \label{ElecHam}
\end{align}
Note that we could therefore also write \rref{MolecHam} as
\begin{align}
    \hat{H}_{\ur{mol}} = \hat{H}_{\ur{el}} + \hat{T}_{\ur{n}} \ . \label{MolecHam2} 
\end{align}
In order to evaluate the variance of the molecular and the electronic upper bound energy for a trial function $\ket{\Psi_{\ur{Trial}}}$, we require the matrix entries of the matrix representations of both Hamiltonians squared, $\textbf{H}_\ur{{mol}}^2$ and $\textbf{H}_\ur{{el}}^2$ respectively. To find the $k^{\ur{th}}$, \, $l^{\ur{th}}$ matrix entry of $\textbf{H}_{\ur{mol}}^2$ and $\textbf{H}_{\ur{el}}^2$, we evaluate 
    $(H_{\ur{mol}}^2)_{kl} = \matrixel{\phi_k}{\hat{H}_{\ur{mol}}^2}{\phi_l}$
and  $(H_{\ur{el}}^2)_{kl} = \matrixel{\phi_k}{\hat{H}_{\ur{el}}^2}{\phi_l}$ 
respectively, where ${\phi_k}$ and ${\phi_l}$ are the $k^{\ur{th}}$ and $l^{\ur{th}}$ ECG  basis functions.

The calculations in this report make use of the Born--Oppenheimer approximation, and so only the matrix elements of the electronic Hamiltonian, $(H_{\ur{el}}^2)_{kl}$, are considered.\\

\subsection{Integrals of the Electronic Hamiltonian} \label{ElecHamInt}
\noindent
To find the required matrix elements, we first square the electronic Hamiltonian as written in \rref{ElecHam}, giving 
\begin{align*}
    \hat{H}_{\ur{el}}^2 &= 
     (\hat{T}_{\ur{e}} + \hat{V}_{\ur{ee}} + \hat{V}_{\ur{nn}} + \hat{V}_{\ur{ne}})(\hat{T}_{\ur{e}} + \hat{V}_{\ur{ee}} + \hat{V}_{\ur{nn}} + \hat{V}_{\ur{ne}}) \\
    &=(\hat{T}_{\ur{e}}\hat{T}_{\ur{e}} + \hat{T}_{\ur{e}}\hat{V}_{\ur{\ur{ee}}} + \hat{T}_{\ur{e}}\hat{V}_{\ur{nn}} + \hat{T}_{\ur{e}}\hat{V}_{\ur{ne}}) \\
    &+(\hat{V}_{\ur{ee}}\hat{T}_{\ur{e}} + \hat{V}_{\ur{ee}}\hat{V}_{\ur{ee}} + \hat{V}_{\ur{ee}}\hat{V}_{\ur{nn}} + \hat{V}_{\ur{ee}}\hat{V}_{\ur{ne}}) \\
    &+(\hat{V}_{\ur{nn}}\hat{T}_{\ur{e}} + \hat{V}_{\ur{nn}}\hat{V}_{\ur{ee}} + \hat{V}_{\ur{nn}}\hat{V}_{\ur{nn}} + \hat{V}_{\ur{nn}}\hat{V}_{\ur{ne}}) \\
    &+(\hat{V}_{\ur{ne}}\hat{T}_{\ur{e}} + \hat{V}_{\ur{ne}}\hat{V}_{\ur{ee}} + \hat{V}_{\ur{ne}}\hat{V}_{\ur{nn}} + \hat{V}_{\ur{ne}}\hat{V}_{\ur{ne}}) \ . \nr \label{ElHamExp}
\end{align*}
We now proceed to find the expectation values of these terms, $\phikl[\hat{H}_{\ur{el}}^2]$, using Eqs. (\ref{eq:EKin}) -- (\ref{eq:NEPot}), which yields the following expressions. Where possible, we have made use of \rref{eq:NablaSum}. 
\begin{align}
    &\matrixel{\phi_k}{\hat{T}_{\ur{e}}\hat{T}_{\ur{e}}}{\phi_l} = \frac{1}{4}\sum_{i=1}^n\sum_{j=1}^n\matrixel{\phi_k}{\nabla_i^2\nabla_j^2}{\phi_l} \ ,  \label{Term1} \\
    &\matrixel{\phi_k}{\hat{T}_{\ur{e}}\hat{V}_{\ur{ee}}}{\phi_l} = -\frac{1}{4}\sum_{i=1}^n\sum_{j=1}^n\sum_{p \ne j}^n\matrixel{\phi_k}{\nabla_i^2\frac{1}{r_{jp}}}{\phi_l} 
     = -\frac{1}{4}\sum_{j=1}^n\sum_{p \ne j}^n\matrixel{\phi_k}{\nabla^2\frac{1}{r_{jp}}}{\phi_l} \ ,
     \label{Term2} \\
    &\matrixel{\phi_k}{\hat{T}_{\ur{e}}\hat{V}_{\ur{ne}}}{\phi_l} = \frac{1}{2}\sum_{i=1}^n\sum_{j=1}^n\sum_{a=1}^NZ_a\matrixel{\phi_k}{\nabla_i^2\frac{1}{r_{ja}}}{\phi_l} 
    = \frac{1}{2}\sum_{j=1}^n\sum_{a=1}^NZ_a\matrixel{\phi_k}{\nabla^2\frac{1}{r_{ja}}}{\phi_l} \ ,
     \label{Term3} \\ 
    &\matrixel{\phi_k}{{\hat{T}_{\ur{e}}\hat{V}_{\ur{nn}}}}{\phi_l} = \matrixel{\phi_k}{{\hat{V}_{\ur{nn}}}{\hat{T}_{\ur{e}}}}{\phi_l} 
    = -\frac{1}{4}\sum_{a=1}^N\sum_{b\ne a}^NZ_aZ_b\frac{1}{r_{ab}}\matrixel{\phi_k}{\nabla^2}{\phi_l} \ ,
     \label{Term4} \\ 
    &\matrixel{\phi_k}{\hat{V}_{\ur{ee}}\hat{T}_{\ur{e}}}{\phi_l} = -\frac{1}{4}\sum_{j=1}^n\sum_{p \ne j}^n\sum_{i=1}^n\matrixel{\phi_k}{\frac{1}{r_{jp}}\nabla_i^2}{\phi_l}
    =-\frac{1}{4}\sum_{j=1}^n\sum_{p \ne j}^n\matrixel{\phi_k}{\frac{1}{r_{jp}}\nabla^2}{\phi_l} \ ,
     \nr \label{Term5} \\
    &\matrixel{\phi_k}{\hat{V}_{\ur{ee}}\hat{V}_{\ur{ee}}}{\phi_l}  = \frac{1}{4}\sum_{i=1}^n\sum_{j \ne i}^n\sum_{p=1}^n\sum_{q \ne p}^n\matrixel{\phi_k}{\frac{1}{r_{ij}}\frac{1}{r_{pq}}}{\phi_l} \ ,  \label{Term6} \\
    &\matrixel{\phi_k}{\hat{V}_{\ur{ee}}\hat{V}_{\ur{nn}}}{\phi_l}  = \matrixel{\phi_k}{\hat{V}_{\ur{nn}}\hat{V}_{\ur{ee}}}{\phi_l}  = \frac{1}{4}\sum_{i=1}^n\sum_{j \ne i}^n\sum_{a=1}^N\sum_{b \ne a}^NZ_aZ_b\matrixel{\phi_k}{\frac{1}{r_{ij}}\frac{1}{r_{ab}}}{\phi_l} \ ,  \label{Term7} \\
    &\matrixel{\phi_k}{\hat{V}_{\ur{ee}}\hat{V}_{\ur{ne}}}{\phi_l} = \matrixel{\phi_k}{\hat{V}_{\ur{ne}}\hat{V}_{\ur{ee}}}{\phi_l} = -\frac{1}{2}\sum_{i=1}^n\sum_{j \ne i}^n\sum_{p=1}^n\sum_{a = 1}^NZ_a\matrixel{\phi_k}{\frac{1}{r_{ij}}\frac{1}{r_{pa}}}{\phi_l} \ , \label{Term8} \\
    &\matrixel{\phi_k}{\hat{V}_{\ur{nn}}\hat{V}_{\ur{nn}}}{\phi_l}= \frac{1}{4}\sum_{a=1}^N\sum_{b \ne a}^N\sum_{c=1}^N\sum_{d \ne c}^NZ_aZ_bZ_cZ_d\frac{1}{r_{ab}}\frac{1}{r_{cd}}\braket{\phi_k}{\phi_l} \ ,  \label{Term11} \\
    &\matrixel{\phi_k}{\hat{V}_{\ur{nn}}\hat{V}_{\ur{ne}}}{\phi_l} = \matrixel{\phi_k}{\hat{V}_{\ur{ne}}\hat{V}_{\ur{nn}}}{\phi_l} = -\frac{1}{2}\sum_{a=1}^N\sum_{b \ne a}^N\sum_{c=1}^N\sum_{i = 1}^nZ_aZ_bZ_c\frac{1}{r_{ab}}\matrixel{\phi_k}{\frac{1}{r_{ci}}}{\phi_l} \ ,  \label{Term12} \\
    &\matrixel{\phi_k}{\hat{V}_{\ur{ne}}\hat{T}_{\ur{e}}}{\phi_l}= \frac{1}{2}\sum_{a=1}^N\sum_{i = 1}^n\sum_{j = 1}^nZ_a\matrixel{\phi_k}{\frac{1}{r_{ia}}\nabla_j^2}{\phi_l} = \sum_{a=1}^N\sum_{i=1}^nZ_a\matrixel{\phi_k}{\frac{1}{r_{ia}}\nabla^2}{\phi_l} \ , \label{Term13}
    \\ 
    &\matrixel{\phi_k}{\hat{V}_{\ur{ne}}\hat{V}_{\ur{ne}}}{\phi_l}= \sum_{a=1}^N\sum_{i = 1}^n\sum_{b=1}^N\sum_{j = 1}^nZ_aZ_b\matrixel{\phi_k}{\frac{1}{r_{ia}}\frac{1}{r_{jb}}}{\phi_l} \ .  \label{Term16}
\end{align}
The indices $i$, $j$, $p$ and $q$ represent electrons, whilst the indices $a$, $b$, $c$ and $d$ represent nuclei. Within the Born--Oppenheimer approximation, the term $\frac{1}{r_{ab}}$ associated with the $\hat{V}_{\ur{nn}}$ operator is a parameter, and so can be removed from the bra-kets. 
\subsection{Expressions for $n=2$} \label{sec:neqtwo}
\noindent
The first application of any lower bound calculations will be to systems consisting of two electrons using the Born--Oppenheimer approximation. This allows for a simplification of some of the integrals of the previous section, solutions to which can be tested on systems such as He, $\ur{H_2}$ or $\ur{H_3^+}$.  As such, we restrict Eqs. (\ref{Term1}) -- (\ref{Term16}) to $n = 2$. Making this restriction, we have
\begin{align}
    &\matrixel{\phi_k}{\hat{T}_{\ur{e}}\hat{T}_{\ur{e}}}{\phi_l} =  \frac{1}{4}\matrixel{\phi_k}{{\nabla_1}^4 + 2{\nabla_1}^2{\nabla_2}^2 + {\nabla_2}^4}{\phi_l} \ , \label{newTerm1} \\ 
    &\matrixel{\phi_k}{\hat{T}_{\ur{e}}\hat{V}_{\ur{ee}}}{\phi_l} = -\frac{1}{2}\left[\matrixel{\phi_k}{\nabla^2\frac{1}{r_{12}}}{\phi_l}\right] \ , \label{newTerm2} \\ 
    &\matrixel{\phi_k}{\hat{T}_{\ur{e}}\hat{V}_{\ur{ne}}}{\phi_l} =
   \frac{1}{2}\sum_{a=1}^NZ_a\left[\matrixel{\phi_k}{\nabla^2\frac{1}{r_{1a}}}{\phi_l}+\matrixel{\phi_k}{\nabla^2\frac{1}{r_{2a}}}{\phi_l}\right] \ ,   \nr \label{newTerm3}\\
    &\matrixel{\phi_k}{\hat{V}_{\ur{ee}}\hat{T}_{\ur{e}}}{\phi_l} = -\frac{1}{2}\left[\matrixel{\phi_k}{\frac{1}{r_{12}}\nabla^2}{\phi_l}\right] \ ,  \label{newTerm5} \\
    &\matrixel{\phi_k}{\hat{V}_{\ur{ee}}\hat{V}_{\ur{ee}}}{\phi_l}  = \matrixel{\phi_k}{\frac{1}{(r_{12})^2}}{\phi_l} \ ,  \label{newTerm6} \\
    &\matrixel{\phi_k}{\hat{V}_{\ur{ee}}\hat{V}_{\ur{nn}}}{\phi_l}  = \matrixel{\phi_k}{\hat{V}_{\ur{nn}}\hat{V}_{\ur{ee}}}{\phi_l}  = \frac{1}{2}\sum_{a=1}^N\sum_{b \ne a}^NZ_aZ_b\frac{1}{r_{ab}}\matrixel{\phi_k}{\frac{1}{r_{12}}}{\phi_l} \ ,  \label{newTerm7} \\
    &\matrixel{\phi_k}{\hat{V}_{\ur{ee}}\hat{V}_{\ur{ne}}}{\phi_l} =  \matrixel{\phi_k}{\hat{V}_{\ur{ne}}\hat{V}_{\ur{ee}}}{\phi_l} = -\sum_{a = 1}^NZ_a\left[\matrixel{\phi_k}{\frac{1}{r_{12}}\frac{1}{r_{1a}}}{\phi_l}+\matrixel{\phi_k}{\frac{1}{r_{12}}\frac{1}{r_{2a}}}{\phi_l}\right] \ ,  \label{newTerm8} \\
    &\matrixel{\phi_k}{\hat{V}_{\ur{nn}}\hat{V}_{\ur{ne}}}{\phi_l} = \matrixel{\phi_k}{\hat{V}_{\ur{ne}}\hat{V}_{\ur{nn}}}{\phi_l} = -\frac{1}{2}\sum_{a=1}^N\sum_{b \ne a}^N\sum_{c=1}^NZ_aZ_bZ_c\frac{1}{r_{ab}}\left[\matrixel{\phi_k}{\frac{1}{r_{1c}}}{\phi_l}+\matrixel{\phi_k}{\frac{1}{r_{2c}}}{\phi_l}\right] \ ,  \label{newTerm12} \\
    &\matrixel{\phi_k}{\hat{V}_{\ur{ne}}\hat{T}_{\ur{e}}}{\phi_l}= 
    \frac{1}{2}\sum_{a=1}^N Z_a\left[\matrixel{\phi_k}{\frac{1}{r_{1a}}\nabla^2}{\phi_l}+\matrixel{\phi_k}{\frac{1}{r_{2a}}\nabla^2}{\phi_l}\right]  \ ,  
    \label{newTerm13} \\
    &\matrixel{\phi_k}{\hat{V}_{\ur{ne}}\hat{V}_{\ur{ne}}}{\phi_l}= 
    \sum_{a=1}^N\sum_{b=1}^NZ_aZ_b\left[\matrixel{\phi_k}{\frac{1}{r_{1a}}\frac{1}{r_{1b}}}{\phi_l}+\matrixel{\phi_k}{\frac{1}{r_{2a}}\frac{1}{r_{1b}}}{\phi_l}+\matrixel{\phi_k}{\frac{1}{r_{1a}}\frac{1}{r_{2b}}}{\phi_l}+\matrixel{\phi_k}{\frac{1}{r_{2a}}\frac{1}{r_{2b}}}{\phi_l}\right] \ .  \label{newTerm16}
\end{align} 
Note that Eqs. (\ref{Term4}) and (\ref{Term11}) take the same form upon restricting the number of electrons to two and so have not been included in the above set of expressions.
These integrals, in addition to Eqs. (\ref{Term4}) and (\ref{Term11}), are required in order to find the matrix elements of the electronic Hamiltonian matrix squared. Fortunately, many of these integrals can be evaluated analytically, providing efficient implementation for the calculation of the matrix elements \cite{Stanke2016}. Many of these integrals can be constructed from `elementary integrals', which are already available in the QUANTEN computer program \cite{MolecularQuantumDynamicsResearchGroup}. These `elementary integrals'  are found by evaluating the Laplacian operators using \rref{eq:NabSqPhi} (see Appendix \ref{appendix:devECG}). However, to complete the calculation of the $\bd{H}_{\ur{el}}^2$ matrix elements, it is necessary to work out the integrals for Eqs. (\ref{newTerm6}) and (\ref{newTerm8}); these derivations form the core of this research. Note that for \rref{newTerm8} we consider the $n$-electron case, \rref{Term8}.

\section{Derivation of Matrix Elements} \label{sec:MatrixEl}
\noindent 
The matrix elements that are examined in this report are the $\phikl[\frac{1}{r_{ij}^2}]$ and $\phikl[\frac{1}{r_{ij} r_{pa}}]$ matrix elements for FECGs, and the $\phikl[\frac{1}{r_{ij} r_{pa}}]$ matrix element for non-floating ECGs. We begin with the $\phikl[\frac{1}{r_{ij}^2}]$ matrix element using FECGs, followed by the $\phikl[\frac{1}{r_{ij} r_{pa}}]$ matrix element using non-floating ECGs, where solutions for both are found analytically. Finally, the matrix element $\phikl[\frac{1}{r_{ij} r_{pa}}]$ is examined using FECGs, where the multi-dimensional integral is reduced to a one-dimensional integral, from which the soution can be obtained via numerical integration. 
\subsection{The Matrix Element of $1/r_{ij}^2$} \label{subsec:1/rij^2}
\noindent
For this matrix element, we begin with 
\begin{flalign*}
    \matrixel{\phi_k}{\frac{1}{r_{ij}^2}}{\phi_l} = \Rintegral\frac{1}{r_{ij}^2}\phi_k\phi_l 
    &= \Rintegral \frac{1}{r_{ij}^2} \exp[-\eta_{kl}]\exp[-\textbf{r}^T\textbf{\underline{A}}_{kl}\textbf{r} +2\textbf{e}^T_{kl}\textbf{r}] \nr \label{rij1} \ ,
\end{flalign*}
where we have used \rref{eq:ProductECG} (see Appendix \ref{appendix:prodECG}). We now try to re-write the denominator so that it appears in the argument of an exponential function. We do this by using the standard integral 
\begin{align}
    \int_{0}^{+\infty} \ur{d}t \,  \exp[-at^2] = \frac{1}{2}\sqrt{\frac{\pi}{a}} \ ,
\end{align}
and letting $a=(\textbf{r}_i-\textbf{r}_j)^2=|\textbf{r}_i-\textbf{r}_j|^2 = r_{ij}^2$, which gives
\begin{align*}
    &\int_{0}^{+\infty} \ur{d}t \,  \exp[-r_{ij}^2t^2] = \frac{1}{2}\frac{\sqrt{\pi}}{r_{ij}} \\
    & \implies \frac{1}{r_{ij}} =\frac{2}{\sqrt{\pi}}\int_{0}^{+\infty} \ur{d}t \, \exp[-r_{ij}^2t^2]. \nr \label{rij2}  
\end{align*}
We can then use this method again, integrating over a different variable ($u$) such that 
\begin{align*}
    \frac{1}{r_{ij}^2} &= \frac{2}{\sqrt{\pi}}\int_{0}^{+\infty} \ur{d}t \, \exp[-r_{ij}^2t^2]\frac{2}{\sqrt{\pi}}\int_{0}^{+\infty} \ur{d}u \, \exp[-r_{ij}^2u^2] \\
    &= \frac{4}{\pi}\int_{0}^{+\infty}\ur{d}t\,\int_{0}^{+\infty}\ur{d}u\,\exp[-r_{ij}^2(t^2+u^2)] \,  , \nr
\end{align*}
and so overall we have
\begin{align*}
    &\matrixel{\phi_k}{\frac{1}{r_{ij}^2}}{\phi_l} = 
    \frac{4}{\pi}\exp[-\eta_{kl}]\int_{0}^{+\infty} \ur{d}t \,\int_{0}^{+\infty} \ur{d}u \, \Rintegral\exp[-r_{ij}^2(t^2+u^2)-\br^T\ubA_{kl}\br+2\be_{kl}^T\br] \, , \nr \label{rij3}
\end{align*}
where we have combined the exponentials together and removed the $\exp[-\eta_{kl}]$ term from the integrals as it is a scalar. We would now like to convert the exponential into a form that allows the use of \rref{lemma:multidgauss}, which can be done by introducing a matrix $\bd{\un{J}}_{ij}$ such that
\begin{align}
    \br^T \uJij \br = (\br_i-\br_j)^2 = r_{ij}^2 \ , \label{Jij} 
\end{align}
for $\bd{\un{J}}_{ij} = \bd{J}_{ij} \otimes \bd{I}_3$. The matrix can be formally defined as 
\begin{align}
(\Jij)_{\alpha\beta} = \delta_{i\alpha}\delta_{i\beta} + \delta_{j\alpha}\delta_{j\beta} -\delta_{i\alpha}\delta_{j\beta} -\delta_{j\alpha}\delta_{i\beta} \ , \label{eq:JijDef}
\end{align}
where $i$, $j$ are the particle indices and $\alpha$, $\beta$ represent the row and column numbers of the matrix entry, respectively. For example, the explicit form of the $\bJ_{12}$ matrix for three particles is
\begin{align*}
    \bd{J}_{12} = 
    \begin{pmatrix*}[r]
     1 &-1 & 0 \\
    -1 & 1 & 0 \\
     0 & 0 & 0 
    \end{pmatrix*} \ . \nr
\end{align*}\\ 
This matrix allows the $-r_{ij}^2(t^2+u^2)$ and $\br^T\ubA_{kl}\br$ terms to be grouped together such that \begin{align*}
    &\matrixel{\phi_k}{\frac{1}{r_{ij}^2}}{\phi_l} = 
    \frac{4}{\pi}\exp[-\eta_{kl}]\int_{0}^{+\infty} \ur{d}t \,\int_{0}^{+\infty} \ur{d}u \, \Rintegral\exp{-\br^T\lrs{\underline{\textbf{J}}_{ij}(t^2+u^2)+\ubA_{kl}}\br+2\be_{kl}^T\br} \ . \nr \label{rij4}
\end{align*}
We now switch to polar coordinates, where we have
\begin{align*}
    t &= \rho\cos{\theta} \ , \nr \\
    u &= \rho \sin{\theta} \ , \nr
\end{align*}
for
\begin{align*}
    \rho &\in [0,\infty) \, , \\ 
    \theta &\in [0, \frac{\pi}{2}] \ ,
\end{align*}
and a Jacobian that equals $\rho$. As such, \rref{rij3} then reads as
\begin{align*}
    &\matrixel{\phi_k}{\frac{1}{r_{ij}^2}}{\phi_l} = 
    \frac{4}{\pi}\exp[-\eta_{kl}]\int_{0}^{\frac{\pi}{2}}d\theta \, \int_{0}^{+\infty}d\rho \, \rho \Rintegral\exp[-\br^T(\rho^2\underline{\textbf{J}}_{ij}+\ubA_{kl})\br + 2 \be_{kl}^T\br] \ . \nr
\end{align*}
We can then integrate in \textbf{r} by using  \rref{mdimgauss}, resulting in
\begin{align*}
    &\matrixel{\phi_k}{\frac{1}{r_{ij}^2}}{\phi_l} = 
    \frac{4}{\pi}\exp[-\eta_{kl}]\int_{0}^{\frac{\pi}{2}}d\theta \, \int_{0}^{+\infty}d\rho \, \rho \frac{\pi^{\frac{3n}{2}}}{{|\ubA_{kl}+\rho^2\underline{\textbf{J}}_{ij}|}^{\frac{1}{2}}}\exp[\be_{kl}^T(\rho^2\underline{{\textbf{J}}}_{ij}+\ubA_{kl})^{-1}\be_{kl}] \ . \nr
\end{align*}
We can simplify this by first writing the determinant as
\begin{align*}
    |\ubA_{kl}+\rho^2\underline{\textbf{J}}_{ij}|^{\frac{1}{2}} \underbrace{=}_{\rref{eq:directdet}} |\rho^2\textbf{J}_{ij}+\bA_{kl}|^{\frac{3}{2}} =  |\rho^2\textbf{J}_{ij}\underbrace{(\bA_{kl}^{-1}\bA_{kl})}_{\textbf{I}}+\bA_{kl}\underbrace{(\bA_{kl}^{-1}\bA_{kl})}_{\textbf{I}}|^{\frac{3}{2}} \ . \nr
\end{align*}
Taking out a factor of $\bA_{kl}$ to the right gives
\begin{align*}
    |\ubA_{kl}+\rho^2\underline{\textbf{J}}_{ij}|^{\frac{1}{2}} = \left|\lrr{\rho^2 \bd{J}_{ij} \bA_{kl}^{-1} + \underbrace{\bA_{kl}\bA_{kl}^{-1}}_{\bd{I}}} \bA_{kl}\right|^{\frac{3}{2}} \ , \nr
\end{align*}
and then using the fact that $\det(\bd{AB}) = \det(\bd{A})\det(\bd{B})$,
\begin{align*}
    |\ubA_{kl}+\rho^2\underline{\textbf{J}}_{ij}|^{\frac{1}{2}} = |\textbf{I}+\rho^2\textbf{J}_{ij}\bA_{kl}^{-1}|^{\frac{3}{2}}|\bA_{kl}|^{\frac{3}{2}} \ . \nr
\end{align*}
Using Lemma (\ref{lemma:productrank}),  $\ur{rank}\lrr{\textbf{J}_{ij}\bA_{kl}^{-1}} = \ur{rank}\lrr{\textbf{J}_{ij}}$ as $\bA_{kl}^{-1}$ is non-singular.  We can see from the definition of the matrix $\textbf{J}_{ij}$ that it is rank-1, as each column is a linear combination of the first. Hence, $\textbf{J}_{ij}\bA_{kl}^{-1}$ is rank-1 and so we can use \rref{eq:lincomdet}, giving
\begin{align}
    |\ubA_{kl}+\rho^2\underline{\textbf{J}}_{ij}|^{\frac{1}{2}} = |\bA_{kl}|^{\frac{3}{2}}\lrr{1+\rho^2\Tr{\textbf{J}_{ij}\bA_{kl}^{-1}}}^{\frac{3}{2}} \ . \label{detsimp}
\end{align}
We can also rewrite the inverse inside the exponent using \rref{lemma:asumbinvunder}, which gives
\begin{align*}
    (\rho^2\underline{{\textbf{J}}}_{kl}+\ubA_{kl})^{-1} &= (\rho^2{\textbf{J}}_{kl}+\bA_{kl})^{-1}\otimes\textbf{I}_3 \\ 
    &=\left(\bA_{kl}^{-1}-\frac{\rho^2\bA_{kl}^{-1} \bJ_{ij} \bA_{kl}^{-1}}{1 +  \rho^2\Tr{\bJ_{ij} \bA_{kl}^{-1}}}\right)\otimes\bI_3 \\
    &=\left(\ubA_{kl}^{-1}-\frac{\rho^2\ubA_{kl}^{-1} \underline{\bJ}_{ij} \ubA_{kl}^{-1}}{1 +  \rho^2\Tr{\bJ_{ij} \bA_{kl}^{-1}}}\right) \ . \nr \label{invsimp}
\end{align*}
Using Eqs. (\ref{detsimp}) and (\ref{invsimp}), we have 
\begin{align*}
    &\matrixel{\phi_k}{\frac{1}{r_{ij}^2}}{\phi_l} = 
\frac{4}{\pi}\exp[-\eta_{kl}]\int_{0}^{\frac{\pi}{2}}d\theta \,  \\
   &\times \int_{0}^{+\infty}d\rho \, \rho \, \pi^{\frac{3n}{2}}|\bA_{kl}|^{-\frac{3}{2}}(1+\rho^2\Tr{\textbf{J}_{ij}\bA_{kl}^{-1}})^{-\frac{3}{2}}\exp[\be_{kl}^T\left(\ubA_{kl}^{-1}-\frac{\rho^2\ubA_{kl}^{-1} \underline{\bJ}_{ij} \ubA_{kl}^{-1}}{1 +  \rho^2 \Tr{\bJ_{ij} \bA_{kl}^{-1}}}\right)\be_{kl}] \\ 
    & =\frac{4}{\pi}\exp[-\eta_{kl}]\int_{0}^{\frac{\pi}{2}}d\theta \, \\
    &\times \int_{0}^{+\infty}d\rho \, \rho \, \pi^{\frac{3n}{2}}|\bA_{kl}|^{-\frac{3}{2}}(1+\rho^2\Tr{\textbf{J}_{ij}\bA_{kl}^{-1}})^{-\frac{3}{2}}\exp[\be_{kl}^T\ubA_{kl}^{-1}\be_{kl}]\exp[-\frac{\rho^2\be_{kl}^T\ubA_{kl}^{-1} \underline{\bJ}_{ij} \ubA_{kl}^{-1}\be_{kl}}{1 +  \rho^2\Tr{\bJ_{ij} \bA_{kl}^{-1}}}] \ . \nr 
\end{align*}
We can simplify the above further by letting $a = \Tr{\textbf{J}_{ij}\bA_{kl}^{-1}}$ and $\beta = \be_{kl}^T\ubA_{kl}^{-1} \underline{\bJ}_{ij} \ubA_{kl}^{-1}\be_{kl}$. We can also remove the factors of $\pi^{\frac{3n}{2}}$, $|\bA_{kl}|^{-\frac{3}{2}}$ and $\exp[\be_{kl}^T\ubA_{kl}^{-1}\be_{kl}]$ from the integrand as they are all independent of $\rho$, and note that $\int_{0}^{\frac{\pi}{2}}d\theta = \frac{\pi}{2}$. This then gives
\begin{align*}
    \matrixel{\phi_k}{\frac{1}{r_{ij}^2}}{\phi_l} = 
    \frac{4}{\pi}\exp[\underbrace{-\eta_{kl}+\be_{kl}^T\ubA_{kl}^{-1}\be_{kl}}_{\gamma_{kl}}]\pi^{\frac{3n}{2}}|\bA_{kl}|^{-\frac{3}{2}}{\frac{\pi}{2}}\int_{0}^{+\infty}d\rho \, \rho \, (1+\rho^2a)^{-\frac{3}{2}}\exp[-\frac{\rho^2\beta}{1+\rho^2a}] \ . \nr
\end{align*}
From \rref{eq:Overlap}, we can see that
\begin{align*}
    \textbf{S}_{kl} = \exp[\gamma_{kl}]\pi^{\frac{3n}{2}}|\bA_{kl}|^{-\frac{3}{2}} \ , \nr
\end{align*}
which is the definition of the overlap matrix elements. This means our matrix element can be written as
\begin{align}
    \matrixel{\phi_k}{\frac{1}{r_{ij}^2}}{\phi_l} = 2\textbf{S}_{kl}\int_{0}^{+\infty}d\rho \, \frac{\rho}{(1+\rho^2a)^{\frac{3}{2}}}\exp[-\frac{\rho^2\beta}{1+\rho^2a}].  \label{rij5}
\end{align}
We then make the substitution $w = 1+\rho^2a$, resulting in the following terms and limits; 
\begin{align*}
    \ur{d}\rho \, \rho &= \frac{1}{2a}\ur{d}w \, , \nr \\
    -\rho^2 &=\frac{1-w}{a} \, , \nr \\
    w = 1+\rho^2a &\implies 
    \begin{cases}
        \rho \rightarrow \infty, & w \rightarrow \infty \\
        \rho = 0, & w = 1 \ .
    \end{cases}  \nr 
\end{align*} 
Applying these substitutions to \rref{rij5} yields
\begin{align*}
    \matrixel{\phi_k}{\frac{1}{r_{ij}^2}}{\phi_l} = \frac{\textbf{S}_{kl}}{a}\int_{1}^{+\infty} \ur{d}w \,  \frac{1}{w^{\frac{3}{2}}} \exp[\frac{(\frac{1-w}{a})\beta}{w}] \ , \nr
\end{align*}
where the exponential can be written as 
\begin{align*}
    \exp[\frac{(\frac{1-w}{a})\beta}{w}] = \exp[\frac{\beta}{wa}-\frac{\beta}{a}] = \exp[\frac{\beta}{wa}]\exp[-\frac{\beta}{a}] \ . \nr
\end{align*}
We can then remove the $\exp[-\frac{\beta}{a}]$ term from the integrand, giving
\begin{align}
    \matrixel{\phi_k}{\frac{1}{r_{ij}^2}}{\phi_l} = \exp[-\frac{\beta}{a}]\frac{\textbf{S}_{kl}}{a}\int_{1}^{+\infty}\ur{d}w \,  \frac{1}{w^\frac{3}{2}} \exp[\left(\frac{\beta}{a}\right)\frac{1}{w}] \ . \label{possibleInt}
\end{align}
Finally, we make the substitution $w=\frac{\beta}{ay^2}$ from which the following terms and limits result;
\begin{align*}
    \ur{d}w &= -\frac{2\beta}{ay^3}\ur{d}y \, , \nr \\
    \frac{1}{w^{\frac{3}{2}}} &= \left(\frac{a}{\beta}\right)^{\frac{3}{2}}y^3 \, , \nr \\
    y = \left(\frac{\beta}{aw}\right)^{\frac{1}{2}} &\implies
    \begin{cases}
        w \rightarrow \infty , & y \rightarrow 0 \\
        w = 1 , & y = \sqrt{\frac{\beta}{a}} \nr \label{Substitutions} \, .
    \end{cases}
\end{align*}
Applying these substitutions to \rref{possibleInt} results in
\begin{align*}
    \matrixel{\phi_k}{\frac{1}{r_{ij}^2}}{\phi_l} &= \exp[-\frac{\beta}{a}]\frac{\textbf{S}_{kl}}{a} \int_{\sqrt{\frac{\beta}{a}}}^0\ur{d}y \, \left(-2\frac{\beta}{ay^3}\right)\left(\frac{a}{\beta}\right)^{\frac{3}{2}}y^3\exp(y^2) \\
    & = -2\exp[-\frac{\beta}{a}] \frac{\textbf{S}_{kl}}{\sqrt{a\beta}}\int_{\sqrt{\frac{\beta}{a}}}^0 \ur{d}y \, \exp(y^2) \ . \nr \label{ErfiInt}
\end{align*}
We then recognise the integral of \rref{ErfiInt} to be related to the complex error function such that
\begin{align}
    \int_{a}^{b} \ur{d}x \, \exp(x^2) = \frac{\sqrt{\pi}}{2}\left[\erfi(x)\right]_{a}^{b} \, . \label{Erfi} 
\end{align}
When the argument of the complex error function is equal to zero, the function itself is zero. Hence, we have
\begin{align*}
   \matrixel{\phi_k}{\frac{1}{r_{ij}^2}}{\phi_l} &= -2\textbf{S}_{kl}\frac{\exp[-\frac{\beta}{a}]}{\sqrt{a\beta}}\frac{\sqrt\pi}{2}\left[0 - \erfi\left[\sqrt{\frac{\beta}{a}}\right]\right] \, , \nr
\end{align*}
meaning that the final result is
\begin{align}
    \matrixel{\phi_k}{\frac{1}{r_{ij}^2}}{\phi_l} = \textbf{S}_{kl} \frac{\sqrt{\pi}\exp[-\frac{\beta}{a}]}{\sqrt{a\beta}}\erfi\left[\sqrt{\frac{\beta}{a}}\right]. \label{r_ij^2Final}
\end{align}
The complex error function is a real function for real arguments. Moreover, the parameters $\beta$ and $a$ are real and positive, so there is no complex part in \rref{r_ij^2Final}.


\subsection{The Matrix Element of $1/(r_{ij} r_{pa})$} \label{sec:rijrpa}
\subsubsection{Zero Shift Case} \label{subsec:zeroshift}
\noindent
We begin by setting the shift vector in our ECG basis functions, \rref{ECGEQN}, to zero, meaning that the $k^{\ur{th}}$ basis function can be written as
\begin{align}
    \phi_k = \exp[-\br^T\ubA_k\br] \ . \label{noshiftbasis}
\end{align}
We can then write the matrix element $\phikl[\frac{1}{r_{ij}r_{pa}}]$ for a non-floating ECG basis set as $\phikl[\frac{1}{r_{ij}r_{pa}}]|_{\bs = 0}$ such that
\begin{align}
    \phikl[\frac{1}{r_{ij}r_{pa}}]|_{\bs = 0} = \Rintegral\frac{1}{r_{ij}r_{pa}}\phi_k\phi_l = \Rintegral\frac{1}{r_{ij}r_{pa}}\exp[- \br^T\ubA_{kl}\br] \ , \label{rijrpa1}
\end{align}
where we have combined the arguments of the two exponentials together and used $\ubA_k + \ubA_l = \ubA_{kl}$. The next step is to re-write the denominator such that it appears in the argument of the exponential. This can be done using the same approach used in Sec. \ref{subsec:1/rij^2}, by writing the terms in the denominator as
\begin{align*}
    \frac{1}{r_{ij}}= \frac{2}{\sqrt{\pi}}\int_{0}^{+\infty}\ur{d}t \, \exp[-\br^T\uJij\br t^2] \, , \nr
\end{align*}
and 
\begin{align*}
    \frac{1}{r_{pa}}= \frac{2}{\sqrt{\pi}}\int_{0}^{+\infty} \ur{d}u \, \exp[-\br^T\underline{\textbf{J}}_{pa}\br u^2] \ , \nr
\end{align*}
meaning that \rref{rijrpa1} now takes the form
\begin{align*}
    \phikl[\frac{1}{r_{ij}r_{pa}}]|_{\bs = 0} = \frac{4}{\pi}\int_{0}^{+\infty}\ur{d}t \, \int_{0}^{+\infty}\ur{d}u \, \Rintegral \exp[-\br^T(\ubA_{kl}+t^2\uJij +u^2\underline{\textbf{J}}_{pa})\br] \, . \nr 
\end{align*}
From here, we apply \rref{nonfloatmultidgauss}, giving
\begin{align*}
    \phikl[\frac{1}{r_{ij}r_{pa}}]|_{\bs = 0} = \frac{4}{\pi}\pi^{\frac{3n}{2}} \int_{0}^{+\infty} \ur{d}t \,  \int_{0}^{+\infty}\ur{d}u \, |\bA_{kl}+t^2\Jij + u^2\textbf{J}_{pa}|^{-\frac{3}{2}} \ . \nr \label{eq:rijrpajunc}
\end{align*}
By re-writing the determinant above as 
\begin{align*}
    |\bA_{kl}+t^2\Jij + u^2\textbf{J}_{pa}|^{-\frac{3}{2}} = |\bI + t^2 \Jij \bA_{kl}^{-1}+ u^2 \textbf{J}_{pa}\bA_{kl}^{-1}|^{-\frac{3}{2}}|\bA_{kl}|^{-\frac{3}{2}} \ , \nr
\end{align*}
we can use the rank-2 determinant expression of  \rref{eq:rank2det}. Therefore, by letting $\textbf{H}_1 =t^2\Jij \bA_{kl}^{-1}$ and $\textbf{H}_2 = u^2\textbf{J}_{pa}\bA_{kl}^{-1}$,
the integral reads as
\begin{align*}
  \phikl[\frac{1}{r_{ij}r_{pa}}]|_{\bs = 0} = \frac{4}{\pi}\pi^{\frac{3n}{2}}|\bA_{kl}|^{-\frac{3}{2}} \int_{0}^{+\infty}\ur{d}t \, \int_{0}^{+\infty}\ur{d}u \, \left[1+t^2a + u^2b + u^2t^2(ab - c)\right]^{-\frac{3}{2}} \ , \nr \label{eq:rijrparnk2}
\end{align*}
where we have defined 
\begin{align*}
    a &= \Tr{\Jij\bA_{kl}^{-1}} \ , \nr \label{eq:aTr} \\ 
    b &= \Tr{\Jpa\bA_{kl}^{-1}} \ , \nr \label{eq:bTr} \\  
    c &= \Tr{\Jij\bA_{kl}^{-1}\Jpa\bA_{kl}^{-1}} \ . \nr \label{eq:cTr}
\end{align*}
We can then see that
\begin{align*}
    1+t^2a + u^2b + u^2t^2(ab-c) &= 1+t^2a + u^2b +u^2bt^2a - u^2t^2c \\
    &=(1+t^2a) + u^2b (1+t^2a) - u^2t^2c \\
    &= (1+t^2a)\left(1 + u^2b - u^2\frac{t^2c}{1+t^2a}\right) \\
    &= (1+t^2a)\left[1+ u^2 \left(b - \frac{t^2c}{1+t^2a}\right)\right] \ , \nr
\end{align*}
which gives
\begin{align}
    \phikl[\frac{1}{r_{ij}r_{pa}}]|_{\bs = 0} = \frac{4}{\pi}\pi^{\frac{3n}{2}}|\bA_{kl}|^{-\frac{3}{2}} \int_{0}^{+\infty}\ur{d}t \, \int_{0}^{+\infty}\ur{d}u \,(1+t^2a)^{-\frac{3}{2}}
    \left[1+u^2 \left( b-\frac{t^2c}{1+t^2a} \right) \right]^{-\frac{3}{2}} \ . \label{eq:rijrparnk2_2}
\end{align}
Considering \rref{eq:rijrparnk2_2}, we can make the following substitutions;
\begin{align*}
    &w = \frac{u^2}{1+u^2\left(b-\frac{t^2c}{1+t^2a}\right)}  \nr \\ 
    \implies &\frac{\diff{w}}{\diff{u}} = 
    \left(\frac{2u}{1+u^2\left(b-\frac{t^2c}{1+t^2a}\right)} - \frac{2u^3\left(b-\frac{t^2c}{1+t^2a}\right)}{\left(1+u^2\left(b-\frac{t^2c}{1+t^2a}\right)\right)^2}\right) \\
    \implies &\ur{d}u = \ur{d} w \frac{\left(1+u^2\left(b-\frac{t^2c}{1+t^2a}\right)\right)^2}{2u} \ . \nr \label{eq:rijrpasub1}
\end{align*}
The limits also change as 
\begin{align*}
    w = \frac{u^2}{1+u^2\left(b-\frac{t^2c}{1+t^2a}\right)} \implies 
    \begin{cases}
        u \rightarrow \infty, & w \rightarrow \left(b-\frac{t^2c}{1+t^2a}\right)^{-1} \\
        u = 0, & w = 0 \ .
    \end{cases}  \nr \label{eq:rijrpalimits2}
\end{align*}
Making these changes, \rref{eq:rijrparnk2_2} now reads as
\begin{align*}
 \phikl[\frac{1}{r_{ij}r_{pa}}]|_{\bs = 0} &= \frac{4}{\pi}\pi^{\frac{3n}{2}}|\bA_{kl}|^{-\frac{3}{2}} 
 \int_{0}^{+\infty}\ur{d}t \int_{0}^{\left(b-\frac{t^2c}{1+t^2a}\right)^{-1}} \ur{d} w \,(1+t^2a)^{-\frac{3}{2}} \frac{\left(1+u^2\left(b-\frac{t^2c}{1+t^2a}\right)\right)^2}{\left(1+u^2 \left( b-\frac{t^2c}{1+t^2a} \right) \right)^{-\frac{3}{2}}}\frac{1}{2u} \ . \nr
\end{align*}
We then note that 
\begin{align*}
    \frac{\left(1+u^2 \left( b-\frac{t^2c}{1+t^2a} \right) \right)^{\frac{1}{2}}}{2u} = \frac{1}{2w^{\frac{1}{2}}} \ , \nr
\end{align*}
and so
\begin{align*}
 \phikl[\frac{1}{r_{ij}r_{pa}}]|_{\bs = 0} = \frac{2}{\pi}\pi^{\frac{3n}{2}} |\bA_{kl}|^{-\frac{3}{2}}
 \int_{0}^{+\infty}\ur{d}t \, \int_{0}^{\left(b-\frac{t^2c}{1+t^2a}\right)^{-1}} \ur{d} w \,(1+t^2a)^{-\frac{3}{2}}\frac{1}{w^{\frac{1}{2}}} \ . \nr
\end{align*}
We can then evaluate the integral in $w$, which gives
\begin{align*}
    \phikl[\frac{1}{r_{ij}r_{pa}}]|_{\bs = 0} &=  \frac{2}{\pi}\pi^{\frac{3n}{2}} |\bA_{kl}|^{-\frac{3}{2}}
 \int_{0}^{+\infty}\ur{d}t \, (1+t^2a)^{-\frac{3}{2}} 2\left[w^{\frac{1}{2}}\right]_{0}^{\left(b-\frac{t^2c}{1+t^2a}\right)^{-1}} \\
 &=\frac{4}{\pi}\pi^{\frac{3n}{2}} |\bA_{kl}|^{-\frac{3}{2}}
 \int_{0}^{+\infty}\ur{d}t \, (1+t^2a)^{-\frac{3}{2}} \left(b-\frac{t^2c}{1+t^2a}\right)^{-\frac{1}{2}} \ . \nr \label{eq:rijrpaIntermediate}
\end{align*}
We then make another substitution, 
\begin{align*}
    z &= \left(\frac{t^2c}{1+t^2a}\right)^{\frac{1}{2}} \, , \nr
\end{align*}
which, by comparison of this substitution with that of \rref{eq:rijrpasub2}, we see that
\begin{align*}
    z &= \left(\frac{t^2c}{1+t^2a}\right)^{\frac{1}{2}} \\
    &\implies \frac{\diff{z}}{\diff{t}} = 
    \frac{1}{2} \underbrace{\left( \frac{t^2c}{1+t^2a}\right)^{-\frac{1}{2}}}_{z^{-1}}\frac{2tc}{(1+t^2a)^2} \\
    &\implies \frac{\diff{z}}{\diff{t}} = \frac{1}{2z}\frac{2tc}{(1+t^2a)^2} \\
    &\implies \ur{d}t = \ur{d} z \frac{z(1+t^2a)^2}{tc} \ . \nr \label{eq:rijrpasub2}
\end{align*}
The limits also change as 
\begin{align*}
    z = \left(\frac{t^2c}{1+t^2a}\right)^{\frac{1}{2}} \implies 
    \begin{cases}
        t \rightarrow \infty, & z \rightarrow \sqrt{\frac{c}{a}} \\
        t = 0, & z = 0 \ .
    \end{cases}  \nr \label{eq:rijrpalimits3}
\end{align*}
Substituting these expression into \rref{eq:rijrparnk2_2} then gives
\begin{align*}
    \phikl[\frac{1}{r_{ij}r_{pa}}]|_{\bs = 0} = \frac{4}{\pi}\pi^{\frac{3n}{2}} |\bA_{kl}|^{-\frac{3}{2}}
    \int_{0}^{\sqrt{\frac{c}{a}}}\ur{d}z \, \frac{(1+t^2a)^2}{(1+t^2a)^{\frac{3}{2}}}\frac{z}{tc}\frac{1}{(b-z^2)^{\frac{1}{2}}} \ , \nr
\end{align*}
where we note that 
\begin{align*}
    \frac{(1+t^2a)^2}{(1+t^2a)^{\frac{3}{2}}} \frac{z}{tc} = z \underbrace{\lrr{\frac{(1+t^2a)^{\frac{1}{2}}}{tc}}}_{\frac{1}{z\sqrt{c}}}  
    =\frac{1}{\sqrt{c}} \ , \nr
\end{align*}
and so
\begin{align}
    \phikl[\frac{1}{r_{ij}r_{pa}}]|_{\bs = 0} = \frac{4}{\pi}\frac{\pi^{\frac{3n}{2}} |\bA_{kl}|^{-\frac{3}{2}}}{\sqrt{c}}
    \int_{0}^{\sqrt{\frac{c}{a}}}\ur{d}z \, \frac{1}{(b-z^2)^{\frac{1}{2}}} \ .
\end{align}
This can be solved by making one final substitution; we first define $z = \sqrt{b}\sin{(\rho)}$, meaning
\begin{align*}
    \frac{\diff{z}}{\diff{\rho}} &= \sqrt{b}\cos{(\rho)} \\
    \implies \ur{d} z &= \ur{d}\rho \,  \sqrt{b}\cos{(\rho)} \ . \nr \label{eq:rijrpasub3}
\end{align*}
The new limits are such that\footnote{We use $\sin^{-1}{(x)}$ to refer to $\arcsin{(x)}$, not $\frac{1}{\sin{(x)}}$.} 
\begin{align*}
    \rho = \sin^{-1}\left(\frac{z}{\sqrt{b}}\right) \implies 
    \begin{cases}
        z \rightarrow \sqrt{\frac{c}{a}}, & \rho \rightarrow \sin^{-1}\left(\sqrt{\frac{c}{ab}}\right)  \\
        z = 0, & \rho = 0 \, .
    \end{cases} \  \nr \label{eq:rijrpalimits4}
\end{align*}
Using Eqs. (\ref{eq:rijrpasub3}) and (\ref{eq:rijrpalimits4}), the integral becomes

\begin{align*}
    \phikl[\frac{1}{r_{ij}r_{pa}}]|_{\bs = 0} &= \frac{4}{\pi}\frac{\pi^{\frac{3n}{2}} |\bA_{kl}|^{-\frac{3}{2}}}{\sqrt{c}}
    \int_{0}^{\sin^{-1}\left(\sqrt{\frac{c}{ab}}\right)}\ur{d}\rho \, \frac{\sqrt{b}\cos{(\rho)}}{(b-b\sin^2{(\rho)})^{\frac{1}{2}}} \\
    &=\frac{4}{\pi}\frac{\pi^{\frac{3n}{2}} |\bA_{kl}|^{-\frac{3}{2}}}{\sqrt{c}}
    \int_{0}^{\sin^{-1}\left(\sqrt{\frac{c}{ab}}\right)} \ur{d} \rho \, \frac{ \sqrt{b} \cos{(\rho)}}{\sqrt{b}(\underbrace{1 - \sin^2(\rho)}_{\cos^2(\rho)})^{\frac{1}{2}} } \\
    &= \frac{4}{\pi}\frac{\pi^{\frac{3n}{2}} |\bA_{kl}|^{-\frac{3}{2}}}{\sqrt{c}} \int_{0}^{\sin^{-1}\left(\sqrt{\frac{c}{ab}}\right)} \ur{d}\rho \,  \frac{\sqrt{b}\cos{(\rho)}}{\sqrt{b}\cos{(\rho)}} \\
    &= \frac{4}{\pi}\frac{\pi^{\frac{3n}{2}} |\bA_{kl}|^{-\frac{3}{2}}}{\sqrt{c}}
    \int_{0}^{\sin^{-1}\left(\sqrt{\frac{c}{ab}}\right)}\ur{d}\rho \ . \nr \label{eq:rijrpa4}
\end{align*}
Hence, evaluation of \rref{eq:rijrpa4} leads to the final expression 
\begin{align*}
    \phikl[\frac{1}{r_{ij}r_{pa}}]|_{\bs = 0} = \frac{4}{\pi}\frac{\pi^{\frac{3n}{2}} |\bA_{kl}|^{-\frac{3}{2}}}{\sqrt{c}} \sin^{-1}\left(\sqrt{\frac{c}{ab}}\right) \ . \nr \label{rijrpafinals0}
\end{align*}
The condition for this expression to be true is $ab \ge c$, as otherwise the argument of the $\sin^{-1}$ function is outwith its domain, and so the matrix element is undefined. This condition is certainly met when the matrices $\bA_{kl}$, $\Jij$ and $\Jpa$ are $2\times2$ matrices, which is the $n = 2$ case. This can be proven using the Cholesky decomposition which states that as $\bd{A}_{kl}$ is a positive definite matrix, it can be written as a product of a lower-triangular matrix $\bd{L}$ and its transpose, $\bd{L}^T$; if we define $\bd{L}$ to be 
\begin{align*}
    \bd{L} = 
    \begin{pmatrix*}[r] 
     L_{11} & 0 \\
     L_{21} & L_{22}  \\
    \end{pmatrix*} \ , \nr
\end{align*}\\ 
then
\begin{align*}
    \bd{A}_{kl} = \bd{L}\bd{L}^T = 
        \begin{pmatrix*}
        {L_{11}}^2 & L_{11}L_{21} \\
        L_{21}L_{11} & \lrr{{L_{21}}^2 + {L_{22}}^2} \\ 
        \end{pmatrix*} \ , \nr
\end{align*}
which has an inverse
\begin{align*}
    \bd{A}_{kl}^{-1} = 
        \begin{pmatrix*}
        \frac{\lrr{{L_{21}}^2 + {L_{22}}^2}}{{L_{11}}^2{L_{22}}^2} & -\frac{L_{21}}{L_{11}{L_{22}}^2} \\
         -\frac{L_{21}}{L_{11}{L_{22}}^2} & \frac{1}{{L_{22}}^2} \\
        \end{pmatrix*} \ . \nr
\end{align*}
We can then simplify the $\bd{J}_{pa}$ matrix by shifting the centre of the coordinate system to $\bd{a}$, meaning that the $\bd{J}_{pa}$ matrix is redefined as the $\bd{J}_{pp}$ matrix, where
\begin{align}
    (\bJ_{pp})_{\alpha \beta}=\delta_{p \alpha} \delta_{p \beta} \ . \label{eq:JppDef} 
\end{align}
In other words, all entries of $\bJ_{pp}$ are zero  apart from the $p$-th diagonal entry (where $p = \alpha = \beta)$ which is equal to one. We then generate the $\bd{J}_{ij}$ and $\bd{J}_{pp}$ matrices using Eqs. (\ref{eq:JijDef}) and (\ref{eq:JppDef}) respectively for $n=2$, noting that $\bd{J}_{12} = \bd{J}_{21}$. These are defined as
\begin{align*}
    \bd{J}_{12} = \bd{J}_{21} = 
        \begin{pmatrix*}
        1 & -1 \\
        -1 & 1
        \end{pmatrix*} \ , \nr
\end{align*}
and 
\begin{align*}
    \bd{J}_{11} = 
        \begin{pmatrix*}
        1 & 0 \\
        0 & 0
        \end{pmatrix*} \ , \  \, \bd{J}_{22} = 
                \begin{pmatrix*}
        0 & 0 \\
        0 & 1
        \end{pmatrix*} \ . \nr
\end{align*}
We then use these matrices to find $a$, $b$ and $c$ as defined by Eqs. (\ref{eq:aTr}), (\ref{eq:bTr}) and (\ref{eq:cTr}) respectively. In order to prove that $ab \ge c$, we can show that $ab - c \ge 0$. By calculating $a$, $b$ and $c$, we find that
\begin{align*}
    ab - c = \frac{1}{{L_{11}}^2{L_{22}}^2} \ . \nr
\end{align*}
Examining the above expression, we note that the matrix $\bd{L}$ is lower triangular, and so its determinant is the product of its diagonal elements. Moreover, its transpose has the same determinant as it is upper triangular. As such,
\begin{align*}
    {L_{11}}^2{L_{22}}^2 = ({L_{11}}{L_{22}})({L_{11}}{L_{22}}) = \det(\bd{L})\det(\bd{L}^T) = \det(\bd{L}\bd{L}^T) \ . \nr
\end{align*}
We then recall that $\bd{A}_{kl} = \bd{L}\bd{L}^T$, and so
\begin{align*}
    ab - c = \frac{1}{\det(\bd{A}_{kl})}. \nr
\end{align*}
As $\bd{A}_{kl}$ is positive-definite, its determinant is always positive. Hence, for $n = 2$, it is true that $ab \ge c$.
The relationship between $a$, $b$ and $c$ for any number of particles is currently being investigated.
\subsubsection{General Case} \label{subsec:gencase}
\noindent
Now using floating ECGs, we begin with
\begin{align}
    \matrixel{\phi_k}{\frac{1}{r_{ij}r_{pa}}}{\phi_l}=\int_{-\infty}^{+\infty}d\textbf{r} \, \phi_k\phi_l\frac{1}{r_{ij}r_{pa}} \ .
\end{align}
Applying \rref{eq:ProductECG} then yields
\begin{align}
    \matrixel{\phi_k}{\frac{1}{r_{ij} r_{pa}}}{\phi_l} = \exp\lrr{- \eta_{kl}} \int_{-\infty}^{\infty} \mbox{d} \br \, \frac{1}{r_{ij} r_{pa}} \exp \left[ - \br^T \ubA_{kl} \br + 2 \be^T \br \right] \ . \label{initialintrijrpa}
\end{align}
In order to simplify the denominator, the center of the coordinate system is shifted to $\ba$, giving
\begin{align*}
    \matrixel{\phi_k}{\frac{1}{r_{ij} r_{pa}}}{\phi_l} =  \exp\lrr{- \eta_a} \int_{-\infty}^{\infty} \mbox{d} \br \, \frac{1}{r_{ij} r_p} \exp \left[ - \br^T \ubA_{kl} \br + 2 \be_a^T \br \right], \nr
\end{align*}
where we have introduced the following notation;
\begin{align}
    \bs_k^a &= \bs_k -\ba \ , \\
    \be_k^a &=\bA \bs_k^a \ , \\
    \be_a &= \be_k^a+ \be_l^a \ , \\
    \eta_a &= \left({\bf s}_k^a\right)^T \underline{{\bf A}}_{k}^{-1} {\bf s}_{k}^a+\left({\bf s}_{l}^a\right)^T \underline{{\bf A}}_{l}^{-1} {\bf s}_{l}^a \ .
\end{align}
We now need to convert the denominator of the integrand into a Gaussian. This can be done by using \rref{onedg} but over the range $ x \in [0,\infty)$,
\begin{align*}
    2\int_{0}^{\infty} dx \, \exp[-ax^2] = \sqrt{\frac{\pi}{a}} \ . \nr
\end{align*}
If $a = (\textbf{r}_i-\textbf{r}_j)^2$, then
\begin{align*}
   \sqrt{\frac{\pi}{(\textbf{r}_i-\textbf{r}_j)^2}} =  2\int_{0}^{\infty} \ur{d}t \, \exp[-(\textbf{r}_i-\textbf{r}_j)^2t^2]
\end{align*}
Then, using the fact that $(\textbf{r}_i-\textbf{r}_j)^2 = |\textbf{r}_{i} - \textbf{r}_{j}|^2$ $=$ $r_{ij}^2$, we find that 
\begin{align}
    \frac{1}{r_{ij}} = \frac{2}{\sqrt{\pi}}\int_{0}^{\infty}\ur{d}t \, \exp[-(\textbf{r}_i-\textbf{r}_j)^2t^2] \ . \label{TransExp}
\end{align}
We can then introduce a matrix $\Jij$ such that
\begin{align}
    \br^T \uJij \br = (\br_i-\br_j)^2 \ . \label{Jij2} 
\end{align}
Hence, we can re-write \rref{TransExp} as
\begin{align}
     \frac{1}{r_{ij}} = \frac{2}{\sqrt{\pi}}\int_{0}^{\infty}\ur{d}t \, \exp[-\br^T \uJij \br t^2] \ . \label{rij}
\end{align}
We then use the same method for the $\frac{1}{r_{p}}$ term, though we instead use the definition of $(\bJ_{pp})_{\alpha \beta}$, as by \rref{eq:JppDef}. This allows us to write
\begin{align}
    \bd{r}^T \bd{J}_{pp}\bd{r} = r_{p} \ , \label{eq:rpdef}
\end{align}
and so
\begin{align}
    \frac{1}{r_{p}} = \frac{2}{\sqrt{\pi}}\int_{0}^{\infty}\ur{d}u \, \exp[-\br^T \uJpp \br u^2] \ . \label{Newrp}
\end{align}
Applying Eqs. (\ref{rij}) and (\ref{Newrp}), we arrive at 
\begin{align*}
    \matrixel{\phi_k}{\frac{1}{r_{ij} r_{pa}}}{\phi_l} &= \frac{4}{\pi}  \exp\lrr{- \eta_a} 
    \int_0^\infty \mbox{d} u \int_0^\infty \mbox{d}t  \int_{-\infty}^{\infty} \mbox{d} \br \, \exp \left[ - \br^T(\ubA_{kl}+t^2\textbf{\underline{J}}_{ij} +u^2\textbf{\underline{J}}_{pp})\br + 2 \be_a^T \br \right] \ . \nr 
\end{align*} 
We can now evaluate the integral in \textbf{r} using \rref{mdimgauss} which gives,
\begin{align*}
    \matrixel{\phi_k}{\frac{1}{r_{ij}r_{pa}}}{\phi_l} &= \frac{4}{\pi} \exp\lrr{- \eta_a} \int_0^{+\infty}\ur{d}u 
    \int_0^{+\infty}\ur{d}t \,\frac{\pi^{\frac{3n}{2}}}{|\bA_{kl}+t^2\textbf{J}_{ij}+u^2\textbf{J}_{pp}|^{\frac{3}{2}}}\exp[\be_a^T(\ubA_{kl}+t^2\textbf{\underline{J}}_{ij} +u^2\textbf{\underline{J}}_{pp})^{-1}\be_a] \ .  \nr \label{Newrijrpa}
\end{align*} \\
Examining \rref{Newrijrpa}, we can introduce 
\begin{align*}
    \ubB_{klpp}(u) = \ubA_{kl}+u^2\ubJ_{pp} \ , \nr
\end{align*}
giving
\begin{align*}
    \phikl[\frac{1}{r_{ij}r_{pa}}] &= \frac{4}{\pi} \exp\lrr{- \eta_a} \int_{0}^{+\infty} \ur{d}u  \int_{0}^{+\infty} \ur{d}t\, \frac{\pi^{\frac{3n}{2}}}{|\bB_{klpp}(u)+ t^2\textbf{J}_{ij}|^{\frac{3}{2}}}\exp\left[\be_a^T\left(\ubB_{klpp}(u)+t^2\ubJ_{ij}\right)^{-1}\be_a\right] \ . \label{eq:Newrijrpa.2} \nr
\end{align*}
We then makes use of Lemma \ref{lemma:productrank}, and Eqs. (\ref{eq:lincomdet}) and (\ref{lemma:asumbinvunder}) to rewrite the determinant and inverse of \rref{eq:Newrijrpa.2}, leading to
\begin{align*}
    \phikl[\frac{1}{r_{ij}r_{pa}}] &= \frac{4}{\pi} \exp\lrr{- \eta_a} \int_{0}^{+\infty} \ur{d}u \\ 
    &\times\int_{0}^{+\infty} \ur{d}t\,  \frac{\pi^{\frac{3n}{2}}}{\left(1+t^2\Tr{\textbf{J}_{ij}\ubB_{klpp}^{-1}(u)}\right)^{\frac{3}{2}}|\textbf{B}_{klpp}(u)|^{\frac{3}{2}}} \\
    &\times \exp\left[\be_a^T\left(\ubB_{klpp}^{-1}(u) - t^2\frac{\ubB_{klpp}^{-1}(u)\uJij\ubB_{klpp}^{-1}(u)}{1+t^2\Tr{\textbf{J}_{ij}\textbf{B}_{klpp}^{-1}(u)}}\right)\be_a\right] \\ 
    \\
    &=\frac{4}{\pi} \exp\lrr{- \eta_a} \int_{0}^{+\infty} \ur{d}u \\ &\times\int_{0}^{+\infty} \ur{d}t\, \frac{\pi^{\frac{3n}{2}}}{\left(1+t^2\Tr{\textbf{J}_{ij}\ubB_{klpp}^{-1}(u)}\right)^{\frac{3}{2}}|\textbf{B}_{klpp}(u)|^{\frac{3}{2}}} \\
    &\times\exp[\be_a^T\ubB_{klpp}^{-1}(u)\be_a]\exp\left[\left(-t^2\frac{\be_a^T\ubB_{klpp}^{-1}(u)\uJij\ubB_{klpp}^{-1}(u)\be_a}{1+t^2\Tr{\textbf{J}_{ij}\textbf{B}_{klpp}^{-1}(u)}}\right)\right] \ . \nr
\end{align*}
Before we integrate in $t$, we introduce the following notation which allows us to remove factors which are independent of $t$ and express the integral in a more manageable form;
\begin{align}
    T_{kl}^a(u) &=  \exp \left( -\eta_a \right) \exp \left( \be_a^T \ubB_{klpp}^{-1}(u) \be_a \right) \frac{\pi^{3n/2}}{|\bB_{klpp}(u)|^{3/2}} \ , \\
    \delta(u) &= \be_a^T \ubB_{klpp}^{-1}(u)\uJij\ubB_{klpp}^{-1}(u) \be_a \ , \\
    b(u) &= \Tr{\bB_{klpp}^{-1}(u)\Jij} \ , 
\end{align}
which gives
\begin{align*}
    \phikl[\frac{1}{r_{ij}r_{pa}}] = \frac{4}{\pi} \int_{0}^{+\infty}\ur{d}u \, T_{kl}^a(u) \int_{0}^{+\infty} \ur{d}t \, \left(1+t^2b(u)\right)^{-\frac{3}{2}}\exp[\frac{-t^2\delta(u)}{1+t^2b(u)}] \ . \nr \label{intermediaterijrpa1}
\end{align*}
Now we make the substitution 
\begin{align*}
    z &= \frac{t^2\delta(u)}{1+t^2b(u)} \nr \label{zsub} \\
    &\implies \frac{\diff{z}}{\diff{t}} = \frac{2t\delta(u)}{1+t^2b(u)} - \frac{2t^3\delta(u)b(u)}{(1+t^2b(u))^2} \\
    &\implies \frac{\diff{z}}{\diff{t}} = \frac{2t\delta(u)}{(1+t^2b(u))^2} \\
    &\implies \ur{d}t = \ur{d} z\frac{(1+t^2b(u))^2}{2t\delta(u)} \ . \nr \label{dzsub}
\end{align*}
Then,
\begin{align*}
    \ur{d}t \left(1+t^2b(u)\right)^{-\frac{3}{2}} &= \ur{d}z \frac{(1+t^2b(u))^2}{(1+t^2b(u))^{\frac{3}{2}}} \frac{1}{2t\delta(u)} = \frac{(1+t^2b(u))^{\frac{1}{2}}}{2t\delta(u)} \\
    &= \ur{d}z \frac{1}{2\delta^{\frac{1}{2}}(u)}\frac{1}{z^{\frac{1}{2}}} \nr \label{zsub2} \ , 
\end{align*}
with the limits altering as
\begin{align*}
    z = \frac{t^2\delta(u)}{1+t^2b(u)} \implies
    \begin{cases}
        t \rightarrow \infty, & z \rightarrow \frac{\delta(u)}{b(u)}  \\
        t = 0, & z = 0 \ .
    \end{cases} \nr \label{eq:zsublim}
\end{align*}
Making the substitutions in Eqs. (\ref{zsub}) - (\ref{eq:zsublim}), we arrive at
\begin{align*}
    \phikl[\frac{1}{r_{ij}r_{pa}}] = \frac{2}{\pi} \exp\lrr{- \eta_a} \int_{0}^{+\infty}\ur{d}u \, \frac{T_{kl}^a(u)}{\delta(u)^{\frac{1}{2}}} \int_{0}^{\frac{\delta(u)}{b(u)}} \ur{d}z \, \frac{1}{z^{\frac{1}{2}}}\exp(-z) \nr \label{gammaint} \ .
\end{align*}
We now recognise the integrand of \rref{gammaint} to be related to that of the lower incomplete gamma function,
\begin{align*}
    \gamma(s,x) = \int_{0}^{x} \ur{d}z \, z^{s-1} \exp(-z) \ , \nr \label{loingamma}
\end{align*}
where in our case, $s=\frac{1}{2}$ and $x = \frac{\delta(u)}{b(u)}$ . Then,
\begin{align*}
     \phikl[\frac{1}{r_{ij}r_{pa}}] = \frac{2}{\pi} \int_{0}^{+\infty}\ur{d}u \, \frac{T_{kl}^a(u)}{\delta^{\frac{1}{2}}(u)} \gamma\left(\frac{1}{2},\frac{\delta(u)}{b(u)}\right) \ . \nr 
\end{align*}
When $s=\frac{1}{2}$, the lower incomplete gamma function is related to the error function as 
\begin{align*}
    \gamma\left(\frac{1}{2}, x \right) =\pi^{\frac{1}{2}}\erf\left(x^{\frac{1}{2}}\right) \nr \label{erf} \ .
\end{align*}
Making the substitution in \rref{erf} gives the integral
\begin{align}
    &\matrixel{\phi_k}{\frac{1}{r_{ij} r_{pa}}}{\phi_l} = \frac{2}{\pi^{1/2}}   \int_0^{+\infty} \mbox{d} u \,  \frac{T_{kl}^a(u) }{\sqrt{\delta(u)}} \erf\left(\sqrt{\frac{\delta(u)}{b(u)}}\right) \ , \label{eq:intrijrpa}
\end{align}
where we can see the explicit dependence on $u$ by  applying \rref{eq:lincomdet},
\begin{align*}
    |\bB_{klpp}(u)|= |\bA_{kl} +u^2 \bJ_{pp}| = |\bA_{kl}| |\bI + u^2 \bA_{kl}^{-1}   \bJ_{pp}| = |\bA_{kl}|  \left[ 1+ u^2 \Tr\left(\bA_{kl}^{-1} \bJ_{pp}\right)  \right] \ , \nr
\end{align*}
and \rref{lemma_asumbinv},
\begin{align*}
    \ubB_{klpp}^{-1}(u)= \left(\ubA_{kl} +u^2 \uJpp \right)^{-1} = \ubA_{kl}^{-1}-u^2 \frac{\ubA_{kl}^{-1} \uJpp \ubA_{kl}^{-1}}{1+u^2\Tr\left(\bA_{kl}^{-1}\bJ_{pp}\right)} \ . \nr 
\end{align*}
By using the following substitutions,
\begin{align}
    a_{ij} &= \Tr\left(\bA_{kl}^{-1}\bJ_{ij}\right)  \ , \\
    a_{pp} &= \Tr\left(\bA_{kl}^{-1}\bJ_{pp}\right) \ , \\
    c &= \Tr\left(\bA_{kl}^{-1}\bJ_{pp} \bA_{kl}^{-1}\bJ_{ij}\right) \ , \\
    \gamma_a &= \be_a^T \ubA_{kl}^{-1} \be_a - \eta_a \ , \\
    \beta_a &= \be_a^T \ubA_{kl}^{-1}\uJij\ubA_{kl}^{-1}\be_a \ , \\
    \mu_a &= \be_a^T \ubA_{kl}^{-1}\uJpp\ubA_{kl}^{-1}\be_a \ , \\
    \epsilon_a &= \be_a^T \ubA_{kl}^{-1} \uJpp \ubA_{kl}^{-1} \uJij\ubA_{kl}^{-1} \be_a  +  \be_a^T \ubA_{kl}^{-1} \uJij\ubA_{kl}^{-1} \uJpp \ubA_{kl}^{-1} \be_a  \ , \\
    \omega_a &= \be_a^T \ubA_{kl}^{-1} \uJpp \ubA_{kl}^{-1} \uJij\ubA_{kl}^{-1} \uJpp \ubA_{kl}^{-1}  \be_a \ , \nr 
\end{align}
we arrive at 
\begin{align*}
    T_{kl}^a(u) &= \exp\left(\gamma_a\right)\exp\left(-\frac{u^2 \mu_a}{1+u^2 a_{pp}}\right ) \frac{\pi^{3n/2}}{|\bA_{kl}|^{\frac{3}{2}} \left(1+u^2a_{pp} \right)^{\frac{3}{2}}} \ , \\
    \delta (u) &= \beta_a - \frac{u^2 \epsilon_a }{1+u^2 a_{pp}} + \frac{u^4 \omega_a}{\left( 1+u^2 a_{pp} \right)^2} \ , \\
    b(u) &= a_{ij}-\frac{u^2 c }{1+u^2 a_{pp}} \ . \nr 
\end{align*}
This allows \rref{eq:intrijrpa} to be written as
\begin{align*}
    \matrixel{\phi_k}{\frac{1}{r_{ij} r_{pa}}}{\phi_l} &= \frac{2}{\pi^{\frac{1}{2}}} \exp(\gamma_a)\frac{\pi^{\frac{3n}{2}}}{|\bA_{kl}|^{\frac{3}{2}}} \\ &\times \int_{0}^{+\infty} \ur{d}u \, \left(1+u^2a_{pp}\right)^{-\frac{3}{2}} \frac{\exp\left[\frac{-u^2\mu_a}{1+u^2_{pp}}\right]}{\left(\beta_a - \frac{u^2\epsilon_a}{1+u^2a_{pp}} + \frac{u^4\epsilon_a}{(1+u^2a_{pp})^2}\right)^{\frac{1}{2}}} \\  &\times \erf\left[\left(\frac{\beta_a - \frac{u^2\epsilon_a}{1+u^2a_{pp}} + \frac{u^4\epsilon_a}{(1+u^2a_{pp})^2}}{a_{ij}-\frac{u^2c}{1+u^2a_{pp}}}\right)^{\frac{1}{2}}\right] \ . \nr \label{fullrijrpa}
\end{align*}
In order to further simplify the integral in \rref{fullrijrpa}, we introduce the substitution 
\begin{align*}
    x&= \frac{u^2 \mu_a}{1+u^2 a_{pp}} \ . \nr
\end{align*}
This is analogous to the substitution made in \rref{zsub}, meaning we can use Eqs. (\ref{zsub}), (\ref{dzsub}) and (\ref{zsub2}) and the corresponding limits, where
\begin{align*}
    z &\rightarrow x \, , \\
    t &\rightarrow u \, , \\
    \delta(u) &\rightarrow \mu_a \, , \\ 
    b(u) &\rightarrow a_{pp} \ .
\end{align*}
This then leads to the following integral expression;
\begin{align*}
    \matrixel{\phi_k}{\frac{1}{r_{ij} r_{pa}}}{\phi_l} &= 2\exp(\gamma_a) \, \frac{\pi^{\frac{1}{2}(3n - 1)}|\bd{A}_{kl}|^{-\frac{3}{2}}}{\sqrt{\mu_a}} \\
    &\times \int_0^{\mu_a/a_{pp}} \mbox{d} x \, \frac{e^{ -x} }{\sqrt{x}}(\beta_a- \frac{\epsilon_a}{\mu_a} x + \frac{\omega_a}{\mu_a^2}x^2)^{-\frac{1}{2}} \erf \left[\left( \frac{\beta_a- \frac{\epsilon_a}{\mu_a} x + \frac{\omega_a}{\mu_a^2}x^2}{a_{ij}- \frac{c}{\mu_a} x}\right)^{\frac{1}{2}}\right] \ . \nr \label{Initalrijrpa}
\end{align*}
There is currently no known analytical solution to \rref{Initalrijrpa}. 

\section{Conclusions and Outlooks} \label{sec:conc}
\noindent
The initial aim of this research was to compute lower bounds to the exact energy using the variance of the upper bound energy provided by the variational method. The variance involves calculating the matrix elements of the Hamiltonian squared. The necessary integrals were determined for the Hamiltonian squared within the Born--Oppenheimer approximation. 

It was decided that this approach would be tested on two-electron systems within the Born--Oppenheimer approximation, from which the integrals with no known solutions were identified and solved. This allowed for analytical solutions to the integrals $\phikl[\frac{1}{r_{ij}^2}]$ (using FECGs) and $\phikl[\frac{1}{r_{ij}r_{pa}}]$ (for the zero shift case of ECGs) to be found. The multi-dimensional integral $\phikl[\frac{1}{r_{ij}r_{pa}}]$ using FECGs was simplified to a one-dimensional integral for which no analytic solution was found, but a solution can be calculated using numerical integration. However, singularities are present in integrand, which may require special quadrature and/or integration techniques. Having a numerical solution to this matrix element will allow for lower bounds to be calculated for two-electron systems, with FECGs extending our approach applicability to molecules or polyatomic ions. After this, the integral $\phikl[\frac{1}{r_{ij}^2}]$ would then be generalised to any number of electrons, which would mean solving the $\phikl[\frac{1}{r_{ij}r_{pq}}]$ integral, where all indices represent electrons. This would likely take a form similar to the $\phikl[\frac{1}{r_{ij}r_{pa}}]$  integral, which is why a numerical solution to this integral will be very useful.

The integral expressions derived in Sec. \ref{sec:MatrixEl} have been implemented using FORTRAN in the QUANTEN computer. The first test case for these lower bound expressions will be the helium atom within the Born--Oppenheimer approximation using non-floating ECG basis functions. 

Whilst it would be beneficial to have a solution for the matrix element $\phikl[\frac{1}{r_{ij}r_{pa}}]$ for general FECG basis functions, the solution for $\bd{s} = 0$ is useful not only for the use of (non-floating) ECGs but also for the use of complex ECGs (CECGs) \cite{Muolo2019}, first proposed in 2006 by S. Bubin and L. Adamowicz and used to calculate the ground-state and excited-state energies of the helium atom \cite{Bubin2006}. A CECG basis function takes the form
\begin{align*}
    {\phi_k} = \exp[-\bd{r}^T \lrr{\bd{C}_k \otimes \bd{I}_3}\bd{r}], \nr \label{eq:CECG}
\end{align*}
where $\bd{C}_k = \bd{A}_k + i \bd{B}_k$ is a complex $N_{\ur{p}} \times N_{\ur{p}}$ matrix for $N_{\ur{p}}$ particles. The matrices $\bd{A}_k$ and $\bd{B}_k$ are symmetric and real. 

The matrix elements of Sec \ref{ElecHamInt} not only have use in calculating lower bounds, but also in relativistic quantum mechanics, specifically in mass-velocity corrections. The matrix element in question is the $\phikl[\nabla^4]$ element. In a method now colloquially known as `Drachmanisation', R. J. Drachman showed in 1981 that the expectation value of the single-particle $\nabla_i^4$ operator could be re-written in terms of the trial wave function and expectation energy $E$ \cite{Drachman1981, Pachucki2005a}
\begin{align}
    \matrixel{\Psi_{\ur{Trial}}}{\sum_i \nabla_i^4}{\Psi_{\ur{Trial}}} = 4\matrixel{\Psi_{\ur{Trial}}}{\lrr{E - \hat{V}}^2}{\Psi_{\ur{Trial}}}  - 2\sum_{i \ge j} \braket{\nabla_i^2\Psi_{\ur{Trial}}}{\nabla_j^2\Psi_{\ur{Trial}}}. \label{eq:mvcor}
\end{align}
This expression has better numerical properties for non-exact trial functions compared to the original expression containing the expectation value of the $\nabla_i^4$ operator. This technique can be applied to calculations to compute mass-velocity corrections in relativistic quantum problems, and the expectation value $\expval{\lrr{E-\hat{V}}^2}$  gives rise to the $\matrixel{\Psi_{\ur{Trial}}}{\frac{1}{r_{ij}r_{pa}}}{\Psi_{\ur{Trial}}}$ matrix element discussed in this report. Hence, finding a numerical solution to this integral has use not only in calculating lower bounds, but also in relativistic corrections to the non-relativistic problem. 

Another application of the matrix element $\phikl[\frac{1}{r_{ij}r_{pa}}]$ to non-relativistic problems is their use in so-called rECG functions, which are ECG functions with $|\bd{r}|$ prefactors \cite{Puchalski2017, Puchalski2019}. The prefactor  ensures that the particle-particle cusp condition is satisfied, and so enhances the numerical convergence of the results. 

\section*{Acknowledgements}
\noindent
This work was originally written as an entry for the 35${^{\ur{th}}}$ Orsz\'agos Tudom\'anyos Di\'akk\"ori Konferencia (OTDK) \cite{Ireland2021}. I would like to thank those working within the School of Physics and Astronomy and the School of Chemistry at the University of Glasgow who have helped me during my four years studying chemical physics, as it is their knowledge and experience that has made my ERASMUS+ scholarship at ELTE possible. I am also grateful to my friends at the University of Glasgow and my family in Scotland; without their encouragement and support this opportunity would not have been possible. I am thankful to my friends and colleagues within the Molecular Quantum Dynamics research group at ELTE for the kindness they have shown since I first arrived in Budapest. I would like to pay particular thanks to my supervisors, Edit M\'atyus and P\'eter Jeszenszki, for their guidance throughout the short time I have spent at ELTE.



\bibliography{ECG_MEL} 

\begin{thebibliography}{59}%
\makeatletter
\providecommand \@ifxundefined [1]{%
 \@ifx{#1\undefined}
}%
\providecommand \@ifnum [1]{%
 \ifnum #1\expandafter \@firstoftwo
 \else \expandafter \@secondoftwo
 \fi
}%
\providecommand \@ifx [1]{%
 \ifx #1\expandafter \@firstoftwo
 \else \expandafter \@secondoftwo
 \fi
}%
\providecommand \natexlab [1]{#1}%
\providecommand \enquote  [1]{``#1''}%
\providecommand \bibnamefont  [1]{#1}%
\providecommand \bibfnamefont [1]{#1}%
\providecommand \citenamefont [1]{#1}%
\providecommand \href@noop [0]{\@secondoftwo}%
\providecommand \href [0]{\begingroup \@sanitize@url \@href}%
\providecommand \@href[1]{\@@startlink{#1}\@@href}%
\providecommand \@@href[1]{\endgroup#1\@@endlink}%
\providecommand \@sanitize@url [0]{\catcode `\\12\catcode `\$12\catcode
  `\&12\catcode `\#12\catcode `\^12\catcode `\_12\catcode `\%12\relax}%
\providecommand \@@startlink[1]{}%
\providecommand \@@endlink[0]{}%
\providecommand \url  [0]{\begingroup\@sanitize@url \@url }%
\providecommand \@url [1]{\endgroup\@href {#1}{\urlprefix }}%
\providecommand \urlprefix  [0]{URL }%
\providecommand \Eprint [0]{\href }%
\providecommand \doibase [0]{https://doi.org/}%
\providecommand \selectlanguage [0]{\@gobble}%
\providecommand \bibinfo  [0]{\@secondoftwo}%
\providecommand \bibfield  [0]{\@secondoftwo}%
\providecommand \translation [1]{[#1]}%
\providecommand \BibitemOpen [0]{}%
\providecommand \bibitemStop [0]{}%
\providecommand \bibitemNoStop [0]{.\EOS\space}%
\providecommand \EOS [0]{\spacefactor3000\relax}%
\providecommand \BibitemShut  [1]{\csname bibitem#1\endcsname}%
\let\auto@bib@innerbib\@empty
\bibitem [{\citenamefont {Ritz}(1909)}]{Ritz1909}%
  \BibitemOpen
  \bibfield  {author} {\bibinfo {author} {\bibfnamefont {W.}~\bibnamefont
  {Ritz}},\ }\bibfield  {title} {\bibinfo {title} {{{\"{U}}ber eine neue
  Methode zur L{\"{o}}sung gewisser Variationsprobleme der mathematischen
  Physik.}},\ }\href {https://doi.org/10.1515/crll.1909.135.1} {\bibfield
  {journal} {\bibinfo  {journal} {J. Reine Angew. Math.}\ }\textbf {\bibinfo
  {volume} {1909}},\ \bibinfo {pages} {1} (\bibinfo {year} {1909})}\BibitemShut
  {NoStop}%
\bibitem [{\citenamefont {MacDonald}(1933)}]{MacDonalnd1933}%
  \BibitemOpen
  \bibfield  {author} {\bibinfo {author} {\bibfnamefont {J.~K.~L.}\
  \bibnamefont {MacDonald}},\ }\bibfield  {title} {\bibinfo {title}
  {{Successive Approximations by the Rayleigh-Ritz Variation Method}},\ }\href
  {https://doi.org/10.1103/PhysRev.43.830} {\bibfield  {journal} {\bibinfo
  {journal} {Phys. Rev.}\ }\textbf {\bibinfo {volume} {43}},\ \bibinfo {pages}
  {830} (\bibinfo {year} {1933})}\BibitemShut {NoStop}%
\bibitem [{\citenamefont {Weinstein}(1934)}]{Weinstein1934}%
  \BibitemOpen
  \bibfield  {author} {\bibinfo {author} {\bibfnamefont {D.~H.}\ \bibnamefont
  {Weinstein}},\ }\bibfield  {title} {\bibinfo {title} {{Modified Ritz
  method}},\ }\href {https://doi.org/10.1073/pnas.20.9.529} {\bibfield
  {journal} {\bibinfo  {journal} {Proc. Natl. Acad. Sci. U.S.A}\ }\textbf
  {\bibinfo {volume} {20}},\ \bibinfo {pages} {529} (\bibinfo {year}
  {1934})}\BibitemShut {NoStop}%
\bibitem [{\citenamefont {Temple}(1928)}]{Temple2007}%
  \BibitemOpen
  \bibfield  {author} {\bibinfo {author} {\bibfnamefont {G.}~\bibnamefont
  {Temple}},\ }\bibfield  {title} {\bibinfo {title} {{The theory of Rayleigh's
  principle as applied to continuous systems}},\ }\href
  {https://doi.org/https://doi.org/10.1098/rspa.1928.0098} {\bibfield
  {journal} {\bibinfo  {journal} {Proc. R. Soc. Lond. A}\ }\textbf {\bibinfo
  {volume} {119}},\ \bibinfo {pages} {276} (\bibinfo {year}
  {1928})}\BibitemShut {NoStop}%
\bibitem [{\citenamefont {Suzuki}\ and\ \citenamefont
  {Varga}(1998)}]{suzuki_stochastic_1998}%
  \BibitemOpen
  \bibfield  {author} {\bibinfo {author} {\bibfnamefont {Y.}~\bibnamefont
  {Suzuki}}\ and\ \bibinfo {author} {\bibfnamefont {K.}~\bibnamefont {Varga}},\
  }\href {https://www.springer.com/gp/book/9783540651529} {{\selectlanguage
  {English}\emph {\bibinfo {title} {Stochastic {Variational} {Approach} to
  {Quantum}-{Mechanical} {Few}-{Body} {Problems}}}}}\ (\bibinfo  {publisher}
  {Springer},\ \bibinfo {year} {1998})\BibitemShut {NoStop}%
\bibitem [{\citenamefont {{Zsuzsanna T{\'{o}}th}}(2018)}]{ZsuzsannaToth2018}%
  \BibitemOpen
  \bibfield  {author} {\bibinfo {author} {\bibnamefont {{Zsuzsanna
  T{\'{o}}th}}},\ }\emph {\bibinfo {title} {{Calculating lower bound via
  perturbation theory methods}}},\ \href@noop {} {Ph.D. thesis},\ \bibinfo
  {school} {E{\"{o}}tv{\"{o}}s Lor{\'{a}}nd University} (\bibinfo {year}
  {2018})\BibitemShut {NoStop}%
\bibitem [{\citenamefont {Stanke}\ \emph {et~al.}(2016)\citenamefont {Stanke},
  \citenamefont {Palikot}, \citenamefont {Kedziera},\ and\ \citenamefont
  {Adamowicz}}]{Stanke2016}%
  \BibitemOpen
  \bibfield  {author} {\bibinfo {author} {\bibfnamefont {M.}~\bibnamefont
  {Stanke}}, \bibinfo {author} {\bibfnamefont {E.}~\bibnamefont {Palikot}},
  \bibinfo {author} {\bibfnamefont {D.}~\bibnamefont {Kedziera}},\ and\
  \bibinfo {author} {\bibfnamefont {L.}~\bibnamefont {Adamowicz}},\ }\bibfield
  {title} {\bibinfo {title} {{Orbit-orbit relativistic correction calculated
  with all-electron molecular explicitly correlated Gaussians}},\ }\href
  {https://doi.org/10.1063/1.4971376} {\bibfield  {journal} {\bibinfo
  {journal} {J. Chem. Phys.}\ }\textbf {\bibinfo {volume} {145}},\ \bibinfo
  {pages} {224111} (\bibinfo {year} {2016})}\BibitemShut {NoStop}%
\bibitem [{\citenamefont {Zaklama}\ \emph {et~al.}(2020)\citenamefont
  {Zaklama}, \citenamefont {Zhang}, \citenamefont {Rowan}, \citenamefont
  {Schatzki}, \citenamefont {Suzuki},\ and\ \citenamefont
  {Varga}}]{Zaklama2020}%
  \BibitemOpen
  \bibfield  {author} {\bibinfo {author} {\bibfnamefont {T.}~\bibnamefont
  {Zaklama}}, \bibinfo {author} {\bibfnamefont {D.}~\bibnamefont {Zhang}},
  \bibinfo {author} {\bibfnamefont {K.}~\bibnamefont {Rowan}}, \bibinfo
  {author} {\bibfnamefont {L.}~\bibnamefont {Schatzki}}, \bibinfo {author}
  {\bibfnamefont {Y.}~\bibnamefont {Suzuki}},\ and\ \bibinfo {author}
  {\bibfnamefont {K.}~\bibnamefont {Varga}},\ }\bibfield  {title} {\bibinfo
  {title} {{Matrix elements of one dimensional explicitly correlated Gaussian
  basis functions}},\ }\href {https://doi.org/10.1007/s00601-019-1539-3}
  {\bibfield  {journal} {\bibinfo  {journal} {Few-Body Syst.}\ }\textbf
  {\bibinfo {volume} {61}},\ \bibinfo {pages} {6} (\bibinfo {year}
  {2020})}\BibitemShut {NoStop}%
\bibitem [{\citenamefont {Ferenc}\ \emph {et~al.}(2020)\citenamefont {Ferenc},
  \citenamefont {Jeszenszki}, \citenamefont {Ireland},\ and\ \citenamefont
  {M{\'{a}}tyus}}]{Ferenc2020}%
  \BibitemOpen
  \bibfield  {author} {\bibinfo {author} {\bibfnamefont {D.}~\bibnamefont
  {Ferenc}}, \bibinfo {author} {\bibfnamefont {P.}~\bibnamefont {Jeszenszki}},
  \bibinfo {author} {\bibfnamefont {R.~T.}\ \bibnamefont {Ireland}},\ and\
  \bibinfo {author} {\bibfnamefont {E.}~\bibnamefont {M{\'{a}}tyus}},\
  }\href@noop {} {\emph {\bibinfo {title} {{ECG integral Notes}}}},\ \bibinfo
  {type} {Tech. Rep.}\ (\bibinfo  {institution} {E{\"{o}}tv{\"{o}}s
  Lor{\'{a}}nd University},\ \bibinfo {address} {Budapest},\ \bibinfo {year}
  {2020})\BibitemShut {NoStop}%
\bibitem [{\citenamefont {{D. Ferenc}}\ \emph {et~al.}()\citenamefont {{D.
  Ferenc}}, \citenamefont {{P. Jeszenszki}}, \citenamefont {{I.
  Horny{\'{a}}k}}, \citenamefont {{R. T. Ireland}},\ and\ \citenamefont {{E.
  M{\'{a}}tyus}}}]{MolecularQuantumDynamicsResearchGroup}%
  \BibitemOpen
  \bibfield  {author} {\bibinfo {author} {\bibnamefont {{D. Ferenc}}}, \bibinfo
  {author} {\bibnamefont {{P. Jeszenszki}}}, \bibinfo {author} {\bibnamefont
  {{I. Horny{\'{a}}k}}}, \bibinfo {author} {\bibnamefont {{R. T. Ireland}}},\
  and\ \bibinfo {author} {\bibnamefont {{E. M{\'{a}}tyus}}},\ }\href
  {www.compchem.hu} {\bibinfo {title} {{QUANTEN, a computer program for the
  QUANTum mechanical description of Electrons and Nuclei}}}\BibitemShut
  {NoStop}%
\bibitem [{\citenamefont {Atkins}\ \emph {et~al.}(2018)\citenamefont {Atkins},
  \citenamefont {Paula},\ and\ \citenamefont {Keeler}}]{Atkins2018}%
  \BibitemOpen
  \bibfield  {author} {\bibinfo {author} {\bibfnamefont {P.}~\bibnamefont
  {Atkins}}, \bibinfo {author} {\bibfnamefont {J.~D.}\ \bibnamefont {Paula}},\
  and\ \bibinfo {author} {\bibfnamefont {J.}~\bibnamefont {Keeler}},\ }\href
  {https://global.oup.com/academic/product/atkins-physical-chemistry-9780198769866?cc=gb&lang=en&]}
  {\emph {\bibinfo {title} {{Atkins' Physical Chemistry}}}},\ \bibinfo
  {edition} {eleventh}\ ed.,\ Vol.~\bibinfo {volume} {1}\ (\bibinfo
  {publisher} {Oxford University Press},\ \bibinfo {address} {Oxford},\
  \bibinfo {year} {2018})\BibitemShut {NoStop}%
\bibitem [{\citenamefont {Horn}\ and\ \citenamefont
  {Johnson}(1985)}]{Horn1985}%
  \BibitemOpen
  \bibfield  {author} {\bibinfo {author} {\bibfnamefont {R.~A.}\ \bibnamefont
  {Horn}}\ and\ \bibinfo {author} {\bibfnamefont {C.~A.}\ \bibnamefont
  {Johnson}},\ }\href {https://doi.org/10.1017/CBO9780511810817} {\emph
  {\bibinfo {title} {{Matrix Analysis}}}},\ \bibinfo {edition} {2nd}\ ed.\
  (\bibinfo  {publisher} {Cambridge University Press},\ \bibinfo {address}
  {Cambridge},\ \bibinfo {year} {1985})\ pp.\ \bibinfo {pages}
  {176--180}\BibitemShut {NoStop}%
\bibitem [{\citenamefont {Bransden}\ and\ \citenamefont
  {Joachain}(2000)}]{Bransden2000}%
  \BibitemOpen
  \bibfield  {author} {\bibinfo {author} {\bibfnamefont {B.~H.}\ \bibnamefont
  {Bransden}}\ and\ \bibinfo {author} {\bibfnamefont {C.~J.}\ \bibnamefont
  {Joachain}},\ }\href
  {https://www.pearson.com/uk/educators/higher-education-educators/product/9780582356917.html}
  {\emph {\bibinfo {title} {{Quantum Mechanics}}}},\ \bibinfo {edition} {2nd}\
  ed.\ (\bibinfo  {publisher} {Pearsons Education},\ \bibinfo {address}
  {Edinburgh},\ \bibinfo {year} {2000})\ pp.\ \bibinfo {pages}
  {469--511}\BibitemShut {NoStop}%
\bibitem [{\citenamefont {Mayer}(2003)}]{Mayer2003}%
  \BibitemOpen
  \bibfield  {author} {\bibinfo {author} {\bibfnamefont {I.}~\bibnamefont
  {Mayer}},\ }\href {https://doi.org/10.1007/978-1-4757-6519-9} {\emph
  {\bibinfo {title} {Simple {Theorems}, {Proofs}, and {Derivations} in
  {Quantum} {Chemistry}}}},\ edited by\ \bibinfo {editor} {\bibfnamefont
  {P.~G.}\ \bibnamefont {Mezey}},\ Mathematical and {Computational}
  {Chemistry}\ (\bibinfo  {publisher} {Springer US},\ \bibinfo {address}
  {Boston, MA},\ \bibinfo {year} {2003})\BibitemShut {NoStop}%
\bibitem [{\citenamefont {Schmidt}(1907)}]{Schmidt1907}%
  \BibitemOpen
  \bibfield  {author} {\bibinfo {author} {\bibfnamefont {E.}~\bibnamefont
  {Schmidt}},\ }\bibfield  {title} {\bibinfo {title} {{Zur Theorie der linearen
  und nichtlinearen Integralgleichungen}},\ }\href
  {https://doi.org/10.1007/BF01449770} {\bibfield  {journal} {\bibinfo
  {journal} {Math. Ann.}\ }\textbf {\bibinfo {volume} {63}},\ \bibinfo {pages}
  {433} (\bibinfo {year} {1907})}\BibitemShut {NoStop}%
\bibitem [{\citenamefont {James}\ and\ \citenamefont {James}()}]{James1992}%
  \BibitemOpen
  \bibfield  {author} {\bibinfo {author} {\bibfnamefont {R.~C.}\ \bibnamefont
  {James}}\ and\ \bibinfo {author} {\bibfnamefont {G.}~\bibnamefont {James}},\
  }\href {https://www.springer.com/gp/book/9780412990410} {\emph {\bibinfo
  {title} {{Mathematics Dictionary}}}},\ \bibinfo {edition} {5th}\ ed.\
  (\bibinfo  {publisher} {Springer Netherlands})\BibitemShut {NoStop}%
\bibitem [{\citenamefont {Mitroy}\ \emph {et~al.}(2013)\citenamefont {Mitroy},
  \citenamefont {Bubin}, \citenamefont {Horiuchi}, \citenamefont {Suzuki},
  \citenamefont {Adamowicz}, \citenamefont {Cencek}, \citenamefont {Szalewicz},
  \citenamefont {Komasa}, \citenamefont {Blume},\ and\ \citenamefont
  {Varga}}]{Mitroy2013}%
  \BibitemOpen
  \bibfield  {author} {\bibinfo {author} {\bibfnamefont {J.}~\bibnamefont
  {Mitroy}}, \bibinfo {author} {\bibfnamefont {S.}~\bibnamefont {Bubin}},
  \bibinfo {author} {\bibfnamefont {W.}~\bibnamefont {Horiuchi}}, \bibinfo
  {author} {\bibfnamefont {Y.}~\bibnamefont {Suzuki}}, \bibinfo {author}
  {\bibfnamefont {L.}~\bibnamefont {Adamowicz}}, \bibinfo {author}
  {\bibfnamefont {W.}~\bibnamefont {Cencek}}, \bibinfo {author} {\bibfnamefont
  {K.}~\bibnamefont {Szalewicz}}, \bibinfo {author} {\bibfnamefont
  {J.}~\bibnamefont {Komasa}}, \bibinfo {author} {\bibfnamefont
  {D.}~\bibnamefont {Blume}},\ and\ \bibinfo {author} {\bibfnamefont
  {K.}~\bibnamefont {Varga}},\ }\bibfield  {title} {\bibinfo {title} {{Theory
  and application of explicitly correlated Gaussians}},\ }\href
  {https://doi.org/10.1103/RevModPhys.85.693} {\bibfield  {journal} {\bibinfo
  {journal} {Rev. Mod. Phys.}\ }\textbf {\bibinfo {volume} {85}},\ \bibinfo
  {pages} {693} (\bibinfo {year} {2013})}\BibitemShut {NoStop}%
\bibitem [{\citenamefont {Lieb}(2003)}]{Lieb2005}%
  \BibitemOpen
  \bibfield  {author} {\bibinfo {author} {\bibfnamefont {E.~H.}\ \bibnamefont
  {Lieb}},\ }\href {https://doi.org/10.4310/cdm.2005.v2005.n1.a3} {\emph
  {\bibinfo {title} {{Quantum mechanics, the stability of matter and quantum
  electrodynamics}}}},\ \bibinfo {type} {Tech. Rep.}\ (\bibinfo  {institution}
  {Departments of Mathematics and Physics, Princeton University},\ \bibinfo
  {address} {Princeton},\ \bibinfo {year} {2003})\ \Eprint
  {https://arxiv.org/abs/0209034} {arXiv:0209034 [math-ph]} \BibitemShut
  {NoStop}%
\bibitem [{\citenamefont {Pollak}\ and\ \citenamefont
  {Martinazzo}(2020)}]{Pollak2020}%
  \BibitemOpen
  \bibfield  {author} {\bibinfo {author} {\bibfnamefont {E.}~\bibnamefont
  {Pollak}}\ and\ \bibinfo {author} {\bibfnamefont {R.}~\bibnamefont
  {Martinazzo}},\ }\bibfield  {title} {\bibinfo {title} {{Self-consistent
  theory of lower bounds for eigenvalues}},\ }\href
  {https://doi.org/10.1063/5.0009436} {\bibfield  {journal} {\bibinfo
  {journal} {J. Chem. Phys.}\ }\textbf {\bibinfo {volume} {152}},\ \bibinfo
  {pages} {244110} (\bibinfo {year} {2020})}\BibitemShut {NoStop}%
\bibitem [{\citenamefont {Goedecker}\ and\ \citenamefont
  {Maschke}(1990)}]{Goedecker1991}%
  \BibitemOpen
  \bibfield  {author} {\bibinfo {author} {\bibfnamefont {S.}~\bibnamefont
  {Goedecker}}\ and\ \bibinfo {author} {\bibfnamefont {K.}~\bibnamefont
  {Maschke}},\ }\bibfield  {title} {\bibinfo {title} {{Comment on `Criterion
  for a good variational wave function'}},\ }\href
  {https://doi.org/https://doi.org/10.1103/PhysRevB.42.6835} {\bibfield
  {journal} {\bibinfo  {journal} {Phys, Rev. B}\ }\textbf {\bibinfo {volume}
  {42}},\ \bibinfo {pages} {365} (\bibinfo {year} {1990})}\BibitemShut
  {NoStop}%
\bibitem [{\citenamefont {Pollak}(2019{\natexlab{a}})}]{Pollak2019}%
  \BibitemOpen
  \bibfield  {author} {\bibinfo {author} {\bibfnamefont {E.}~\bibnamefont
  {Pollak}},\ }\bibfield  {title} {\bibinfo {title} {{An improved lower bound
  to the ground-state energy}},\ }\href
  {https://doi.org/10.1021/acs.jctc.9b00128} {\bibfield  {journal} {\bibinfo
  {journal} {J. Chem. Theory Comput.}\ }\textbf {\bibinfo {volume} {15}},\
  \bibinfo {pages} {1498} (\bibinfo {year} {2019}{\natexlab{a}})}\BibitemShut
  {NoStop}%
\bibitem [{\citenamefont {Nakashima}\ and\ \citenamefont
  {Nakatsuji}(2008)}]{Nakashima2008}%
  \BibitemOpen
  \bibfield  {author} {\bibinfo {author} {\bibfnamefont {H.}~\bibnamefont
  {Nakashima}}\ and\ \bibinfo {author} {\bibfnamefont {H.}~\bibnamefont
  {Nakatsuji}},\ }\bibfield  {title} {\bibinfo {title} {How accurately does the
  free complement wave function of a helium atom satisfy the schr\"odinger
  equation?},\ }\href {https://doi.org/10.1103/PhysRevLett.101.240406}
  {\bibfield  {journal} {\bibinfo  {journal} {Phys. Rev. Lett.}\ }\textbf
  {\bibinfo {volume} {101}},\ \bibinfo {pages} {240406} (\bibinfo {year}
  {2008})}\BibitemShut {NoStop}%
\bibitem [{\citenamefont {Stevenson}(1938)}]{Stevenson1938}%
  \BibitemOpen
  \bibfield  {author} {\bibinfo {author} {\bibfnamefont {A.~F.}\ \bibnamefont
  {Stevenson}},\ }\bibfield  {title} {\bibinfo {title} {{On the Lower Bounds of
  Weinstein and Romberg in Quantum Mechanics}},\ }\href
  {https://doi.org/10.1103/PhysRev.53.199.2} {\bibfield  {journal} {\bibinfo
  {journal} {Phys. Rev.}\ }\textbf {\bibinfo {volume} {53}},\ \bibinfo {pages}
  {199} (\bibinfo {year} {1938})}\BibitemShut {NoStop}%
\bibitem [{\citenamefont {Stevenson}\ and\ \citenamefont
  {Crawford}(1938)}]{Stevenson1938b}%
  \BibitemOpen
  \bibfield  {author} {\bibinfo {author} {\bibfnamefont {A.~F.}\ \bibnamefont
  {Stevenson}}\ and\ \bibinfo {author} {\bibfnamefont {M.~F.}\ \bibnamefont
  {Crawford}},\ }\bibfield  {title} {\bibinfo {title} {{A lower limit for the
  theoretical energy of the normal state of helium}},\ }\href
  {https://doi.org/10.1103/PhysRev.54.375} {\bibfield  {journal} {\bibinfo
  {journal} {Phys. Rev.}\ }\textbf {\bibinfo {volume} {54}},\ \bibinfo {pages}
  {375} (\bibinfo {year} {1938})}\BibitemShut {NoStop}%
\bibitem [{\citenamefont {Kleindienst}\ and\ \citenamefont
  {Altmann}(1976)}]{Kleindienst1976}%
  \BibitemOpen
  \bibfield  {author} {\bibinfo {author} {\bibfnamefont {H.}~\bibnamefont
  {Kleindienst}}\ and\ \bibinfo {author} {\bibfnamefont {W.}~\bibnamefont
  {Altmann}},\ }\bibfield  {title} {\bibinfo {title} {{I. Lineare
  Fehlerminimisierung Ein Verfahren zur Eigenwertberechnung bei
  Schr{\"{o}}dinger-Operatoren}},\ }\href
  {https://doi.org/10.1002/qua.560100516} {\bibfield  {journal} {\bibinfo
  {journal} {Int. J. Quantum Chem.}\ }\textbf {\bibinfo {volume} {10}},\
  \bibinfo {pages} {873} (\bibinfo {year} {1976})}\BibitemShut {NoStop}%
\bibitem [{\citenamefont {Bazley}(1959)}]{Bazley}%
  \BibitemOpen
  \bibfield  {author} {\bibinfo {author} {\bibfnamefont {N.~W.}\ \bibnamefont
  {Bazley}},\ }\bibfield  {title} {\bibinfo {title} {{Lower bounds for
  eigenvalues with application to the helium atom}},\ }\href
  {https://doi.org/https://doi.org/10.1073/pnas.45.6.850} {\bibfield  {journal}
  {\bibinfo  {journal} {Proc. Natl. Acad. Sci. U.S.A}\ }\textbf {\bibinfo
  {volume} {45}},\ \bibinfo {pages} {850} (\bibinfo {year} {1959})}\BibitemShut
  {NoStop}%
\bibitem [{\citenamefont {Hill}(1979)}]{Hill1979}%
  \BibitemOpen
  \bibfield  {author} {\bibinfo {author} {\bibfnamefont {R.~N.}\ \bibnamefont
  {Hill}},\ }\bibfield  {title} {\bibinfo {title} {{Tight lower bounds to
  eigenvalues of the Schr{\"{o}}dinger equation}},\ }\href
  {https://doi.org/10.1063/1.524700} {\bibfield  {journal} {\bibinfo  {journal}
  {J. Math. Phys.}\ }\textbf {\bibinfo {volume} {21}},\ \bibinfo {pages} {2182}
  (\bibinfo {year} {1979})}\BibitemShut {NoStop}%
\bibitem [{\citenamefont {Pollak}(2019{\natexlab{b}})}]{Pollak2019a}%
  \BibitemOpen
  \bibfield  {author} {\bibinfo {author} {\bibfnamefont {E.}~\bibnamefont
  {Pollak}},\ }\bibfield  {title} {\bibinfo {title} {{A tight lower bound to
  the ground-state energy}},\ }\href {https://doi.org/10.1021/acs.jctc.9b00344}
  {\bibfield  {journal} {\bibinfo  {journal} {J. Chem. Theory Comput.}\
  }\textbf {\bibinfo {volume} {15}},\ \bibinfo {pages} {4079} (\bibinfo {year}
  {2019}{\natexlab{b}})}\BibitemShut {NoStop}%
\bibitem [{\citenamefont {Lehmann}(1949)}]{Lehmann1949}%
  \BibitemOpen
  \bibfield  {author} {\bibinfo {author} {\bibfnamefont {N.~J.}\ \bibnamefont
  {Lehmann}},\ }\bibfield  {title} {\bibinfo {title} {{Beitr{\"{a}}ge zur
  numerischen L{\"{o}}sung linearer Eigenwertprobleme. I}},\ }\href
  {https://doi.org/10.1002/zamm.19502911005} {\bibfield  {journal} {\bibinfo
  {journal} {J. Appl. Math. Mech.}\ }\textbf {\bibinfo {volume} {29}},\
  \bibinfo {pages} {341} (\bibinfo {year} {1949})}\BibitemShut {NoStop}%
\bibitem [{\citenamefont {Lehmann}(1950)}]{Lehmann1950}%
  \BibitemOpen
  \bibfield  {author} {\bibinfo {author} {\bibfnamefont {N.~J.}\ \bibnamefont
  {Lehmann}},\ }\bibfield  {title} {\bibinfo {title} {{Beitr{\"{a}}ge zur
  numerischen L{\"{o}}sung linearer Eigenwertprobleme.}},\ }\href
  {https://doi.org/10.1002/zamm.19500300101} {\bibfield  {journal} {\bibinfo
  {journal} {J. Appl. Math. Mech.}\ }\textbf {\bibinfo {volume} {30}},\
  \bibinfo {pages} {1} (\bibinfo {year} {1950})}\BibitemShut {NoStop}%
\bibitem [{\citenamefont {Ovtchinnikov}(2011)}]{Ovtchinnikov2011}%
  \BibitemOpen
  \bibfield  {author} {\bibinfo {author} {\bibfnamefont {E.~E.}\ \bibnamefont
  {Ovtchinnikov}},\ }\bibfield  {title} {\bibinfo {title} {{Lehmann bounds and
  eigenvalue error estimation}},\ }\href {https://doi.org/10.1137/100793062}
  {\bibfield  {journal} {\bibinfo  {journal} {SIAM J. Numer. Anal.}\ }\textbf
  {\bibinfo {volume} {49}},\ \bibinfo {pages} {2078} (\bibinfo {year}
  {2011})}\BibitemShut {NoStop}%
\bibitem [{\citenamefont {{Per-Olov L{\"{o}}wdin}}(1965)}]{Per-OlovLowdin1965}%
  \BibitemOpen
  \bibfield  {author} {\bibinfo {author} {\bibnamefont {{Per-Olov
  L{\"{o}}wdin}}},\ }\bibfield  {title} {\bibinfo {title} {{Studies in
  perturbation theory. X. Lower bounds to energy eigenvalues in
  perturbation-theory ground state}},\ }\href
  {https://doi.org/https://doi.org/10.1103/PhysRev.139.A357} {\bibfield
  {journal} {\bibinfo  {journal} {Phys. Rev.}\ }\textbf {\bibinfo {volume}
  {139}},\ \bibinfo {pages} {A357} (\bibinfo {year} {1965})}\BibitemShut
  {NoStop}%
\bibitem [{\citenamefont {Szabados}\ and\ \citenamefont
  {T{\'{o}}th}(2014)}]{Szabados2014}%
  \BibitemOpen
  \bibfield  {author} {\bibinfo {author} {\bibfnamefont {{\'{A}}.}~\bibnamefont
  {Szabados}}\ and\ \bibinfo {author} {\bibfnamefont {Z.}~\bibnamefont
  {T{\'{o}}th}},\ }\bibfield  {title} {\bibinfo {title} {{L{\"{o}}wdin's
  bracketing function revisited}},\ }\href
  {https://doi.org/10.1007/s10910-014-0379-0} {\bibfield  {journal} {\bibinfo
  {journal} {J. Math. Chem.}\ }\textbf {\bibinfo {volume} {52}},\ \bibinfo
  {pages} {2210} (\bibinfo {year} {2014})}\BibitemShut {NoStop}%
\bibitem [{\citenamefont {L{\"{o}}wdin}(1962)}]{Lowdin1962}%
  \BibitemOpen
  \bibfield  {author} {\bibinfo {author} {\bibfnamefont {P.}~\bibnamefont
  {L{\"{o}}wdin}},\ }\bibfield  {title} {\bibinfo {title} {{Studies in
  perturbation theory. IV. Solution of eigenvalue problem by projection
  operator formalism}},\ }\href {https://doi.org/10.1063/1.1724312} {\bibfield
  {journal} {\bibinfo  {journal} {J. Math. Phys.}\ }\textbf {\bibinfo {volume}
  {3}},\ \bibinfo {pages} {969} (\bibinfo {year} {1962})}\BibitemShut {NoStop}%
\bibitem [{\citenamefont {Atkins}\ \emph {et~al.}(2013)\citenamefont {Atkins},
  \citenamefont {de~Paula},\ and\ \citenamefont {Friedman}}]{Atkins2013}%
  \BibitemOpen
  \bibfield  {author} {\bibinfo {author} {\bibfnamefont {P.}~\bibnamefont
  {Atkins}}, \bibinfo {author} {\bibfnamefont {J.}~\bibnamefont {de~Paula}},\
  and\ \bibinfo {author} {\bibfnamefont {R.}~\bibnamefont {Friedman}},\ }\href
  {https://global.oup.com/academic/product/physical-chemistry-9780199609819?q=quanta}
  {\emph {\bibinfo {title} {{Physical Chemistry: Quanta, Matter and
  Change}}}},\ \bibinfo {edition} {2nd}\ ed.\ (\bibinfo  {publisher} {Oxford
  University Press},\ \bibinfo {address} {Oxford},\ \bibinfo {year}
  {2013})\BibitemShut {NoStop}%
\bibitem [{\citenamefont {M{\'{a}}tyus}(2019)}]{Matyus2019}%
  \BibitemOpen
  \bibfield  {author} {\bibinfo {author} {\bibfnamefont {E.}~\bibnamefont
  {M{\'{a}}tyus}},\ }\bibfield  {title} {\bibinfo {title}
  {{Pre-Born–Oppenheimer molecular structure theory}},\ }\href
  {https://doi.org/10.1080/00268976.2018.1530461} {\bibfield  {journal}
  {\bibinfo  {journal} {Mol. Phys.}\ }\textbf {\bibinfo {volume} {117}},\
  \bibinfo {pages} {590} (\bibinfo {year} {2019})},\ \Eprint
  {https://arxiv.org/abs/1801.05885} {arXiv:1801.05885} \BibitemShut {NoStop}%
\bibitem [{\citenamefont {Mátyus}\ and\ \citenamefont
  {Reiher}(2012)}]{Matyus2012}%
  \BibitemOpen
  \bibfield  {author} {\bibinfo {author} {\bibfnamefont {E.}~\bibnamefont
  {Mátyus}}\ and\ \bibinfo {author} {\bibfnamefont {M.}~\bibnamefont
  {Reiher}},\ }\bibfield  {title} {\bibinfo {title} {Molecular structure
  calculations: {A} unified quantum mechanical description of electrons and
  nuclei using explicitly correlated {Gaussian} functions and the global vector
  representation},\ }\href {https://doi.org/10.1063/1.4731696} {\bibfield
  {journal} {\bibinfo  {journal} {J. Chem. Phys.}\ }\textbf {\bibinfo {volume}
  {137}},\ \bibinfo {pages} {024104} (\bibinfo {year} {2012})}\BibitemShut
  {NoStop}%
\bibitem [{\citenamefont {Hartree}(1927)}]{Hartree1927}%
  \BibitemOpen
  \bibfield  {author} {\bibinfo {author} {\bibfnamefont {D.~R.}\ \bibnamefont
  {Hartree}},\ }\bibfield  {title} {\bibinfo {title} {{The wave mechanics of an
  atom with a non-Coulomb central field Part II. Some results and
  discussion}},\ }\href {https://doi.org/10.1017/S0305004100011920} {\bibfield
  {journal} {\bibinfo  {journal} {Math. Proc. Camb. Philos. Soc.}\ }\textbf
  {\bibinfo {volume} {24}},\ \bibinfo {pages} {111} (\bibinfo {year}
  {1927})}\BibitemShut {NoStop}%
\bibitem [{\citenamefont {Slater}(1930)}]{Slater1930a}%
  \BibitemOpen
  \bibfield  {author} {\bibinfo {author} {\bibfnamefont {J.~C.}\ \bibnamefont
  {Slater}},\ }\bibfield  {title} {\bibinfo {title} {{Atomic shielding
  constants}},\ }\href {https://doi.org/10.1103/PhysRev.36.57} {\bibfield
  {journal} {\bibinfo  {journal} {Phys. Rev.}\ }\textbf {\bibinfo {volume}
  {36}},\ \bibinfo {pages} {57} (\bibinfo {year} {1930})}\BibitemShut {NoStop}%
\bibitem [{\citenamefont {Slater}(1928)}]{Slater1930}%
  \BibitemOpen
  \bibfield  {author} {\bibinfo {author} {\bibfnamefont {J.~C.}\ \bibnamefont
  {Slater}},\ }\bibfield  {title} {\bibinfo {title} {The self consistent field
  and the structure of atoms},\ }\href {https://doi.org/10.1103/PhysRev.32.339}
  {\bibfield  {journal} {\bibinfo  {journal} {Phys. Rev.}\ }\textbf {\bibinfo
  {volume} {32}},\ \bibinfo {pages} {339} (\bibinfo {year} {1928})}\BibitemShut
  {NoStop}%
\bibitem [{\citenamefont {{J. Rychlewski}}(2003)}]{J.Rychlewski2003}%
  \BibitemOpen
  \bibfield  {author} {\bibinfo {author} {\bibfnamefont {E.}~\bibnamefont {{J.
  Rychlewski}}},\ }\href {https://www.springer.com/gp/book/9781402016745}
  {\emph {\bibinfo {title} {{Explicitly Correlated Wave Functions in Chemistry
  and Physics}}}},\ \bibinfo {edition} {first edit}\ ed.\ (\bibinfo
  {publisher} {Springer Netherlands},\ \bibinfo {year} {2003})\BibitemShut
  {NoStop}%
\bibitem [{\citenamefont {Bubin}\ \emph {et~al.}(2013)\citenamefont {Bubin},
  \citenamefont {Pavanello}, \citenamefont {Tung}, \citenamefont {Sharkey},\
  and\ \citenamefont {Adamowicz}}]{Bubin2013}%
  \BibitemOpen
  \bibfield  {author} {\bibinfo {author} {\bibfnamefont {S.}~\bibnamefont
  {Bubin}}, \bibinfo {author} {\bibfnamefont {M.}~\bibnamefont {Pavanello}},
  \bibinfo {author} {\bibfnamefont {W.~C.}\ \bibnamefont {Tung}}, \bibinfo
  {author} {\bibfnamefont {K.~L.}\ \bibnamefont {Sharkey}},\ and\ \bibinfo
  {author} {\bibfnamefont {L.}~\bibnamefont {Adamowicz}},\ }\bibfield  {title}
  {\bibinfo {title} {{Born-Oppenheimer and non-Born-Oppenheimer, atomic and
  molecular calculations with explicitly correlated Gaussians}},\ }\href
  {https://doi.org/10.1021/cr200419d} {\bibfield  {journal} {\bibinfo
  {journal} {Chem. Rev.}\ }\textbf {\bibinfo {volume} {113}},\ \bibinfo {pages}
  {36} (\bibinfo {year} {2013})}\BibitemShut {NoStop}%
\bibitem [{\citenamefont {Hylleraas}(1928)}]{Hylleraas1}%
  \BibitemOpen
  \bibfield  {author} {\bibinfo {author} {\bibfnamefont {E.~A.}\ \bibnamefont
  {Hylleraas}},\ }\bibfield  {title} {\bibinfo {title} {{On the ground state of
  the helium atom}},\ }\href
  {https://doi.org/https://doi.org/10.1007/978-3-642-75471-5_16} {\bibfield
  {journal} {\bibinfo  {journal} {Z. Phys.}\ }\textbf {\bibinfo {volume}
  {48}},\ \bibinfo {pages} {81} (\bibinfo {year} {1928})}\BibitemShut {NoStop}%
\bibitem [{\citenamefont {Hylleraas}(1929{\natexlab{a}})}]{Hylleraas2}%
  \BibitemOpen
  \bibfield  {author} {\bibinfo {author} {\bibfnamefont {E.~A.}\ \bibnamefont
  {Hylleraas}},\ }\bibfield  {title} {\bibinfo {title} {{Die
  Ionisierungsspannungen von Atomkonfigurationen mit zwei Elektronen}},\ }\href
  {https://doi.org/https://doi.org/10.1007/BF01506430} {\bibfield  {journal}
  {\bibinfo  {journal} {Naturwissenschaften}\ }\textbf {\bibinfo {volume}
  {17}},\ \bibinfo {pages} {982} (\bibinfo {year}
  {1929}{\natexlab{a}})}\BibitemShut {NoStop}%
\bibitem [{\citenamefont {Hylleraas}(1929{\natexlab{b}})}]{Hylleraas3}%
  \BibitemOpen
  \bibfield  {author} {\bibinfo {author} {\bibfnamefont {E.~A.}\ \bibnamefont
  {Hylleraas}},\ }\bibfield  {title} {\bibinfo {title} {{New calculation of the
  energy of helium in the ground state, as well as the lowest term of
  ortho-helium}},\ }\href
  {https://doi.org/https://doi.org/10.1142/9789812795762_0006} {\bibfield
  {journal} {\bibinfo  {journal} {Z. Phys.}\ }\textbf {\bibinfo {volume}
  {54}},\ \bibinfo {pages} {347} (\bibinfo {year}
  {1929}{\natexlab{b}})}\BibitemShut {NoStop}%
\bibitem [{\citenamefont {Hylleraas}(1930)}]{Hylleraas4}%
  \BibitemOpen
  \bibfield  {author} {\bibinfo {author} {\bibfnamefont {E.~A.}\ \bibnamefont
  {Hylleraas}},\ }\bibfield  {title} {\bibinfo {title} {{Bemerkungen zu meiner
  Arbeit: Die Elektronenaffinit{\"{a}}t des Wasserstoffatoms nach der
  Wellenmechanik}},\ }\href {https://doi.org/10.1007/BF01421744} {\bibfield
  {journal} {\bibinfo  {journal} {Z. Phys.}\ }\textbf {\bibinfo {volume}
  {63}},\ \bibinfo {pages} {291} (\bibinfo {year} {1930})}\BibitemShut
  {NoStop}%
\bibitem [{\citenamefont {Hylleraas}\ and\ \citenamefont
  {Undheim}(1930)}]{Hylleraas5}%
  \BibitemOpen
  \bibfield  {author} {\bibinfo {author} {\bibfnamefont {E.~A.}\ \bibnamefont
  {Hylleraas}}\ and\ \bibinfo {author} {\bibfnamefont {B.}~\bibnamefont
  {Undheim}},\ }\bibfield  {title} {\bibinfo {title} {{Undheim, B. Numerische
  Berechnung der 2S-Terme von Ortho- und Par-Helium}},\ }\href
  {https://doi.org/10.1007/BF01397263} {\bibfield  {journal} {\bibinfo
  {journal} {Z. Phys.}\ }\textbf {\bibinfo {volume} {65}},\ \bibinfo {pages}
  {759} (\bibinfo {year} {1930})}\BibitemShut {NoStop}%
\bibitem [{\citenamefont {Boys}(1960)}]{Boys1960}%
  \BibitemOpen
  \bibfield  {author} {\bibinfo {author} {\bibfnamefont {S.~F.}\ \bibnamefont
  {Boys}},\ }\bibfield  {title} {\bibinfo {title} {{The integral formulae for
  the variational solution of the molecular many-electron wave equation in
  terms of Gaussian functions with direct electronic correlation}},\ }\href
  {https://doi.org/https://doi.org/10.1098/rspa.1960.0195} {\bibfield
  {journal} {\bibinfo  {journal} {Proc. R. Soc. Lond. A}\ }\textbf {\bibinfo
  {volume} {258}},\ \bibinfo {pages} {402} (\bibinfo {year}
  {1960})}\BibitemShut {NoStop}%
\bibitem [{\citenamefont {Singer}(1960)}]{Singer1960}%
  \BibitemOpen
  \bibfield  {author} {\bibinfo {author} {\bibfnamefont {K.}~\bibnamefont
  {Singer}},\ }\bibfield  {title} {\bibinfo {title} {{The use of Gaussian
  (exponential quadratic) wave functions in molecular problems - I . General
  formulae for the evaluation of integrals}},\ }\href
  {https://doi.org/https://doi.org/10.1098/rspa.1960.0196} {\bibfield
  {journal} {\bibinfo  {journal} {Proc. R. Soc. Lond. A}\ }\textbf {\bibinfo
  {volume} {258}},\ \bibinfo {pages} {412} (\bibinfo {year}
  {1960})}\BibitemShut {NoStop}%
\bibitem [{\citenamefont {Kato}(1957)}]{Kato1957}%
  \BibitemOpen
  \bibfield  {author} {\bibinfo {author} {\bibfnamefont {T.}~\bibnamefont
  {Kato}},\ }\bibfield  {title} {\bibinfo {title} {{On the eigenfunctions of
  many-particle systems in quantum mechanics}},\ }\href
  {https://doi.org/10.1002/cpa.3160100201} {\bibfield  {journal} {\bibinfo
  {journal} {Commun. Pure Appl. Math.}\ }\textbf {\bibinfo {volume} {10}},\
  \bibinfo {pages} {151} (\bibinfo {year} {1957})}\BibitemShut {NoStop}%
\bibitem [{\citenamefont {Nakamura}(2010)}]{Nakamura_2010}%
  \BibitemOpen
  \bibfield  {author} {\bibinfo {author} {\bibfnamefont {K.}~\bibnamefont
  {Nakamura}},\ }\bibfield  {title} {\bibinfo {title} {{Review of Particle
  Physics}},\ }\href {https://doi.org/10.1088/0954-3899/37/7a/075021}
  {\bibfield  {journal} {\bibinfo  {journal} {J . Phys. G}\ }\textbf {\bibinfo
  {volume} {37}},\ \bibinfo {pages} {75021} (\bibinfo {year}
  {2010})}\BibitemShut {NoStop}%
\bibitem [{\citenamefont {Born}\ and\ \citenamefont
  {Oppenheimer}(1927)}]{Born1927}%
  \BibitemOpen
  \bibfield  {author} {\bibinfo {author} {\bibfnamefont {M.}~\bibnamefont
  {Born}}\ and\ \bibinfo {author} {\bibfnamefont {R.}~\bibnamefont
  {Oppenheimer}},\ }\bibfield  {title} {\bibinfo {title} {{Zur Quantentheorie
  der Molekeln}},\ }\href {https://doi.org/10.1002/andp.19273892002} {\bibfield
   {journal} {\bibinfo  {journal} {Ann. Phys.}\ }\textbf {\bibinfo {volume}
  {389}},\ \bibinfo {pages} {457} (\bibinfo {year} {1927})}\BibitemShut
  {NoStop}%
\bibitem [{\citenamefont {Muolo}\ \emph {et~al.}(2019)\citenamefont {Muolo},
  \citenamefont {M{\'{a}}tyus},\ and\ \citenamefont {Reiher}}]{Muolo2019}%
  \BibitemOpen
  \bibfield  {author} {\bibinfo {author} {\bibfnamefont {A.}~\bibnamefont
  {Muolo}}, \bibinfo {author} {\bibfnamefont {E.}~\bibnamefont
  {M{\'{a}}tyus}},\ and\ \bibinfo {author} {\bibfnamefont {M.}~\bibnamefont
  {Reiher}},\ }\bibfield  {title} {\bibinfo {title} {{{$\mathrm{H}_{3}^+$} as a
  five-body problem described with explicitly correlated Gaussian basis
  sets}},\ }\href {https://doi.org/10.1063/1.5121318} {\bibfield  {journal}
  {\bibinfo  {journal} {J. Chem. Phys.}\ }\textbf {\bibinfo {volume} {151}},\
  \bibinfo {pages} {154110} (\bibinfo {year} {2019})},\ \Eprint
  {https://arxiv.org/abs/1907.10168} {arXiv:1907.10168} \BibitemShut {NoStop}%
\bibitem [{\citenamefont {Bubin}\ and\ \citenamefont
  {Adamowicz}(2006)}]{Bubin2006}%
  \BibitemOpen
  \bibfield  {author} {\bibinfo {author} {\bibfnamefont {S.}~\bibnamefont
  {Bubin}}\ and\ \bibinfo {author} {\bibfnamefont {L.}~\bibnamefont
  {Adamowicz}},\ }\bibfield  {title} {\bibinfo {title} {{Matrix elements of
  N-particle explicitly correlated Gaussian basis functions with complex
  exponential parameters}},\ }\href {https://doi.org/10.1063/1.2204605}
  {\bibfield  {journal} {\bibinfo  {journal} {J. Chem. Phys.}\ }\textbf
  {\bibinfo {volume} {124}},\ \bibinfo {pages} {224317} (\bibinfo {year}
  {2006})}\BibitemShut {NoStop}%
\bibitem [{\citenamefont {Drachman}(1981)}]{Drachman1981}%
  \BibitemOpen
  \bibfield  {author} {\bibinfo {author} {\bibfnamefont {R.~J.}\ \bibnamefont
  {Drachman}},\ }\bibfield  {title} {\bibinfo {title} {{A new global operator
  for two-particle delta functions}},\ }\href
  {https://doi.org/https://iopscience.iop.org/article/10.1088/0022-3700/14/16/003/meta}
  {\bibfield  {journal} {\bibinfo  {journal} {J. Phys. B At. Mol. Phys.}\
  }\textbf {\bibinfo {volume} {14}},\ \bibinfo {pages} {2733} (\bibinfo {year}
  {1981})}\BibitemShut {NoStop}%
\bibitem [{\citenamefont {Pachucki}\ \emph {et~al.}(2005)\citenamefont
  {Pachucki}, \citenamefont {Cencek},\ and\ \citenamefont
  {Komasa}}]{Pachucki2005a}%
  \BibitemOpen
  \bibfield  {author} {\bibinfo {author} {\bibfnamefont {K.}~\bibnamefont
  {Pachucki}}, \bibinfo {author} {\bibfnamefont {W.}~\bibnamefont {Cencek}},\
  and\ \bibinfo {author} {\bibfnamefont {J.}~\bibnamefont {Komasa}},\
  }\bibfield  {title} {\bibinfo {title} {{On the acceleration of the
  convergence of singular operators in Gaussian basis sets}},\ }\href
  {https://doi.org/10.1063/1.1888572} {\bibfield  {journal} {\bibinfo
  {journal} {J. Chem. Phys.}\ }\textbf {\bibinfo {volume} {122}},\ \bibinfo
  {pages} {184101} (\bibinfo {year} {2005})}\BibitemShut {NoStop}%
\bibitem [{\citenamefont {Puchalski}\ \emph {et~al.}(2017)\citenamefont
  {Puchalski}, \citenamefont {Komasa},\ and\ \citenamefont
  {Pachucki}}]{Puchalski2017}%
  \BibitemOpen
  \bibfield  {author} {\bibinfo {author} {\bibfnamefont {M.}~\bibnamefont
  {Puchalski}}, \bibinfo {author} {\bibfnamefont {J.}~\bibnamefont {Komasa}},\
  and\ \bibinfo {author} {\bibfnamefont {K.}~\bibnamefont {Pachucki}},\
  }\bibfield  {title} {\bibinfo {title} {{Relativistic corrections for the
  ground electronic state of molecular hydrogen}},\ }\href
  {https://doi.org/10.1103/PhysRevA.95.052506} {\bibfield  {journal} {\bibinfo
  {journal} {Phys. Rev. A}\ }\textbf {\bibinfo {volume} {95}},\ \bibinfo
  {pages} {1} (\bibinfo {year} {2017})},\ \Eprint
  {https://arxiv.org/abs/1704.07153} {arXiv:1704.07153} \BibitemShut {NoStop}%
\bibitem [{\citenamefont {Puchalski}\ \emph {et~al.}(2019)\citenamefont
  {Puchalski}, \citenamefont {Komasa}, \citenamefont {Spyszkiewicz},\ and\
  \citenamefont {Pachucki}}]{Puchalski2019}%
  \BibitemOpen
  \bibfield  {author} {\bibinfo {author} {\bibfnamefont {M.}~\bibnamefont
  {Puchalski}}, \bibinfo {author} {\bibfnamefont {J.}~\bibnamefont {Komasa}},
  \bibinfo {author} {\bibfnamefont {A.}~\bibnamefont {Spyszkiewicz}},\ and\
  \bibinfo {author} {\bibfnamefont {K.}~\bibnamefont {Pachucki}},\ }\bibfield
  {title} {\bibinfo {title} {{Dissociation energy of molecular hydrogen
  isotopologues}},\ }\href {https://doi.org/10.1103/PhysRevA.100.020503}
  {\bibfield  {journal} {\bibinfo  {journal} {Phys. Rev. A}\ }\textbf {\bibinfo
  {volume} {100}},\ \bibinfo {pages} {1} (\bibinfo {year} {2019})},\ \Eprint
  {https://arxiv.org/abs/1909.01985} {arXiv:1909.01985} \BibitemShut {NoStop}%
\bibitem [{\citenamefont {Ireland}(2021)}]{Ireland2021}%
  \BibitemOpen
  \bibfield  {author} {\bibinfo {author} {\bibfnamefont {R.~T.}\ \bibnamefont
  {Ireland}},\ }\href {https://doi.org/http://hdl.handle.net/10831/57773}
  {\emph {\bibinfo {title} {{Integrals for lower bounds to the exact
  energy}}}},\ \bibinfo {type} {Tech. Rep.}\ (\bibinfo  {institution}
  {E{\"{o}}tv{\"{o}}s Lor{\'{a}}nd University},\ \bibinfo {address}
  {Budapest},\ \bibinfo {year} {2021})\BibitemShut {NoStop}%
\end{thebibliography}%


\appendix
\section{Lemmas} \label{appendix:lemmas}
\noindent
This appendix comes from reference \cite{Ferenc2020}. This note is not publicly available, and so here we have written the most important lemmas.

\subsection{About the eigenvalues of a rank-one matrix \label{lemma:eigvalrank1}}
\noindent
One of the eigenvalues of a rank-one matrix is the trace of this matrix. The remaining eigenvalues are degenerate and their values are zero.\\

\noindent
{\bf Proof}:
Let us consider the rank-one matrix $\bB$ as the direct product of two vectors,
\begin{eqnarray*}
\bB = \bu \otimes \bv^T \ .
\end{eqnarray*}
Then, the non-degenerate eigenvalue can be found easily as,
\begin{eqnarray*}
\bB \bu = \bu \otimes \underbrace{\bv^T \bu}_{\sum_i v_i u_i = \Tr \bB} = \Tr \left(\bB \right) \bu  \ .
\end{eqnarray*}
The remaining eigenvectors can be found by considering the vectors from the orthogonal subspace of $\bv$:
${\bf v}_1, {\bf v}_2, {\bf v}_3, \dots, {\bf v}_{N-1}$. 
\begin{eqnarray*}
\bB  {\bf v}_i = \bu \otimes \underbrace{\bv^T {\bf v}_i}_{0} = 0  \ ,
\end{eqnarray*}
where $i=1,\dots,N-1$. Hence these vectors are eigenvectors with eigenvalues equal to $0$. 

\subsection{About the product of matrices with different ranks \label{lemma:productrank}}

\begin{enumerate}
    \item $\rank \left(\bA \bB\right) \le \rank \left(\bA\right) $
    \item $\rank \left( \bA \bB \right)= \rank \left(\bA \right)$ if $\bB$ is a non-singular matrix \\
\end{enumerate} 

\noindent
{\bf Proof}:
\begin{enumerate}
    \item We recall that the rank is the dimension of the range [$\mathcal{R}()$]:
    \begin{eqnarray*}
\rank \left( \bA \right) = \dim\left[ \mathcal{R} \left( \bA \right)\right].
    \end{eqnarray*}
    Hence, $\rank \left(\bA \bB\right) \le \rank \left(\bA\right)$ can be written alternatively as 
    \begin{eqnarray*}
       \dim\left[ \mathcal{R} \left( \bA \bB \right)\right] \le  \dim\left[ \mathcal{R} \left( \bA \right)\right] \ .
    \end{eqnarray*}
    Therefore, it is enough to show that 
    \begin{eqnarray}
    \mathcal{R} \left( \bA \bB \right) \in \mathcal{R} \left( \bA \right) \ . \label{rabinra}
    \end{eqnarray}
    Let us consider the vector $\bu \in \mathcal{R} \left( \bA \bB \right)$,
    \begin{eqnarray*}
    \bu = \bA \underbrace{\bB \bv}_\bx \ .
    \end{eqnarray*}
    Hence, $\bu \in \mathcal{R} \left( \bA \right)$ as well, which satisfies the first lemma.
    \item Let us consider the matrix $\bA \bB \bB^{-1}$, where the following relations can be written for the rank using the first lemma:
    \begin{eqnarray*}
    \rank\left( \bA \right) =  \rank\left( \bA \bB \bB^{-1} \right) \le \rank\left( \bA \bB  \right) \le \rank\left( \bA \right) \ .  
    \end{eqnarray*}
    This can only be true if
    \begin{eqnarray*}
    \rank\left( \bA \bB \right) = \rank\left( \bA \right) \ .
    \end{eqnarray*}
\end{enumerate}

\subsection{About the determinant of the linear combination of the identity matrix and rank-one matrix
\label{lemma:detrank}}
\begin{align*}
\det\left( \bI + a \bB \right) = 1+ a \Tr \bB \ , \nr \label{eq:lincomdet}
\end{align*}
where $\rank \left(\bB \right)=1$. \\

\noindent
{\bf Proof}:
Let us diagonalise the matrix $\bB$,
\begin{eqnarray*}
\bB = \bL \bB_d \bL^{-1} ,
\end{eqnarray*}
where $\bB_d$ is a diagonal matrix with eigenvalues equal to its diagonal entries. According to Lemma \ref{lemma:eigvalrank1}, one of the eigenvalues is $\Tr{\bB}$ and the rest are equal to zero. Applying the similarity transformation to identity operator in the determinant, $\bd{L}\bd{I}\bd{L}^{-1}$, we have
\begin{eqnarray*}
\det\left[ \bL \left(\bI + a \bB_d \right) \bL^{-1} \right]= \underbrace{\det\left( \bL \bL^{-1} \right)}_1 \underbrace{\det\left( \bI + a \bB_d \right)}_{1+ a \Tr \bB} = 1+ a \Tr \bB \ ,
\end{eqnarray*}
which proves the lemma.
\subsection{About the inverse of a sum of matrices \label{lemma:suminv}}

\begin{eqnarray}
\left( \bA + \bB \right)^{-1}=\bA^{-1}-\frac{\bA^{-1} \bB \bA^{-1}}{1 +  \Tr{\bB \bA^{-1}}} \ ,
\label{lemma_asumbinv}
\end{eqnarray}
where $\rank \left(\bB \right)=1$. \\

\noindent
{\bf Proof}:
 Let us multiply \rref{lemma_asumbinv} with $(\bA + \bB)$ from the left, giving
\begin{eqnarray*}
\bI=\left( \bA + \bB \right)\left( \bA^{-1} - \frac{\bA^{-1} \bB \bA^{-1}}{1 + \Tr{\bB \bA^{-1}}}\right) \ ,
\end{eqnarray*}
Expanding the expression on right hand side and multiplying it with $1 + \Tr{\bB \bA^{-1}}$, we see that
\begin{eqnarray}
\bB \bA^{-1} \bB \bA^{-1} = \Tr \left( \bB \bA^{-1} \right) \bB \bA^{-1} \ . \label{invlemma}
\end{eqnarray}
Equation (\ref{invlemma}) can be proven using Lemmas \ref{lemma:eigvalrank1} and \ref{lemma:productrank}.  Firstly, as $\rank \left(\bB \right)=1$, we can write $\bd{B}$ as a direct product of two vectors,
\begin{eqnarray*}
\bB &=& \bu \otimes \bv^{T} \ .
\end{eqnarray*}
Hence, 
\begin{eqnarray}
\bB \bA^{-1} \bB = \bu \otimes \underbrace{\bv^{T} \bA^{-1} \bu}_{\beta} \otimes \bv^{T} = \beta \bB \ . \label{expressingBAinvB}
\end{eqnarray}
The scalar $\beta$ can be expressed in terms of the trace,
\begin{eqnarray}
\beta = \sum_{ij} v_i \left(A^{-1} \right)_{ij} u_j = \sum_{ij} \underbrace{u_j v_i}_{B_{ji}} \left(A^{-1} \right)_{ij} = \Tr \left( \bB \bA^{-1} \right) \ . \label{expressingbeta}
\end{eqnarray}
Using \rrefsa{expressingBAinvB} and \rrefsb{expressingbeta}, \rref{invlemma} is satisfied, which also satisfies this lemma.

We can also consier matrices of the form $\un{\bd{B}} = \bd{I} \otimes \bd{B}$, where $\rank \left( \bI \otimes \bB\right)>1$ and $\rank \left( \bB\right)=1$. Then, inverse can be calculated in the following way:
\begin{align*}
    &\lrr{\un{\bd{A}} + \un{\bd{B}}}^{-1} = \left( \bI \otimes \bA +\bI \otimes \bB \right)^{-1}= \bI^{-1} \otimes \underbrace{\left( \bA + \bB \right)^{-1}}_{\rref{lemma_asumbinv}} = \bI \otimes  \bA^{-1}-\frac{\bI \otimes \left(\bA^{-1} \bB \bA^{-1} \right) }{1 +  \Tr{\bB \bA^{-1}}} \\
    \implies &\lrr{\un{\bd{A}} + \un{\bd{B}}}^{-1} = \un{\bA}^{-1}-\frac{\left(\un{\bA}^{-1} \un{\bB} \un{\bA}^{-1} \right) }{1 +  \Tr{\bB \bA^{-1}}} \ . \nr \label{lemma:asumbinvunder}
\end{align*}

\subsection{About the determinant of a sum of three matrices} \label{lemma:rank2det}

\begin{align*}
    |\bd{G} + \bd{H}_1 + \bd{H}_2| &= |\bd{G}| \lrr{1 + \Tr{\bd{H}_1\bd{G}^{-1}} 
    + \Tr{\bd{H}_2\bd{G}^{-1}} 
    + \Tr{\bd{H}_1\bd{G}^{-1}}\Tr{\bd{H}_2\bd{G}^{-1}} 
    - \Tr{\bd{H}_1\bd{G}^{-1}\bd{H}_2\bd{G}^{-1}}} \nr \label{eq:rank2det} \ ,
\end{align*}
where rank($\bd{H}_1$) = rank($\bd{H}_2$) = 1. \\

\noindent
{\bf Proof}: We can write

\begin{align*}
    |\bd{G} + \bd{H}_1 + \bd{H}_2| &= |\bd{C}_2 +  \bd{H}_2| \\
    &=|\bd{C}_2\underbrace{\bd{C}_2^{-1}\bd{C}_2}_{\bd{I}} + \bd{H}_2\underbrace{\bd{C}_2^{-1}\bd{C}_2}_{\bd{I}}| \\
    &= |\underbrace{\bd{C}_2\bd{C}_2^{-1}}_{\bd{I}} + \bd{H}_2\bd{C}_2^{-1}||\bd{C}_2| \\
    &= |\bd{C}_2||\bd{I} + \bd{H}_2\bd{C}_2^{-1}| \ .
\end{align*}
We can then apply Lemma \ref{lemma:detrank}, giving

\begin{align*}
    |\bd{G} + \bd{H}_1 + \bd{H}_2| =|\bd{C}_2|\lrr{1 + \Tr{\bd{H}_2\bd{C}_2^{-1}}} \ .
\end{align*}
From here, we use the fact that $\bd{C}_2 = \bd{G}+\bd{H}_1$, meaning that

\begin{align*}
    |\bd{C}_2| = |\bd{G}+\bd{H}_1| = |\bd{G}\bd{G}^{-1}\bd{G}+\bd{H}_1\bd{G}^{-1}\bd{G}| \underbrace{=}_{\ref{lemma:detrank}} |\bd{G}|\lrr{1 + \Tr{\bd{H}_1\bd{G}^{-1}}} \ ,
\end{align*}
and
\begin{align*}
    \bd{C}_2^{-1} = \lrr{\bd{G}+\bd{H}_1}^{-1} \underbrace{=}_{\rref{lemma_asumbinv}} \bd{G}^{-1} - \frac{\bd{G}^{-1}\bd{H}_1\bd{G}^{-1}}{1 + \Tr{\bd{H}_1\bd{G}^{-1}}} \ .
\end{align*}
\noindent
Overall, we now have
\begin{align*}
    |\bd{G} + \bd{H}_1 + \bd{H}_2| &= |\bd{G}|\lrr{1 + \Tr{\bd{H}_1\bd{G}^{-1}}}\lrr{ 1+ \Tr{ \bd{H}_2 \lrr{\bd{G}^{-1} - \frac{\bd{G}^{-1}\bd{H}_1\bd{G}^{-1}}{1 + \Tr{\bd{H}_1\bd{G}^{-1}}}}}} \\
    &= |\bd{G}|\lrr{1 + \Tr{\bd{H}_1\bd{G}^{-1}}}\lrr{1+ \Tr{\bd{H}_2\bd{G}^{-1}}-\frac{\Tr{\bd{H}_2\bd{G}^{-1}\bd{H}_1\bd{G}^{-1}}}{1+\Tr{\bd{H}_1\bd{G}^{-1}}}} \ ,
\end{align*}
where we have removed the $1+\Tr{\bd{H}_1\bd{G}^{-1}}$ term from the trace in the second set of round brackets as it is a scalar term. Then, by expanding the brackets and using the fact that $\Tr{\bd{H}_2\bd{G}^{-1}\bd{H}_1\bd{G}^{-1}} = \Tr{\bd{H}_1\bd{G}^{-1}\bd{H}_2\bd{G}^{-1}}$, we have
\begin{align*}
    |\bd{G} + \bd{H}_1 + \bd{H}_2|  &= |\bd{G}|\lrr{1 + \Tr{\bd{H}_1\bd{G}^{-1}}}  \\
    &+|\bd{G}|\lrr{\Tr{\bd{H}_2\bd{G}^{-1}}}   \\
    &+|\bd{G}|\lrr{\Tr{\bd{H}_1\bd{G}^{-1}}\Tr{\bd{H}_2\bd{G}^{-1}}} \\ 
    &-|\bd{G}|\lrr{\frac{\Tr{\bd{H}_1\bd{G}^{-1}\bd{H}_2\bd{G}^{-1}} \lrr{1 + \Tr{\bd{H}_1\bd{G}^{-1}}}}{1 + \Tr{\bd{H}_1\bd{G}^{-1}} } }  \ ,  
\end{align*}
and so 
\begin{align*}
    |\bd{G} + \bd{H}_1 + \bd{H}_2| = |\bd{G}| \lrs{1 + \Tr{\bd{H}_1\bd{G}^{-1}} 
    + \Tr{\bd{H}_2\bd{G}^{-1}} 
    + \Tr{\bd{H}_1\bd{G}^{-1}}\Tr{\bd{H}_2\bd{G}^{-1}} 
    - \Tr{\bd{H}_1\bd{G}^{-1}\bd{H}_2\bd{G}^{-1}}}
\end{align*}
which is the required result. \\

\subsection{About the derivative of an inverse matrix}
\begin{eqnarray}
{\pdv{}{x}} {\bf M}^{-1} &=& -{\bf M}^{-1} \left( {\pdv{}{x}} {\bf M} \right){\bf M}^{-1} \ . \label{invder}
\end{eqnarray}

\noindent
{\bf Proof}: Let us start with
\begin{eqnarray*}
{\bf I} &=& {\bf M}^{-1} {\bf M} \ ,
\end{eqnarray*}
and take the derivative of both sides with respect to $x$,
\begin{align}
\implies {\bf 0}={\pdv{}{x}}{\bf I} = {\pdv{}{x}} \left({\bf M}^{-1} {\bf M} \right)=  {\pdv{}{x}} \left({\bf M}^{-1}  \right) {\bf M} +  {\bf M}^{-1} {\pdv{}{x}} \left( {\bf M} \right) \ . \nr \label{lemma:derint}
\end{align} 
By multiplying \rref{lemma:derint} with ${\bf M}^{-1}$ from the right, we can obtain back \rref{invder}.

\subsection{About the derivative of the determinant}

\begin{eqnarray}
{\pdv{}{x}} \det\left( {\bf M} \right) &=&  \det\left( {\bf M} \right) \Tr \left( {\bf M}^{-1} {\pdv{}{x}} {\bf M}\right) \\ . \label{detder}
\end{eqnarray}

\noindent
{\bf Proof}: Using Jacobi's formula,
\begin{eqnarray*}
{\pdv{}{x}} \det\left( {\bf M} \right) &=&   \Tr \left( \adj\left({\bf M}\right) {\pdv{}{x}} {\bf M}\right) \ ,
\end{eqnarray*}
and the following expression for the inverse of the matrix $\bd{M}$,
\begin{eqnarray*}
{\bf M}^{-1} &=& \frac{\adj\left({\bf M}\right)}{\det\left( {\bf M} \right)} \ ,
\end{eqnarray*}
we obtain
\begin{eqnarray*}
{\pdv{}{x}} \det\left( {\bf M} \right) &=&   \Tr \left( \det\left( {\bf M} \right) {\bf M}^{-1}  {\pdv{}{x}} {\bf M}\right) \ .
\end{eqnarray*}
Then, by factorising out the determinant, the lemma is proven.

Note that if $\bd{M}$ is such that
\begin{eqnarray*}
{\pdv{}{x}} \left( {\bf M} \right) = {\bf M_{x}}, \quad
{\pdv{}{y}} \left( {\bf M} \right) = {\bf M_{y}}, \quad
{\pdv{}{x}{y}} \left( {\bf M} \right) = {\bf M_{x, y}} \ ,
\end{eqnarray*} then
\begin{eqnarray*}
\pdv{}{x}{y} \det\left( {\bf M} \right) = \det\left( {\bf M} \right) \left[ \Tr \left( {\bf M}^{-1} {\bf M_{x}}\right) \Tr \left( {\bf M}^{-1} {\bf M_{y}}\right)
 - \Tr \left( {\bf M}^{-1} {\bf M_{x}} {\bf M}^{-1} {\bf M_{y}}\right) + \Tr \left( {\bf M}^{-1}
{\bf M_{x, y}} \right) \right] \ .
\end{eqnarray*}

\subsection{About the evaluation of a one-dimensional Gaussian integral
\label{lemma:onedgauss}}
\begin{align}
\int_{-\infty}^{\infty} dx \, e^{-ax^2} = \sqrt{\frac{\pi}{a}}  \label{onedg}
\end{align}

\noindent
{\bf Proof}: We begin with the following integral,
\begin{align*}
\left( \int_{-\infty}^{\infty} dx \, e^{-ax^2} \right)^2 = \int_{-\infty}^{\infty} dx \, e^{-ax^2} \int_{-\infty}^{\infty} dy \, e^{-ay^2} = \int_{-\infty}^{\infty} \int_{-\infty}^{\infty} dx \, dy \,  e^{-a(x^2+y^2)} \, .
\end{align*}
The integral can be transformed into polar coordinates, with the usual definition $r^2=x^2+y^2$ and $\theta \in [ 0,2\pi), \ r \in [0 ,\infty )$. Then,
\begin{align*}
    \int_{-\infty}^{\infty} \int_{-\infty}^{\infty} dx \, dy \,  e^{-a(x^2+y^2)} = \int_0^{2\pi} d \theta \int_0^\infty dr \, r \, e^{-ar^2} = 2\pi \int_0^\infty dr \, r \, e^{-ar^2} \ .
\end{align*}
We change the integration variable again by defining $s=-r^2$, and so $dr=-\frac{ds}{2r}$. Note that the non-zero integration limit also changes to $-\infty$, as $r \rightarrow \infty \implies \ s \rightarrow -\infty$. Making these changes, 
\begin{align*}
    2\pi \int_0^\infty dr \, r \, e^{-ar^2} = - \pi \int_0^{-\infty} ds \, e^{-as} = -\pi \left[ \frac{1}{a} e^{as} \right]_{0}^{-\infty} = -\frac{\pi}{a}\left( e^{-\infty} -e^{0} \right) = \frac{\pi}{a} \ .
\end{align*}
Taking the square root of this gives the initial expression, proving the lemma.

\subsection{About the evaluation of a multidimensional Gaussian integral
\label{lemma:multidgauss}}
\begin{align}
    \int_{-\infty}^{\infty} d^n \mx{x} \, \exp\left[ -\mx{x}^T\mx{M}\mx{x} +\mx{y}^T \mx{x} \right] = \frac{\pi^{n/2}}{| \mx{M} |^{1/2} } \exp \left[ \frac{1}{4} \mx{y}^T\mx{M}^{-1} \mx{y} \right] \label{mdimgauss} \ ,
\end{align}
where  $ \mx{x},\mx{y} \in \mathbb{R}^n, \  \mx{M} \in \mathbb{R}^{n \times n},\  \mx{M}^T=\mx{M} $. The integral is convergent only if $\mx{M}$ is positive-definite i.e. $\mx{v}^T\mx{M}\mx{v}>0 , \ \forall \ \mx{v}\in \mathbb{R}^n $. \\

\noindent
{\bf Proof}: The first step is to eliminate the linear $\mx{y}^T \mx{x}$ term which is achieved by completing the square,
\begin{align*}
     \int_{-\infty}^{\infty} d^n \mx{x} \, \exp\left[ -\mx{x}^T\mx{M}\mx{x} +\mx{y}^T \mx{x} \right] &= 
     \int_{-\infty}^{\infty} d^n \mx{x} \, \exp\left[ -\left(\mx{x}^T-\frac{1}{2}\mx{y}^T\mx{M}^{-1} \right)\mx{M} \left( \mx{x}-\frac{1}{2}\mx{M}^{-1}\mx{y} \right)  \right] \exp\left[ \frac{1}{4} \mx{y}^T\mx{M}^{-1}\mx{y}  \right] \\ 
     &= 
     \exp\left[ \frac{1}{4} \mx{y}^T\mx{M}^{-1}\mx{y}  \right] \int_{-\infty}^{\infty} d^n \tilde{\mx{x}} \, \exp\left[ -\tilde{\mx{x}}^T \mx{M}  \tilde{\mx{x}}\right]  \ ,
\end{align*}
where we define $\tilde{\mx{x}} = \mx{x}-\frac{1}{2}\mx{M}^{-1}\mx{y}  $. Since this only introduces a shift in the integration variable, it does not change the integration measure (i.e. no Jacobian needs to be included).  Real, symmetric matrices like $\mx{M}$ are diagonalisable by orthogonal matrices. Let $\mx{Q}$ be an orthogonal matrix (i.e. $\mx{Q}^T = \mx{Q}^{-1}$), with $\det(\mx{Q})=1$ such that $\bd{Q} $ diagonalises $\mx{M}$. Then,
\begin{align*}
    (\mx{Q} \mx{M} \mx{Q}^T)_{ij} = m_i \delta_{ij} \ ,
\end{align*}
where $m_i$ is the $i^{\ur{th}}$ eigenvalue of $\mx{M}$. The above integral can be rewritten by inserting the identity matrix in the form $\mx{I}=\mx{Q}^T\mx{Q}$, giving
\begin{align*}
     & \exp\left[ \frac{1}{4} \mx{y}^T\mx{M}^{-1}\mx{y}  \right] \int_{-\infty}^{\infty} d^n \tilde{\mx{x}} \, \exp\left[ -\tilde{\mx{x}}^T \mx{M}  \tilde{\mx{x}}\right] =    
    \exp\left[ \frac{1}{4} \mx{y}^T\mx{M}^{-1}\mx{y}  \right] \int_{-\infty}^{\infty} d^n \tilde{\mx{x}} \,
     \exp\left[ -\tilde{\mx{x}}^T \mx{Q}^T\mx{Q} \mx{M} \mx{Q}^T\mx{Q} \tilde{\mx{x}}\right] \ .
\end{align*}
Let us define another variable $\mx{z} = \mx{Q} \tilde{\mx{x}} $. The integration measure changes by the determinant of the Jacobian matrix, $d\mx{\tilde{x}}=\det(\mx{J})d\mx{z}$, where the Jacobian determinant equals 1;
\begin{align*}
    \mx{J}_{ij} = \frac{\partial \tilde{\mx{x}}_i }{\partial \mx{z}_j} =
    \frac{\partial \sum_k \mx{Q}_{ki} \mx{z}_k}{\partial \mx{z}_j} = \mx{Q}_{ji} \implies  \det(\mx{J}) = \det(\mx{Q}^T)=1
\end{align*}
Now, we rewrite the integral in terms of the new variables and the diagonalised $\mx{M}$ matrix, writing the summation explicitly. This gives
\begin{align*}
    &\exp\left[ \frac{1}{4} \mx{y}^T\mx{M}^{-1}\mx{y}  \right] \int_{-\infty}^{\infty} d^n \mx{z} \,
     \exp\left[ - \sum_{ij}^n \mx{z}_i m_i \delta_{ij}  \mx{z}_j \right] = 
     \exp\left[ \frac{1}{4} \mx{y}^T\mx{M}^{-1}\mx{y}  \right] \int_{-\infty}^{\infty} d^n \mx{z} \,
     \exp\left[ - \sum_{i}  m_i \mx{z}^2_i \right] \ .
\end{align*}
The sum in the exponent can then be written as a product of exponential functions,
\begin{align*}
    &\exp\left[ \frac{1}{4} \mx{y}^T\mx{M}^{-1}\mx{y}  \right] \int_{-\infty}^{\infty} d^n \mx{z} \exp\left[ - \sum_{i}  m_i \mx{z}^2_i \right] =
    \exp\left[ \frac{1}{4} \mx{y}^T\mx{M}^{-1}\mx{y}  \right]  \prod_i^n   \int_{-\infty}^{\infty} d \mx{z}_i \exp\left[ -   m_i \mx{z}^2_i \right] \ .
\end{align*}
Each one dimensional integral can be evaluated using Lemma (\ref{lemma:onedgauss}) yielding
\begin{align}
    \exp\left[ \frac{1}{4} \mx{y}^T\mx{M}^{-1}\mx{y}  \right]  \prod_i^n  \frac{\pi^{1/2}}{m_i^{1/2}} = \exp\left[ \frac{1}{4} \mx{y}^T\mx{M}^{-1}\mx{y}  \right]   \frac{\pi^{n/2}}{\det(\mx{M})^{1/2}} \ ,
\end{align} \label{multdigaussresult}
where in the last step we have used the fact that for a diagonal matrix, the product of the diagonal elements is its determinant. This completes the proof of lemma \ref{lemma:multidgauss}. 

Note that if we have regular ECG functions (i.e. no shift vectors $\textbf{s}_k$, \, $\textbf{s}_l$) then in Lemma \ref{lemma:multidgauss}, the vector $\textbf{y}$ would be equal to zero, and so $\tilde{\textbf{x}} = \textbf{x}$. The result of Lemma \rref{lemma:multidgauss} then becomes
\begin{align}
    \int_{-\infty}^{+\infty} \ur{d}^n\bx \, \exp[-\bx^T\bM\bx] = \frac{\pi^{n/2}}{\det(\bM)^{\frac{1}{2}}} \label{nonfloatmultidgauss} \ . 
\end{align}

\section{The Product of two ECGs} \label{appendix:prodECG}
\noindent
This appendix shows the derivation of \rref{eq:ProductECG}. If we have the product of two basis functions, say $\phi_k\phi_l$, we can write this product as
\begin{align*}
    \phi_k\phi_l = \exp[-(\textbf{r}-\textbf{s}_k)^T\textbf{\underline{A}}_k(\textbf{r}-\textbf{s}_k)]\exp[-(\textbf{r}-\textbf{s}_l)^T\textbf{\underline{A}}_l(\textbf{r}-\textbf{s}_l)] \ .
\end{align*}
We then merge the two arguments of the exponentials, giving
\begin{align*}
    \phi_k\phi_l = \exp[-(\textbf{r}-\textbf{s}_k)^T\textbf{\underline{A}}_k(\textbf{r}-\textbf{s}_k)-(\textbf{r}-\textbf{s}_l)^T\textbf{\underline{A}}_l(\textbf{r}-\textbf{s}_l] \ .
\end{align*}
Expanding out the brackets of the argument then gives
\begin{align*}
   \phi_k\phi_l = \exp[\underbrace{-\textbf{r}^T\textbf{\underline{A}}_k\textbf{r} -\textbf{r}^T\textbf{\underline{A}}_l\textbf{r}}_{\textbf{r}^T(\textbf{\underline{A}}_k + \textbf{\underline{A}}_l)\textbf{r}}
    +\underbrace{{\textbf{s}_k^T\textbf{\underline{A}}_k\textbf{r}
    +\textbf{r}^T\textbf{{\underline{A}}}_k\textbf{s}_k}
    +{\textbf{s}_l^T\textbf{\underline{A}}_l\textbf{r}
    +\textbf{r}^T\textbf{{\underline{A}}}_l\textbf{s}_l}}_{2(\textbf{s}_k^T\textbf{\underline{A}}_k+\textbf{s}_l^T\textbf{\underline{A}}_l)}
    -\textbf{s}_k^T\textbf{\underline{A}}_k\textbf{s}_k
    -\textbf{s}_l^T\textbf{\underline{A}}_l\textbf{s}_l] \ .
\end{align*}
 The term $\textbf{s}_k^T\textbf{\underline{A}}_k\textbf{r}$ is a scalar and so is equal to its own transpose, that is
\begin{align*}
    \textbf{s}_k^T\textbf{\underline{A}}_k\textbf{r} = (\textbf{s}_k^T\textbf{\underline{A}}_k\textbf{r})^T = \textbf{r}^T\textbf{\underline{A}}_k\textbf{s}_k \ ,
\end{align*}
where we have used the fact that $\textbf{\underline{A}}_k$ is symmetric. The same is true for the $\textbf{s}_l^T\textbf{\underline{A}}_l\textbf{r}$ term. Using this fact and collecting terms gives
\begin{align*}
    \phi_k\phi_l = \exp[-\textbf{r}^T(\underbrace{\textbf{\underline{A}}_k + \textbf{\underline{A}}_l}_{\textbf{\underline{A}}_{kl}})\textbf{r} + 2(\underbrace{\textbf{s}_k^T\textbf{\underline{A}}_k+\textbf{s}_l^T\textbf{\underline{A}}_l}_{\textbf{e}_{kl}^T})\textbf{r}-(\underbrace{\textbf{s}_k^T\textbf{\underline{A}}_k\textbf{s}_k
    +\textbf{s}_l^T\textbf{\underline{A}}_l\textbf{s}_l}_{\eta_{kl}})] \ .
\end{align*}
Using the notations in Sec.  \ref{sec:notations}, the exponential can be written more concisely as
\begin{align*}
    \phi_k\phi_l = \exp[-\eta_{kl}]\exp[-\textbf{r}^T\textbf{\underline{A}}_{kl}\textbf{r} +2\textbf{e}^T_{kl}\textbf{r}] \ , 
\end{align*}
which is \rref{eq:ProductECG}. This form appears often when determining matrix elements using ECG basis functions, and is the form used in lemma \ref{lemma:multidgauss}.

\newpage

\section{Derivatives of ECGs} \label{appendix:devECG}
\noindent
We often find that we have to determine expressions such as $\nabla\phi_k$ or $\nabla^2\phi_k$. This appendix shows the details for the derivations of Eqs. (\ref{eq:nabphi}) and (\ref{eq:NabSqPhi}). 
We begin with
\begin{align*}
    \nabla\phi_k &= \nabla\exp[-(\textbf{r}-\textbf{s}_k)^T\textbf{\underline{A}}_k(\textbf{r}-\textbf{s}_k)] \\
    &= -\phi_k\nabla[(\textbf{r}-\textbf{s}_k)^T\textbf{\underline{A}}_k(\textbf{r}-\textbf{s}_k)] \\
    & = -\phi_k\nabla[\textbf{r}^T\ubA_k\textbf{r}-\textbf{r}^T\ubA_k\textbf{s}_k-\textbf{s}_k^T\ubA_k\textbf{r}+\textbf{s}_k^T\ubA\textbf{s}_k] \nr \label{midnabphi} \ .
\end{align*}
We then proceed to find $\nabla$ for all of the individual terms of \rref{midnabphi}, noting that we must use the product rule for the first term, that the second and third terms are equal, and that the final term is independent of $\textbf{r}$. As such, we then have
\begin{align}
    \nabla\phi_k = -2\ubA_k(\textbf{r} - \textbf{s}_k)\phi_k \ . 
\end{align}
To find $\nabla^2\phi_k$, we first note that the gradient of a scalar is a vector, and the divergence of a vector is a scalar. Hence, we need to write the Laplacian operator as $\nabla^2 = \nabla^T\nabla$. Then,
\begin{align*}
    \nabla^2\phi_k &= \nabla^T(\nabla\phi_k) = \nabla^T[-2\ubA_k(\textbf{r}-\textbf{s}_k)\phi_k] \\
    &= \nabla^T[-2\ubA_k\textbf{r}\phi_k + 2\ubA_k\textbf{s}_k\phi_k] = -2\nabla^T(\ubA_k\textbf{r}\phi_k) + 2\nabla^T\ubA_k\textbf{s}_k\phi_k \ .
\end{align*}
The term $\ubA_k\textbf{r}$ is a column vector that is a function of the Cartesian coordinates for each particle. We also know that $\phi_k$ is a function of \textbf{r} too, and so we must use the product rule for the $\ubA_k\textbf{r}\phi_k$ term. The term $\ubA_k\textbf{s}_k$ is also a column vector, but it is independent of \textbf{r} and so is not operated upon by $\nabla^T$. The term $\nabla^T\ubA_k\bd{s}_k$ is a scalar, and so equal to its own transpose, meaning that $(\nabla^T\ubA_k\bd{s}_k)\phi_k = (\bd{s}_k^T\ubA_k\nabla)\phi_k$ were we have used the fact that $\bd{A}_k$ is symmetric. This gives
\begin{align*}
    \nabla^2\phi_k = -2(\nabla^T\ubA_k\textbf{r})\phi_k - 2(\nabla\phi_k)^T\ubA_k\textbf{r} +2(\textbf{s}_k^T\ubA_k\nabla)\phi_k \ ,
\end{align*}
where in the first term $\nabla^T$ is operating on the column vector $\ubA_k\bd{r}$ and in the second term it is acting on the scalar $\phi_k$. We note also that $\nabla^T\phi_k = (\nabla\phi_k)^T$. This can be more clearly denoted in the second term by using the fact that $\nabla^T\ubA_k\bd{r}$ is a scalar, and so is equal to  $\bd{r}^T\ubA_k\nabla$. Then, we have
\begin{align*}
    \nabla^2\phi_k = -2(\nabla^T\ubA_k\textbf{r})\phi_k - 2\bd{r}^T\ubA_k(\nabla\phi_k) +2\textbf{s}_k^T\ubA_k(\nabla\phi_k) \ .
\end{align*}
where all three terms still represent scalars. 

For the first term, we should first note that the matrix $\bA_k$ is a $N_{\ur{p}} \times N_{\ur{p}}$ matrix, and $\textbf{r}$ is a column vector with $3N_{\ur{p}}$ entries, where $N_{\ur{p}}$ is the number of particles. These can be written explicitly as
\begin{align*} 
    \bA_k =  
    \begin{pmatrix*} 
     a_{11} & a_{12} & \cdots & a_{1N_{\ur{p}}} \\
     a_{21} & a_{22} & \cdots & a_{2N_{\ur{p}}} \\
     \vdots & \vdots & \ddots & \vdots \\
     a_{N_{\ur{p}}1} & a_{N_{\ur{p}}2} & \cdots & a_{N_{\ur{p}}N_{\ur{p}}}
    \end{pmatrix*}  \ ,  \\ 
\end{align*}
and
\begin{align*}
    \textbf{r} = 
    \begin{pmatrix*}
     x_1 \\
     x_2 \\
     \vdots \\
     x_{N_{\ur{p}}} \\
     y_1 \\
     y_2 \\
     \vdots \\
     y_{N_{\ur{p}}} \\
     z_1 \\
     z_2 \\
     \vdots \\
     z_{N_{\ur{p}}} 
    \end{pmatrix*} \ .  
\end{align*}
The vector $\textbf{r}$ describes the $x,y,z$ components of each of the $N_{\ur{p}}$ particles. The nabla operator therefore must take the form
\begin{align*}
    \nabla = 
    \begin{pmatrix*}
     \partial_{x_{1}} \\
     \vdots \\
     \partial_{x_{N_{\ur{p}}}} \\
     \partial_{y_{1}} \\
     \vdots \\
     \partial_{y_{N_{\ur{p}}}} \\\\
     \partial_{z_{1}} \\
     \vdots \\
     \partial_{z_{N_{\ur{p}}}} \\
     \end{pmatrix*} \ .
\end{align*}
\noindent
For reference, we also have that the column vector $\textbf{s}_k$ can be written for $N_{\ur{p}}$ particles as
\begin{align*}
    \textbf{s}_k =
    \begin{pmatrix*}
    (s_k)_{x_{1}} \\
    (s_k)_{x_{2}} \\
    \vdots \\
    (s_k)_{x_{N_{\ur{p}}}} \\
    (s_k)_{y_{1}} \\
    (s_k)_{y_{2}} \\
    \vdots \\
    (s_k)_{y_{N_{\ur{p}}}} \\
    (s_k)_{z_{1}} \\
    (s_k)_{z_{2}} \\
    \vdots \\
    (s_k)_{z_{N_{\ur{p}}}}
    \end{pmatrix*} \ .
\end{align*}
For simplicity, we will limit the number of particles to 2. We can then see that
\begin{align}
    \ubA_k\textbf{r} = 
    \begin{pmatrix*}
     a_{11} & a_{12} & 0 & 0 & 0 & 0 \\
     a_{21} & a_{22} & 0 & 0 & 0 & 0 \\
     0 & 0 & a_{11} & a_{12} & 0 & 0 \\
     0 & 0 & a_{21} & a_{22} & 0 & 0 \\
     0 & 0 & 0 & 0 & a_{11} & a_{12} \\
     0 & 0 & 0 & 0 & a_{21} & a_{22} \\
    \end{pmatrix*}
    \begin{pmatrix*}
    x_1 \\
    x_2 \\
    y_1 \\
    y_2 \\
    z_1 \\
    z_2 \\
    \end{pmatrix*}
    =
    \begin{pmatrix*}
    a_{11}x_1 + a_{12}x_2 \\
    a_{21}x_1 + a_{22}x_2 \\
    a_{11}y_1 + a_{12}y_2 \\
    a_{21}y_1 + a_{22}y_2 \\
    a_{11}z_1 + a_{12}z_2 \\
    a_{21}z_1 + a_{22}z_2 \\ 
    \end{pmatrix*}  \ , \label{Arproduct}
\end{align}
where $\ubA_k = \bA \otimes \bI_3$. Applying the transpose of the nabla operator then yields
\begin{align*}
    \nabla^T(\ubA_k\textbf{r}) =
    \begin{pmatrix*}
    \partial_{x_{1}}(a_{11}x_1 + a_{12}x_2) \\
    \partial_{x_{2}}(a_{21}x_1 + a_{22}x_2) \\
    \partial_{y_{1}}(a_{11}y_1 + a_{12}y_2) \\
    \partial_{y_{2}}(a_{21}y_1 + a_{22}y_2) \\
    \partial_{z_{1}}(a_{11}z_1 + a_{12}z_2) \\
    \partial_{z_{2}}(a_{21}z_1 + a_{22}z_2) \\
    \end{pmatrix*}
    = (\underbrace{a_{11} + a_{22}}_{\Tr(\bA_k)}) +(a_{11} + a_{22})+(a_{11} + a_{22}) \ , 
\end{align*}
and so 
\begin{align}
    \nabla^T(\ubA_k\textbf{r}) = 3\Tr(\bA_k) \ . \label{eq:trannabla}
\end{align}
Equation \ref{eq:trannabla} holds for any number of particles. So, overall we have
\begin{align*}
    \nabla^2\phi_k = -2\lrs{3\Tr(\bA_k)\phi_k + (\bd{r}^T\ubA_k\nabla)\phi_k - (\textbf{s}_k^T\ubA_k\nabla)\phi_k}.
\end{align*}
Using \rref{eq:nabphi} and then taking out a factor of $\phi_k$ yields 
\begin{align*}
    \nabla^2\phi_k = -2\lrs{3\Tr(\bA_k) -2 \bd{r}^T\ubA_k\ubA_k(\textbf{r}-\textbf{s}_k) +2 \textbf{s}_k^T\ubA_k\ubA_k(\textbf{r}-\textbf{s}_k)}\phi_k \ .
\end{align*}
Finally, we multiply out the $-$2 coefficient, and then take out a factor of $\ubA_k\ubA_k(\bd{r}-\bd{s}_k)$ to the right (ignoring the trace term), resulting in
\begin{align*}
    \nabla^2\phi_k = 4\lrs{(\bd{r}-\bd{s}_k)^T\ubA_k\ubA_k(\bd{r}-\bd{s}_k) - 6\Tr{\bd{A}_k}}\phi_k \ , 
\end{align*}
which is \rref{eq:NabSqPhi}. This expression can then be used to find so-called `elementary integrals', which are integral expressions in terms of the vectors $\bd{r}$ and $\bd{s}_k$, and matrix $\ubA_k$
\newpage

\end{document}